\documentclass[11pt]{article}
\usepackage[utf8]{inputenc}
\usepackage{amsmath,amsfonts,amssymb}
\usepackage{amsthm}
\usepackage{latexsym}
\usepackage[noadjust]{cite}
\usepackage{graphicx}
\usepackage{xcolor}
\usepackage{dirtytalk}
\usepackage{mathtools}
\usepackage[margin=1in]{geometry}
\usepackage{thm-restate}
\usepackage{enumitem}
\usepackage[linesnumbered,ruled]{algorithm2e}
\usepackage[colorlinks=true, allcolors=blue]{hyperref}
\usepackage[nameinlink,capitalise]{cleveref}
\usepackage{tikz}

\newtheorem{theorem}{Theorem}[section]
\newtheorem{corollary}[theorem]{Corollary}
\newtheorem{lemma}[theorem]{Lemma}

\newtheorem{proposition}[theorem]{Proposition}
\newtheorem{claim}[theorem]{Claim}
\theoremstyle{definition}
\newtheorem{definition}[theorem]{Definition}

\newtheorem{fact}[theorem]{Fact}

\newtheorem{conjecture}[theorem]{Conjecture}

\newcommand{\F}{\mathbb{F}}

\newcommand{\E}{\mathbb{E}}

\newcommand{\one}{\mathbf 1}
\newcommand{\norm}[1]{\left\lVert #1\right\rVert}
\newcommand{\abs}[1]{\left\lvert #1\right\rvert}
\newcommand{\Pp}{\mathbb P}
\DeclareMathOperator{\Cov}{Cov}
\DeclareMathOperator{\Var}{Var}
\DeclareMathOperator{\diag}{diag}
\DeclareMathOperator{\dist}{dist}
\DeclareMathOperator{\tr}{tr}
\DeclareMathOperator{\Unif}{Unif}
\DeclareMathOperator{\cc}{cc}

\newcommand{\zo}{\{0, 1\}}

\newcommand{\eps}{\varepsilon}

\newcommand{\Supp}{\mathrm{Supp}}

\newcommand{\R}{\mathbb{R}}
\newcommand{\Z}{\mathbb{Z}}

\newif\ifdraft
\drafttrue

\title{
Deterministic Connectivity Sketches Under Edge Deletions}
\author{
    Sanjeev Khanna\thanks{Courant Institute, New York University. Supported in part by NSF award CCF-2625203 and AFOSR award FA9550-25-1-0107. Contact: \href{mailto:sanjeev.khanna@nyu.edu}{sanjeev.khanna@nyu.edu}.}
    \and
    Aaron Putterman\thanks{School of Engineering and Applied Sciences, Harvard University, Cambridge, Massachusetts, USA. Supported in part by a Jane Street Graduate Research Fellowship, the Simons Investigator Awards to Madhu Sudan and Salil Vadhan, and AFOSR award FA9550-25-1-0112. Email:\texttt{aputterman@g.harvard.edu}} 
    \and   
    Madhu Sudan\thanks{School of Engineering and Applied Sciences, Harvard University, Cambridge, Massachusetts, USA. Supported in part by a Simons Investigator Award, NSF Award CCF 2152413 and AFOSR award FA9550-25-1-0112. Email:\texttt{madhu@cs.harvard.edu}} 
    }
\date{\today}

\begin{document}
\maketitle

\begin{abstract}
How much information about a graph is needed to determine its connectivity after arbitrary valid edge deletions? This question is at the heart of designing deterministic dynamic streaming algorithms for graph connectivity. Towards answering this question, we study the deterministic one-way communication problem in which Alice holds a graph $G_A = (V, E_A)$ on $n$ vertices, Bob holds a graph $G_B = (V, E_B)$ with $E_B \subseteq E_A$, and Alice sends a message $M \in \zo^m$ that allows Bob to decide whether $G_A - G_B$ is connected. Randomized sketches solve this problem using $\tilde O(n)$ bits, whereas deterministic recovery of even one surviving edge requires $\Omega(n^2)$ bits.

We prove an upper bound of $m = \widetilde{O}(n^{7/4})$ bits and a lower bound of $m = \widetilde{\Omega}(n^{3/2})$ bits for deciding connectivity. Thus, connectivity admits a subquadratic deterministic representation despite the quadratic cost of edge recovery, but remains polynomially harder than its randomized counterpart. Our upper bound relies on a careful instantiation of recursive subsampling to estimate set densities and cut sizes, alongside linear sketches that separate zero cuts from sufficiently small nonzero cuts in graph differences. Our lower bound instead relies on positive semidefinite certificates for the (vector) chromatic number of a special family of graphs.

The upper bound immediately yields an $\widetilde{O}(n^{7/4})$-bit deterministic $n^2$-fault tolerant connectivity oracle, which reports the connected components of a graph $G$ after any valid set of edge deletions. We also obtain an $\tilde \Omega(n f^{1/4})$-bit lower bound for deterministic $f$-fault tolerant connectivity oracles for every $f \in [1,n^2]$.

Finally, we show that the separation between connectivity and edge recovery persists in deterministic single-pass dynamic streams with bounded update multiplicity. For every fixed $\eps \in (0,1/3)$, our algorithm reports the connected components of the final simple graph using $\widetilde{O}(n^{15/8 + 3\eps/8})$ bits (subquadratic space), provided each edge undergoes at most $n^{\eps}$ insertion or deletion updates in total. 
\vspace{3cm}

\textbf{AI Disclosure:} Several proofs in the paper were discovered entirely by AI. The authors decoded the proofs and wrote everything, except the appendix, from scratch.
\end{abstract}

\pagenumbering{gobble}
\pagebreak 

\tableofcontents

\pagebreak

\section{Introduction}
\pagenumbering{arabic}

\subsection{Background}

Can a graph be compressed deterministically while preserving its connectivity after every valid set of edge deletions? This question connects one-way communication, fault-tolerant data structures, and single-pass dynamic streaming. It also exposes a distinction between deciding connectivity and recovering the edges that witness it.

We study the following communication problem: Alice holds a simple undirected graph $G_A = (V, E_A)$ on $n$ labeled vertices, and Bob holds a graph $G_B = (V, E_B)$ on the same vertex set, with the promise that $E_B \subseteq E_A$. Alice sends a message $M \in \zo^m$, after which Bob must decide, without error, whether $G_A - G_B$ is connected. The message must work for every valid choice of $G_B$, with no bound on the number of deleted edges. Our goal is to determine the minimum worst-case message length $m$ as a function of $n$. We call this the \emph{deterministic one-way connectivity problem}.

This is a natural variation of the classical \emph{connectivity of the union} model, where there is no containment promise on $E_B$ and Bob must decide whether $G_A \cup G_B$ is connected. In this model, Alice can simply send a spanning forest of $G_A$, using $O(n \log(n))$ bits. We can immediately observe that this strategy does \emph{not} extend to our setting: edges omitted from the spanning forest may become essential for connectivity after Bob's deletions.

Before this work, the only bounds for this deletion problem were essentially trivial, with $\Omega(n \log(n))$ bit lower bounds and an $O(n^2)$ bit upper bound. The upper bound follows readily, by storing the entire graph. For the lower bound, it turns out that a connectivity protocol on $n+1$ vertices can recover the connected components of a graph on $n$ vertices, using the exact same message (albeit, with some extra postprocessing -- see \cref{clm:componentsToConnectivity}). $\Omega(n\log(n))$ bits then follows from the bit complexity of encoding an arbitrary partition of the vertices.

If one relaxes our problem and instead allows \emph{randomization}, it is known that a short message can support not only connectivity decision but even the recovery of edges witnessing connectivity. This dates back to the seminal work of Ahn, Guha, and McGregor \cite{ahn2012analyzing} who gave an $\widetilde{O}(n)$-bit linear sketch, based on $\ell_0$-sampling, from which Bob can recover a spanning forest of $G_A - G_B$, with probability at least $1 - 1/ \mathrm{poly}(n)$ for each fixed valid input pair. Thus, Bob recovers actual edges forming a \emph{spanning tree} for each connected component of $G_A - G_B$.  Subsequent work provides still richer representations, allowing Bob to recover cut sparsifiers \cite{ahn2012graph} and spectral sparsifiers \cite{kapralov2017single} using $\widetilde{O}(n)$ bits of space for constant approximation accuracy.

It is natural to ask whether these stronger operations of reporting spanning forests and sparsifiers are possible in the \emph{deterministic} setting as well, but this is very strongly not the case.
Even deterministic edge recovery, where Bob simply wants to recover \emph{a single edge} from the graph $G_A - G_B$, in a sharp contrast to randomized setting, requires $\Omega(n^2)$ bits. To see this, let us suppose that a message $M$ lets Bob recover one edge of $G_A - G_B$ whenever a surviving edge exists or output empty if there is none. Starting with $G_B = \emptyset \subseteq G_A$, Bob recovers an edge $e_1 \in G_A - G_B$ and updates his graph to $G_B = (V, \{e_1\})$. Bob repeats this procedure, adding each recovered edge to $G_B$. Every query satisfies $E_B \subseteq E_A$, and after $|E_A|$ iterations Bob has reconstructed the entire graph $G_A$. The original message must therefore distinguish all $2^{\binom{n}{2}}$ graphs, giving $m = \Omega(n^2)$. This obstruction also rules out subquadratic deterministic messages for spanning-forest recovery; related connections between connectivity and single-element recovery appear in \cite{ACK21}.

However, very importantly, the same argument does not apply to a message that only decides connectivity. A connectivity answer does not identify an edge to delete next. The central question considered in our work is whether this distinction permits subquadratic deterministic communication.

The question also has a data-structure interpretation. Given a graph $G = (V, E)$ and a parameter $f$, a deterministic $f$-fault tolerant connectivity oracle stores a representation $\mathcal{D}(G)$ that reports the connected components of $G - F$ for every valid set of edge faults $F \subseteq E$, $|F| \leq f$. Alice's graph $G_A$ is $G$, her message $M$ is $\mathcal{D}(G)$, and Bob's graph $G_B$ is the set of deletions $F$. With an unrestricted failure budget, this is exactly the one-way \emph{connected-components} problem; \cref{clm:componentsToConnectivity} relates it to connectivity decision at the cost of one additional vertex.

The sparse connectivity certificates of Nagamochi and Ibaraki \cite{nagamochi1992linear} give a classical $\widetilde{O}(nf)$-bit baseline for this problem. These remained the best known until the recent work of Khanna, Konrad, and Putterman \cite{khanna2026fault} obtained $\widetilde{O}(n \sqrt{f})$-bit deterministic fault-tolerant distance oracles, which also support connectivity queries, and asked whether this bound is optimal for connectivity. For an unrestricted number of failures, conventionally written $f = n^2$, neither bound improves on the trivial sketch of storing the entire graph. This highlights our main question:
\begin{center}
    \emph{What is the one-way communication complexity of deterministic connectivity after arbitrary valid edge deletions? Is quadratic space necessary, or does connectivity admit a substantially smaller representation?}
\end{center}

\subsection{Our Results}

Our main result places the complexity of deterministic one-way connectivity strictly between the nearly-linear randomized bound and the quadratic cost of edge recovery:

\begin{theorem}\label{thm:main}
    There is a protocol solving the one-way deterministic connectivity problem with messages of length $m = \widetilde{O}(n^{1.75})$ bits. Moreover, every deterministic protocol for this problem requires messages of length $\tilde\Omega(n^{1.5})$ bits.
\end{theorem}

The upper bound gives a deterministic subquadratic representation that supports connectivity queries after \emph{every valid set of edge deletions}. Thus, deciding whether the surviving edges connect \emph{all} vertices requires polynomially less communication than recovering \emph{even one} surviving edge. The lower bound also establishes a polynomial separation between the deterministic and randomized communication for the same promised deletion problem.

We prove the upper bound in two stages, first obtaining a $\widetilde{O}(n^{1.8})$-bit protocol in \cref{sec:n18} and then improving it to $\widetilde{O}(n^{1.75})$ bits in \cref{sec:n74}. For the lower bounds, we express the constraints on valid messages as a graph-coloring problem and construct positive semidefinite certificates to lower-bound the number of messages required, following the framework of \cite{lovasz1979shannon}. This yields an $\Omega(n^{1.25})$ lower bound in \cref{sec:genLB} and an $\tilde \Omega(n^{3/2})$ lower bound in \cref{sec:appn32}. The stronger bound uses an analytic inequality, proved in \cref{sec:keyInequalityApp}, to establish positive semidefiniteness of the certificate.\footnote{An accompanying Lean proof is available at \url{https://github.com/Louiealbp/ChromaticInequality}.}

For the widely used class of \emph{vertex-incidence linear sketches}, we obtain a separate proof of an $\tilde \Omega(n^{3/2})$ lower bound without the use of heavy complex analysis machinery used for \cref{sec:keyInequalityApp}. Recall that a \emph{linear sketch} is specified by a matrix $S \in \Z^{m \times \binom{n}{2}}$ with entries bounded by $\mathrm{poly}(n)$, and maps the edge indicator vector $\mathbf{1}_G$ to $S \cdot \mathbf{1}_G$. We say it is a \emph{ vertex-incidence} sketch if every row of $S$ is supported on edges incident to a single vertex.
These are a widely used class of sketches, including, for instance, the sketches from the work of \cite{ahn2012analyzing} (this model has also been used for lower bounds before, see, i.e., \cite{CKL22}).

\begin{theorem}
    Let $S \in \Z^{m\times \binom{n}{2}}$ be a vertex-incidence linear sketch with entries bounded by $\mathrm{poly}(n)$ such that, for every graph $G$, $S \cdot \mathbf{1}_G$ deterministically determines whether $G$ is connected. Then, $m = \widetilde{\Omega}(n^{1.5})$.
\end{theorem}

In the above theorem $m$ counts the number of \emph{measurements} in the linear sketch, whereas \cref{thm:main} measures communication in bits. These are effectively the same, as each measurement requires $O(\log(n))$ bits in a standard explicit representation. The proof, given in \cref{sec:VILS}, thus provides a separate structural explanation for the difficulty of deterministic connectivity sketching.

As mentioned in the abstract, the upper bound from \cref{thm:main} yields a deterministic fault-tolerant connectivity oracle of subquadratic size, \emph{even} when the number of edge failures is \emph{unrestricted}:

\begin{corollary}
    There is a deterministic $n^2$-fault tolerant connectivity oracle of size $\widetilde{O}(n^{1.75})$ bits that reports the connected components of $G - F$ for every $F \subseteq E$.
\end{corollary}

Crucially, this significantly improves the $\widetilde{O}(n \sqrt{f})$ bound of \cite{khanna2026fault} in the large-failure regime, and \emph{disproves} their conjecture lower bound of $\Omega(n^2)$ when $f = n^2$.  At the same time, our lower-bound construction also yields the first (to our knowledge) lower bound with a polynomial dependence on the failure parameter $f$:

\begin{corollary}\label{cor:ftcoLB}
    For every number of vertices $n$ and number of faults $f\in [1,n^2]$, deterministic $f$-fault tolerant connectivity oracles require $\tilde\Omega(n f^{1/4})$ bits of space in the worst case.
\end{corollary}

Finally, we extend deterministic subquadratic compression to \emph{single-pass dynamic streaming},  under a bound on the number of updates to each edge. This requires an additional algorithmic step, as in one-way communication Alice has \emph{unrestricted access} to her entire graph, and unbounded work-space. Alice can use this to construct the shortest possible message to Bob. On the other hand, a streaming algorithm must process interleaved insertions and deletions while keeping its entire working memory small.

For linear sketches, such as those of Ahn, Guha, and McGregor \cite{ahn2012analyzing}, updating the linear sketch is very simple in the streaming model: an insertion or deletion of an edge $e$ is reflected by adding or subtracting $S \cdot \mathbf{1}_e$ from the current sketch. In contrast, our deterministic encoding also contains non-linear information about the graph at hand. To maintain this information, we ``buffer'' updates and repeatedly compress them into small summaries. Controlling the error accumulated  through these compressions yields a tradeoff between update multiplicity and space.

Formally, the stream starts from the empty graph and, at each time step, an edge $e_i$ is inserted or deleted. We say that the stream has \emph{multiplicity} $K$ if each edge undergoes at most $K$ updates in total, counting insertions and deletions together. We still require that the resulting graph $G$ is simple. I.e., after canceling insertions against deletions, each edge must have either $0$ or $1$ copies. This promise needs to hold only for the final graph.

\begin{theorem}\label{thm:introStreaming}
    Fix a number of vertices $n$ and a parameter $\eps \in [0,1/3)$. There is a deterministic single-pass streaming algorithm that, for every dynamic stream with multiplicity at most $n^{\eps}$, reports its connected components using $\widetilde{O}\left (n^{15/8 + 3\eps/8} \right)$ bits of working space. In particular, for $\eps = 1/3 - \Omega(1)$, this requires space $n^{2 - \Omega(1)}$.
\end{theorem}

The streaming implementation builds on the $\widetilde{O}(n^{1.8})$ construction from \cref{sec:n18}. For every fixed $\eps \in (0,1/3)$, this extends the separation between deterministic connectivity and edge recovery to the single-pass streaming setting. Note that our sketches do not extend to \emph{arbitrary} multiplicities because we make use of non-linear sketches of the graph $G$. Thus as the stream length scales, we are required to store more information.

\subsection{Related Work}

\paragraph{Communication complexity of graph connectivity.}
In the standard communication model for graph connectivity, Alice and Bob hold graphs $G_A$ and $G_B$ on the same $n$ vertices and must decide whether $G_A \cup G_B$ is connected \cite{babai1986complexity}. There is no containment promise between their edge sets in this \emph{union model}. However, this does admit a simple, efficient, deterministic communication protocol, as Alice can simply send a spanning forest of $G_A$ using $O(n \log(n))$ bits; together with $G_B$, this lets Bob recover a spanning forest of $G_A \cup G_B$ and hence decide connectivity. Hajnal, Maass, and Tur\'an \cite{hajnal1988communication} proved that the matching $\Omega(n \log(n))$ deterministic lower bound already holds for the \emph{decision problem}, even with unrestricted interaction.

For randomized protocols, the disjointness reduction of \cite{babai1986complexity}, together with the lower bounds of \cite{kalyanasundaram1992probabilistic,razborov1990distributional, bar2004information}, gives an $\Omega(n)$ lower bound for connectivity decision with constant error, even with unrestricted interaction. Sun and Woodruff \cite{sun2015tight} proved an $\Omega(n \log(n))$ lower bound for randomized one-way protocols for this decision problem. Consequently, both connectivity decision and spanning-forest recovery have $\Theta(n \log(n))$ randomized one-way communication complexity in the union model. The same bound is tight for both tasks in single-pass insertion-only streams.

More recent work establishes stronger randomized decision lower bounds when the number of rounds is bounded. Radhakrishnan, Reddy, and Venkat \cite{RRV26} proved an $\Omega(n \log(n))$ bound for two-round protocols. Zhang~\cite{zhang2026Connectivity} proved an $\Omega((n/r) \log(n))$ bound for $r$-round protocols with constant error, giving a tight bound for every constant number of rounds. Zhang also established an optimal $\Omega(n \log(n))$ bound for nondeterministic connectivity decision, improving \cite{raz1995log}.

Importantly, these results concern connectivity of a \emph{union}, which naturally corresponds to the setting of \emph{insertion-only} streams. The spanning-forest upper bound does not extend to our promised deletion setting, as edges omitted from Alice's forest may become essential for connectivity after Bob's deletions.

\paragraph{Randomized dynamic streaming and distributed sketching.}
When both edge insertions and deletions are allowed, Ahn, Guha, and McGregor \cite{ahn2012analyzing,ahn2012graph} gave a randomized linear sketch using $O(n \log^3(n))$ bits from which a spanning forest can be recovered with high probability. This also gives an algorithm for connectivity decision. Nelson and Yu \cite{nelson2019optimal} proved a matching $\Omega(n \log^3(n))$ space lower bound for \emph{spanning-forest recovery}, even at constant success probability. Thus, the space complexity of the recovery problem is understood, while the corresponding decision problem remains less well understood.

A matching lower bound for \emph{connectivity decision} is known in the distributed graph-sketching model. In this model, each vertex sends a single message computed from its neighborhood to a referee, who must decide whether the graph is connected. Yu \cite{yu2021tight} proved an $\Omega(n \log^3(n))$ lower bound on total communication, even with constant error. Decision and spanning-forest recovery therefore have the same asymptotic complexity in this distributed model. For general single-pass dynamic-streaming algorithms, however, no matching lower bound is known for connectivity decision version. To our knowledge, the best general lower bound at constant error remains $\Omega(n \log(n))$, inherited from the insertion-only lower bound of \cite{sun2015tight}.

\paragraph{Query models and deterministic multipass streaming.}
Graph connectivity has also been studied in several query models. Assadi, Chakrabarty, and Khanna \cite{ACK21} study the tradeoff between query complexity and \emph{round complexity} for deterministic spanning-forest recovery. In each round, the algorithm makes a batch of queries that may depend on the answers from previous rounds. They relate spanning-forest recovery to the more basic task of recovering \emph{even a single edge}, and establish upper and lower bounds for several classes of queries. These questions are closely related to the extensive literature on graph reconstruction from queries \cite{alon2004learning,alon2005learning,angluin2008learning, bshouty2009reconstructing,choi2008optimal,abasi2019learning}.

Their query upper bound also has a streaming consequence: for every fixed positive integer $r$, it yields a deterministic $O(r)$-pass dynamic-streaming algorithm that recovers a spanning forest using $\widetilde{O}(n^{1+1/r})$ space. Our streaming result achieves polynomially subquadratic working space in a single pass for reporting connected components, under a bound on per-edge update multiplicity.

\paragraph{Streaming versus linear sketching.}
The relationship between turnstile streaming and linear sketching depends on the class of update sequences on which an algorithm must be correct. Li, Nguyen, and Woodruff~\cite{li2014turnstile} establish a streaming-to-sketching equivalence for algorithms that handle arbitrarily long turnstile streams. Kallaugher and Price \cite{KP20} show that restricting stream length or intermediate frequencies can separate the models, and establish a uniform equivalence for deterministic algorithms on unrestricted streams.

More recently, Jiang, Liu, and Yu~\cite{jiang2026turnstile} extend streaming-to-sketching simulations to sufficiently long polynomial-length unit-update streams. For final vectors with bounded norm, their general simulation converts an $S$-bit streaming algorithm into a linear sketch using $O(S \log(S))$ bits, provided the algorithm handles every stream up to a sufficiently large polynomial length in the dimension and the norm bound. Their framework also includes promise decision problems.

The simulation of \cite{jiang2026turnstile} requires correctness on every update sequence up to a sufficiently large polynomial length. Our streaming theorem guarantees correctness under a different restriction: each edge receives at most $n^{\eps}$ updates. This guarantee does not cover all the update sequences required by their simulation, so the theorem does not directly apply in our parameter regime.

\paragraph{Fault-tolerant oracles and labels.}
P\u{a}tra\c{s}cu and Thorup \cite{PT07} give deterministic data structures that process batches of edge failures and support fast subsequent connectivity queries, using $O(n + |E|)$ words of space. Duan and Pettie \cite{DP20} give a Monte Carlo edge-failure connectivity oracle using $\widetilde{O}(n)$ bits, with efficient batch processing and subsequent connectivity queries. For each fixed failure set and vertex pair, the answer is correct with high probability. More recently, Khanna, Konrad, and Putterman \cite{khanna2026fault} give deterministic $f$-fault-tolerant distance oracles with polylogarithmic stretch using $\widetilde{O}(n \sqrt{f})$ bits; these also support exact connectivity queries. Our oracle bounds concern the total number of bits in a deterministic representation, allowing unrestricted preprocessing and query computation.

Edge-fault-tolerant connectivity labeling schemes measure the maximum label length over vertices and edges. Long, Pettie, and Saranurak \cite{LPS26} give a deterministic scheme with $O(\log^2(n))$-bit labels that supports every failure set $F \subseteq E$. To determine whether two vertices remain connected, the labeling decoder receives their labels and the graph-dependent labels of all failed edges. In our model, Bob knows the identities of the deleted edges, but their labels depend on $G_A$ and cannot be computed from $G_B$ alone. Using this decoder would therefore require Alice's message to enable Bob to recover the relevant labels for every valid choice of $G_B$; a bound on individual label sizes does not provide such a message.

\subsection{Technical Overview}

We now turn to an overview of our techniques. The starting point for both our upper and lower bounds is simple characterization of valid messages. Essentially, graphs that receive the \emph{same message} must remain indistinguishable after every deletion set that is valid for both. We first express this condition in terms of graph differences, then describe the upper- and lower-bound constructions.

As we discussed before, it is essentially equivalent (and much more convenient) to consider sketches which suffice for recovering the \emph{connected components} of $G_A - G_B$ (see \cref{clm:componentsToConnectivity}). For this problem, our key characterization is the following: Alice's encoding scheme supports component reporting after every valid deletion set \emph{if and only if}, whenever $G_A$ and $G'_A$ receive the same message, the graphs $G_A - G_A \cap G'_A$ and $G'_A - G_A \cap G'_A$ have the same connected components.

To see the necessity, we can observe that if Bob sets $G_B = G_A \cap G'_A$, then $G_B$ is a valid set of deletions for both $G_A$ and $G'_A$. Moreover, if $G_A - G_B$ and $G'_A - G_B$ have different connected components, this then forces $G_A$ and $G'_A$ to have different messages. Sufficiency is a bit more subtle.  To start, consider any graph $G_B$ which is a valid set of deletions for $G_A$ and $G'_A$. This then forces $G_B$ to be contained in both $G_A$ and $G'_A$. We have
\[
G_A - G_B = \left ( G_A - (G_A \cap G'_A) \right ) \cup \left ( (G_A \cap G'_A) - G_B \right ),
\]
and likewise
\[
G'_A - G_B = \left ( G'_A - (G_A \cap G'_A) \right ) \cup \left ( (G_A \cap G'_A) - G_B \right ).
\]
Crucially, if the graphs $G_A - G_A \cap G'_A$ and $G'_A - G_A \cap G'_A$ have the same connected components, this then forces $G_A - G_B$ and $G'_A - G_B$ to have the same connected components! Indeed, these two graphs are obtained by adding the \emph{same edges} to the graphs $G_A - (G_A \cap G'_A)$ and $G'_A - (G_A \cap G'_A)$, which already have the same connected components. In general, adding the same edges to graphs with the same connected components preserves equality of their component partitions.

Consequently, Bob may choose any graph $G'_A$ consistent with the received message and containing $G_B$, and report the connected components of $G'_A - G_B$. This characterization permits compression without recovering Alice's graph and provides the basis of both our upper and lower bounds.

\subsubsection{Upper-Bound}

We warm-up by discussing a sufficient set of ideas which allow us to prove an $\widetilde{O}(n^{1.8})$ bit communication scheme. 

\paragraph{A Subset-Density Estimation Sketch}

Our starting point is a \emph{deterministic} sketch of a graph $G$ which allows us to estimate, for every set $S \subseteq V$ of the vertices, the number of edges in the induced subgraph $G[S]$. More formally, we capture this by introducing a parameter $\tau \geq 1$. We show that there is a sketch $f$, which for any graph $G$ on $n$ vertices, uses $\widetilde{O}(n^3 / \tau^2)$ many bits of space, and ensures that \emph{for every} set $S \subseteq V$, it is the case that 
\[
\bigl||E[G[S]]| - f_G(S)\bigr| \leq \tau \cdot |S|. 
\]
In words, we get an \emph{additive approximation} to the number of edges in the induced subgraph $G[S]$, where the additive error scales as $\tau \cdot |S|$. In order to make this sketch obtain any space savings over storing the entire graph, we must choose $\tau \geq n^{0.5}$, and our final choice will indeed turn out to be $\tau = n^{0.6}$. 

To prove that these sketches exist is more non-trivial, and it turns out that we start by analyzing a \emph{randomized} sampling scheme. Indeed, to get sketches of size $\widetilde{O}(n^{1.8})$ with $\tau= n^{0.6}$, we randomly sample the edges of the graph at rate $1 / n^{0.2}$, and define our function $f_G$ to count the number of $G$'s sampled edges inside the set $S$, and then scale this up by $n^{0.2}$. By carefully analyzing the probability of the event $\bigl||E[G[S]]| - f_G(S)\bigr| > n^{0.6} \cdot |S|$ for each $S$, and then taking a union bound over all possible choices of $S$, one can then show that a random sample does indeed preserve these quantities with probability $\geq 1/2$. Unfortunately, this alone is not enough, as we desire such a sketch which works \emph{deterministically} for every possible graph $G$.

Here, we use inspiration from \cite{newman1996public}. If we repeat the sampling scheme above $n^2$ times, then, for every graph $G$, with probability $1 - 2^{-n^2}$ there will \emph{exist} some iteration of random sampling for which 
\[
\bigl||E[G[S]]| - f_G(S)\bigr| \leq \tau \cdot |S|
\]
holds for every $S \subseteq V$. We model this by a sequence of sampled graphs $H_1, \dots H_{n^2}$, where each $H_i$ is understood to be a uniformly random sample of the complete graph $K_n$ at rate $1 / n^{0.2}$. In particular, because the number of graphs on $n$ vertices is bounded by $2^{\binom{n}{2}}$, this means there must exist a \emph{universal sample set}  (note that this means our communication protocol is a type of non-uniform sketch) which we call $H_1, \dots H_{n^2}$ such that, for every single graph $G$, there is some choice of a single index $i \in [n^2]$ for which 
\[
f^{(i)}_G(S) := n^{0.2} \cdot |(G \cap H_i)[S]|
\]
approximates $|E[G[S]]|$ to additive error $\tau \cdot |S|$ for every $S$. For a graph $G$, this then gives us our final sketch. We store the index $i \in [n^2]$ corresponding to which ``global sample'' works for the graph $G$, along with the exact set of edges in this sample $G \cap H_i$. The total size of this sketch is $O(\log(n))$ bits for the index $i \in [n^2]$, along with $O(n^{1.8})$ bits for the stored edges $G \cap H_i$.

\paragraph{Proving an $\widetilde{O}(n^{1.8})$ Bound}

With the above subset-size estimation sketch in hand, the remaining ingredients for the $\widetilde{O}(n^{1.8})$ size sketch can be implemented using known ideas. Indeed, given the graph $G_A$, the second part of the sketch relies on a decomposition of the $[n]$ vertices into $n^{0.8}$ many groups, which we call $C_1, \dots C_{n^{0.8}}$ (each of size $n^{0.2}$). For each such group, and for every vertex $v \in V$, we store the number of edges going from $v$ to $C_i$. This is a total of $n^{1.8}$ many numbers, hence $\widetilde{O}(n^{1.8})$ many bits. 

The final piece of the sketch is a so-called \emph{sparse-recovery} sketch. For every vertex $v \in V$, we store this sketch with a so-called \emph{sparsity threshold} of $k = 2 \cdot n^{0.8}$. The way to understand this sketch is as a matrix $P \in \Z^{O(k) \times n}$, such that, for any distinct vectors $y, y' \in \Z^n$ with $|y'_i|, |y_i| \leq n$, if $|y|_0, |y'|_0 \leq k$, then
\[
P \cdot y \neq P \cdot y'.
\]
This ensures that \emph{for sparse vectors} $y$, given only the sketching information $P \cdot y$ one can \emph{uniquely} recover the original vector $y$. 

To summarize, the message that Alice sends to Bob contains:
\begin{enumerate}
    \item The (\emph{non-linear}) subset-size estimation sketch applied to $G_A$ with parameter $\tau = n^{0.6}$.
    \item For every $v \in V$ and every $C_i: i \in [n^{0.8}]$, the number of edges from $v$ to $C_i$ in $G_A$.
    \item For every vertex $v \in V$, the sparse recovery sketch applied to $v$'s neighborhood in $G_A$.
\end{enumerate}

As discussed above, with this sketching scheme in hand, we must only show that for any two graphs $G_A, G'_A$ that if the sketches above coincide, then it must be that $G_A - G'_A \cap G_A$ and $G'_A - G'_A \cap G_A$ have the same connected components. So, let $G_A, G'_A$ denote two such graphs for which the sketches coincide, and let us suppose that there is some cut $(S, V - S)$ such that no edges in $G_A - G'_A \cap G_A$ cross the cut, while there are some edges in $G'_A - G'_A \cap G_A$ that cross the cut. We will use this supposition in conjunction with the aforementioned sketches to reach a contradiction. 

We omit the entire argument as it is quite meticulous, but summarize the intuition here: because the sparse recovery sketches coincide, for each vertex $v$, we know that the sketch $P \cdot \mathbf{1}_{\Gamma_{G_A}(v)} = P \cdot \mathbf{1}_{\Gamma_{G'_A}(v)}$, and thus the difference vector $P \cdot (\mathbf{1}_{\Gamma_{G_A}(v)} - \mathbf{1}_{\Gamma_{G'_A}(v)}) = 0$. This gives us valuable information; indeed, by the property of our sparse recovery sketch, whenever $P \cdot y = 0$, it must be the case that either $y = 0$, \emph{or} $y$ has at least $k = 2 \cdot n^{0.8}$ non-zero entries. 

To get this into a useful form, we manipulate the above expression and show that for every one of our pieces $C_i$:
\begin{enumerate}
    \item In the graph $G_A - G_A \cap G'_A$, $C_i$  has \emph{no} edges crossing the cut $S$.
    \item In the graph $G'_A - G_A \cap G'_A$, $C_i$  has either (a) no edges crossing the cut $S$, or (b) \emph{at least} $2 \cdot n^{0.8}$ edges crossing the cut $S$.
\end{enumerate}

To reach a contradiction, we then analyze these pieces $C_i$ which have $\geq 2 \cdot n^{0.8}$ edges crossing the cut $S$. The key point is that this property \emph{forces} the density of edges in $C_i \cap S$ in $G'_A - G_A \cap G'_A$ to \emph{significantly} decrease, as it is sending so many edges across the cut. Meanwhile, in $G_A - G_A \cap G'_A$, no such density decrease is possible, as the edges stay within the same side of the set $S$. But, $G_A$ and $G'_A$ received the same sketch for estimating their edge densities! This forces the edge densities in $G_A - G_A \cap G'_A$ and $G'_A - G_A \cap G'_A$ to be close, ultimately yielding a contradiction. In the true argument, this reasoning requires careful book-keeping, and we invite the reader to complete argument in \cref{sec:n18}.

\paragraph{Improving to $\widetilde{O}(n^{1.75})$ Bits}

Our argument to achieve $\widetilde{O}(n^{1.75})$ bits is significantly more intricate than the $\widetilde{O}(n^{1.8})$ bit argument outlined above. Recall that above, the ultimate argument relies on a sort of ``amplification'' in the sense that in the graph $G'_A - G_A \cap G'_A$ for our clusters $C_i$, there were either $0$ or $\geq 2 \cdot n^{0.8}$ many edges crossing the cut $S$. In general, the larger we can make the second piece of this promise (i.e., if we had $ \geq n$ instead of $\geq 2 \cdot n^{0.8}$), the easier it is to conclude the argument, as we ultimately seek to reach a violation of the accuracy of our density sketch. 

The same general principle guides our $\widetilde{O}(n^{1.75})$ bit upper bound. The key new lemma is the following ``one-sided sketch lemma'':

\begin{lemma}[Informal One-Sided Sketch Lemma]
    For any parameters $n$ and $K$, there is a linear sketching matrix $\mathcal{S} \in \F^{m \times \binom{n}{2}}$ with $m = \widetilde{O}(n \sqrt{K})$, such that for any two graphs $G_A, G'_A$ on $n$ vertices, if $\mathcal{S}\cdot \mathbf{1}_{G_A} = \mathcal{S} \cdot \mathbf{1}_{G'_A}$, then:
    \begin{enumerate}
        \item For any cut $(S, V - S)$ with no edges crossing in $G_A - G'_A \cap G_A$, the same cut in $G'_A - G'_A \cap G_A$ must be of size $0$ or size $> K$.
        \item For any cut $(S, V - S)$ with no edges crossing in $G'_A - G'_A \cap G_A$, the same cut in $G_A - G'_A \cap G_A$ must be of size $0$ or size $> K$.
    \end{enumerate}
\end{lemma}

Intuitively, this promises that for any graphs which receive the same sketch, if a cut in one graph is size $0$, the same cut in the second graph must either be of size $0$ or size $> K$. Ultimately, we choose $K \approx n^{1.5}$, and so this provides a significant ``gap'' between cut-sizes whenever one has two graphs with the same sketch, yet different connected components. 

This lemma on its own is not enough to prove the entire result. Similar to the deterministic subset-size estimation sketch implemented above, we build a deterministic \emph{cut estimation sketch}, which estimates the size of all cuts up to additive error $\approx n^{1.5}$. Unfortunately, this sketch cannot be implemented as a single one-time sketch in space $\widetilde{O}(n^{1.75})$ as this would require a sampling rate which is far too small (and thus would blow up our additive error). Instead, we define a recursive subsampling scheme which repeatedly subsamples the graph at rate $1/2$, and stores a sketch at each level of sampling. Whenever two graphs receive the same sketch, this then requires that they receive the same sketch \emph{at all levels of sampling}, which enables us to derive tighter bounds on the additive error in the cut estimation, ultimately giving us additive error $\approx n^{1.5}$ in space $\widetilde{O}(n^{1.75})$ many bits. This suffices for proving that any graphs $G_A, G'_A$ which receive the same one-sided sketch and cut-estimation sketch do indeed have the same components in $G_A - G_A \cap G'_A$ and $G'_A - G_A \cap G'_A$.

\paragraph{Proving the One-Sided Sketch Lemma}

It remains still to prove the above one-sided sketch lemma. This is also quite non-trivial and itself relies on several new (to our knowledge) variants of deterministic linear sketches, which we summarize below. Our first such result is a modified sparse-recovery sketch:

\begin{lemma}\label{lem:sparseRecoveryModifiedIntro}
    For any parameter $t$, there is a matrix $X \in \{\pm 1\}^{n \times \widetilde{O}(t)}$ such that, for any set $U \subseteq [n]$ with $|U| \leq t$:
    \begin{enumerate}
        \item The rows $X|_U$ are linearly independent. 
        \item There exists a vector $h$ such that $(Xh)_v = 0$ when $v \in U$ and $(Xh)_v > 0$ when $v \notin U$.
    \end{enumerate}
\end{lemma}

One way to understand the power of the above sketch is to view it as an even stronger version of the aforementioned sparse recovery sketch. Indeed, let $A = X^T$, and let $z$ be an arbitrary vector such that $z \geq 0$ and $|z|_0 \leq t$. When we apply the sketching matrix to $z$, we obtain the vector $A z = y$. We claim that simply given $y$, we can \emph{uniquely recover} the vector $z$ by solving $Ax = y$, $x\geq 0$. Indeed, the second property of the above lemma implies that for $U = \Supp(z)$, there exists a vector $h$ such that $(Xh)_v = 0$ when $v \in U$ and $(Xh)_v > 0$ when $v \notin U$.

In particular, this means that \[
h^Ty = \sum_v z_v (Xh)_v = 0,
\]
as $(Xh)_v =0 $ for any entry in $U$. By this same equation, it must also then be that \emph{for any other nonnegative vector} $w$ such that $Aw = y$, 
\[
0 = h^Ty = h^T Aw = \sum_v w_v (Xh)_v = 0,
\]
which forces $w_v = 0: v \notin U$. That is to say, \emph{every nonnegative solution} to $Ax = y$ must have its support \emph{entirely} contained within $U$. Because $X|_U$ is linearly independent, we then know that there is only a single possible solution to this equation. This is extremely non-trivial: sparse vectors can not only be recovered, but we know that there are no other nonnegative vectors (even dense ones!) in the universe which produce the same sketch.

As we explore in \cref{sec:n74}, the power extends even beyond what is discussed above. For instance, suppose we are given the same sketch $y = Ax$ for $x$ non-negative. We claim that for any candidate set $U, |U| \leq t$, we can even test whether $\Supp(x) \subseteq U$: indeed, we must only observe that $h^T_U y = 0$ if and only if $\Supp(x) \subseteq U$. Together, this is distinctly stronger than a sparse recovery sketch; ordinarily, sparse recovery only guarantees that when a vector is \emph{promised} to be sparse, one can recovery the vector itself. The aforementioned sketch allows for \emph{verification} of whether a vector is sparse given only its measurement, and subsequently enables \emph{recovery}.

The second key ingredient in proving the one-sided sketch is the \emph{vertex cover} detection sketch:

\begin{lemma}\label{lem:vertexCoverSketchIntro}
    For any parameter $K$, there is matrix $P \in \F_q^{n \times \widetilde{O}(\sqrt{K})}$ (with $q = \mathrm{poly}(n)$) such that, for any skew-symmetric matrix $M \in \{-1,0,1\}^{n \times n}$, if $M$ has at most $K$ non-zero entries above the diagonal, and $P^T M P = 0$, then it must be the case that $\Supp(M)$ has a vertex cover of size $\leq 10\sqrt{K}$.
\end{lemma}

Intuitively, the above supplies a sketching matrix $P$ such that whenever $P$ yields the zero sketch, the matrix it is operating on must have all of its edges supported on a small set of vertices. The size of the stored sketch that we end up using will be $\widetilde{O}(n \sqrt{K})$ many bits, and when equal to $0$, will ensure that the matrix being sketched has a vertex cover of size $\leq 10\sqrt{K}$.

To finally prove the one-sided sketch, we store one copy each of \cref{lem:sparseRecoveryModifiedIntro} and \cref{lem:vertexCoverSketchIntro}, each of size $\widetilde{O}(n \sqrt{K})$ bits. The actual argument still requires significant care, but proceeds by showing that if, for two graphs $G, H$ with adjacency matrices $A_G, A_H$, it is the case that $A_G X = A_H X$ and $A_H P = A_G P$, then one immediately obtains
\[
(A_G - A_H) X = 0 \quad \text{and} \quad (A_G - A_H) P = 0.
\]
When studying an arbitrary cut $S \subseteq V$ as is necessary in proving the one-sided cut sketch, we let $J_S$ denote the diagonal matrix which is the indicator of the set $S$, and define the matrix 
\[
M = J_S (A_G - A_H) - (A_G - A_H) J_S,
\]
whose non-zero edges \emph{exactly capture} all those edges crossing the cut $S$. When one of the graphs $G-G\cap H$ or $H-G\cap H$ does not have any edges crossing $S$, we then know that the entire support of $M$ must be edges from one of the above graphs. By the above, we can also see that 
\[
X^T M X = 0 \quad \text{and} \quad P^TMP = 0.
\]
Ultimately, by combining \cref{lem:sparseRecoveryModifiedIntro} and \cref{lem:vertexCoverSketchIntro}, we show that if $M$ has at most $K$ non-zero entries above the diagonal all of whom are in the support of only one of $G$ or $H$, then the conditions above force $M = 0$. Said another way, whenever $M$ is non-zero and supported only by $G$ or $H$, then it must have $> K$ entries above the diagonal, which is exactly what we sought to prove. 

\subsubsection{Lower Bound}

\paragraph{Viewing the Communication Complexity as the Chromatic Number}

We now briefly present some intuition for the lower bound. As mentioned before, a necessary and sufficient condition for a protocol to be valid for the deterministic connected components problem is if, for any graphs $G_A, G'_A$ where the connected components of $G_A - G_A \cap G'_A$ and $G'_A - G_A \cap G'_A$ differ, the messages for $G_A$ and $G'_A$ differ. One way to capture these constraints is to consider a \emph{universal graph} $P_n$ with $2^{\binom{n}{2}}$ many vertices, one for each possible graph on $n$ vertices. For each pair of graphs $G_A, G'_A$ where the connected components of $G_A - G_A \cap G'_A$ and $G'_A - G_A \cap G'_A$ differ, we then add an edge between $G_A$ and $G'_A$ in $P_n$.

Ultimately, the goal is to assign to every vertex a message (AKA a sketch) such that whenever two vertices share an edge, they receive different messages. The smallest possible number of distinct messages is then \emph{exactly} the chromatic number of this universal graph $P_n$. This reframes our original question: how do we lower bound the chromatic number of $P_n$?

Fortunately, lower bounding chromatic numbers of graphs is an extensively studied question. For our purposes, we adopt Lov\'asz's \cite{lovasz1979shannon} framework. This states that for any graph $G$ on $N$ vertices, one can lower bound the chromatic number of $G$ by taking the maximum over all PSD matrices $M \in \R^{N \times N}$ such that $M_{v, v} = 1$ and $M_{u,v} = 0$ if $u\neq v$ and $(u,v) \notin G$ of the quantity
\[
\frac{\mathbf{1}^T M \mathbf{1}}{N}.
\]
In words, we consider only PSD matrices with $1$'s on the diagonal and whose off-diagonal support is entirely contained in the graph $G$. The conclusion is that the chromatic number must always be \emph{at least} as large as the average entry-wise sum of each row. In the context of our universal graph, $N = 2^{\binom{n}{2}}$, and our goal is to find a PSD matrix with a large entry-wise sum, but whose off-diagonal support is entirely on edges in $P_n$.

\paragraph{Lower-Bounding the Chromatic Number}

It remains only to construct the PSD matrix $M$ whose off-diagonal support is contained in $P_n$. Before doing so, we make a few minor adjustments. Rather than considering all possible graphs on $n$ vertices, we break the $n$ vertices into two groups of $k = n/2$ vertices, and consider only all the possible subgraphs of $K_{k \times k}$. We index these graphs by $\mathcal{X}_k = \zo^{k \times k}$.
For $x \in \mathcal{X}_k$ we let $G_x$ denote the graph indexed by $x$, and we let $H_k$ be the restriction of the universal graph $P_n$ to the graphs $G_x:x \in \mathcal{X}_k$.

As mentioned, the off-diagonal support of our ultimate choice of matrix $M$ will be contained on the graphs $H_k$, but we will additionally require a parameterization in terms of $\theta \in (0, \pi/2)$. Letting $s = \sin \theta$ and writing 
\[
A_+ = \begin{bmatrix}
    1 & 2s \\
    0 & 1
\end{bmatrix},
\]
and 
\[
A_- = \begin{bmatrix}
    1 & 0 \\
    2s & 1
\end{bmatrix},
\]
our final choice of $M$ is then exactly defined as
\[
M_{k, \theta} =\E_{\xi \sim \{\pm 1\}^k, \zeta\sim \{\pm 1\}^k} \left [\bigotimes_{i,j \in [k]} A_{\xi_i \cdot  \zeta_j}\right ].
\]
The way to understand this matrix is that if we denote $L$ and $R$ to be the left and right sides of the bipartite graphs we consider, we then sample uniformly random signs to all vertices in $L$ and $R$, and then take a tensor product over all edges between $L$ and $R$, where the assigned matrix for each edge slot $(i,j)$ is $A_{\xi_i \cdot  \zeta_j}$. $M_{k, \theta}$ is the \emph{expectation} of this entire operation. 

Some properties of this matrix $M_{k, \theta}$ are not hard to see; for instance, its diagonal is trivially equal to $1$ as it is the tensor product of matrices whose diagonal is $1$. It is also of the right dimension, and one can observe that $\frac{\mathbf{1}^T M_{k, \theta} \mathbf{1}}{2^{k^2}} = \frac{(2 + 2 \sin \theta)^{k^2}}{2^{k^2}} = (1 + \sin \theta)^{k^2}$ by simply taking the product of the entry-wise sums of all of the tensored matrices. Ultimately, we will select $\theta \approx 1 / k^{3/4}$, at which point $(1 + \sin \theta)^{k^2} = 2^{\Theta(k^{5/4})}$, thereby giving us our desired lower bound. 

That said, there are still two more subtle conditions which must show if we want to use this matrix in the framework of \cite{lovasz1979shannon}: namely (a) that $M_{k, \theta}$ is PSD, and (b) that its off-diagonal support is contained in the edges of $P_n$. Proving this first condition turns out to be quite intricate, and relies on a sequence of careful inequalities, which we defer until \cref{sec:genLB}. The latter point (b) follows much more readily, and so we sketch the ideas here. Roughly speaking, if some entry $(x, y)$ for distinct $x, y \in \zo^{k \times k}$ is non-zero in $M_{k, \theta}$, we want to show that this means $G_x - G_x \cap G_y$ and $G_y - G_x \cap G_y$ have different connected components. 
To start, we can observe that if $(M_{k, \theta})_{x, y} > 0$ then there is some choice of $\xi, \zeta$ such that 
\[
\left [\bigotimes_{i,j \in [k]} A_{\xi_i \cdot  \zeta_j}\right ]_{x,y} > 0.
\]
The only way this is possible is if every factor in the product contributing to entry $(x, y)$ is positive. 
Because we want to understand the edges in $G_x - G_x \cap G_y$ and $G_y - G_x \cap G_y$, we focus our attention exactly on these edges, or equivalently, those edges in $G_x \oplus G_y$. For any such edge $e = (i,j)$, there are two possible cases:
\begin{enumerate}
    \item If $e \in G_x - G_x \cap G_y$, then $x_{ij} = 1$ and $y_{ij} = 0$. In order for entry $x,y$ to be positive in $\left [\bigotimes_{i,j \in [k]} A_{\xi_i \cdot  \zeta_j}\right ]_{x,y}$, this forces the product to contain $A_-$ in the tensor product at that term, as it is the only matrix which is non-zero in entry $(1, 0)$, and thus $\xi_i \zeta_j = -1$.
    \item If $e \in G_y - G_x \cap G_y$, then $x_{ij} = 0$ and $y_{ij} = 1$. In order for entry $x,y$ to be positive, this forces the product to contain $A_+$, as it is the only matrix which is non-zero in entry $(0, 1)$ and thus $\xi_i \zeta_j = 1$.
\end{enumerate}

Now, let us consider all vertices $S$ which received sign $+1$. For any edge $(u,v) \in G_x - G_x \cap G_y$, it must be the case that $\xi_u \zeta_v = -1$; equivalently, $u, v$ are on opposite side of the cut defined by $S$. On the other hand,  for any edge $(u,v) \in G_y - G_x \cap G_y$, it must be the case that $\xi_u \zeta_v = 1$; equivalently, $u, v$ are on \emph{same} side of the cut defined by $S$. This is to say that \emph{every edge} in $G_x - G_x \cap G_y$ crosses $(S, \bar{S})$ while \emph{no} edge in $G_y - G_x \cap G_y$ crosses the cut. This then forces $G_x - G_x \cap G_y$ and $G_y - G_x \cap G_y$ to have different connected components. 

We take a similar approach to proving the $\tilde{\Omega}(n^{3/2})$ lower bound, still relying on the framework of \cite{lovasz1979shannon}. The only difference in that setting is that we slightly modify the class of graphs and matrices under consideration, and require an even more complex inequality for the PSD proof.

\subsection{Open Questions}

We are particularly excited about the open questions which this work suggests. We discuss several of these below:

\paragraph{1. Tight Bounds for One-Way Deterministic Connectivity} Taken together, our work shows that the complexity of a deterministic one-way protocol for establishing connectivity lies somewhere in the interval $[\Omega(n^{1.5}), \widetilde{O}(n^{1.75})]$.  We think it is an extremely interesting question to pin down the exact complexity further within the interval $[\tilde\Omega(n^{1.5}), \widetilde{O}(n^{1.75})]$, and currently conjecture that the optimal complexity will be $\approx n^{1.5}$:

\begin{conjecture}
    The communication complexity of the one-way deterministic connectivity problem is $\tilde \Theta(n^{3/2})$ bits.
\end{conjecture}

\paragraph{2. Understanding the Vector Chromatic Number} Both the $\Omega(n^{1.25})$ and $\widetilde{\Omega}(n^{3/2})$ lower bounds rely on \cite{lovasz1979shannon}'s PSD matrix characterization of the chromatic number. Crucially however, this lower bound also applies to a more general quantity called the \emph{vector chromatic number} of a graph \cite{karger1998approximate}. We therefore ask: can one show that the vector chromatic number of $P_n$ is bounded by $2^{O(n^{3/2})}$? This would show that one cannot hope to prove a stronger lower bound using the techniques presented here. 

\paragraph{3. Tight Bounds for Linear Sketches} As mentioned above in the discussion of \cref{thm:introStreaming}, while our sketches enable us to solve connectivity in dynamic streams with bounded multiplicity, they do not extend to arbitrary length dynamic streams. In order to build sketches in this regime, one requires \emph{linear sketches} \cite{li2014turnstile, jiang2026turnstile}. What is the optimal complexity of deterministic linear sketches that solve the connectivity problem? I.e., this must be a sketching matrix $\mathcal{S}$ such that for any graphs $G, G'$ if $\mathcal{S} \cdot \mathbf{1}_G = \mathcal{S} \cdot \mathbf{1}_{G'}$, then $G$ and $G'$ must both either be connected or not connected. We conjecture that the right complexity for purely linear sketches will be $\tilde \Omega(n^2)$:

\begin{conjecture}
    Let $\mathcal{S}$ be an $m \times \binom{n}{2}$ linear sketching matrix with polynomially bounded entries. Suppose further that for any graphs $G, G'$, $\mathcal{S}\mathbf{1}_G \neq \mathcal{S}\mathbf{1}_{G'}$ whenever exactly one of $G, G'$ is connected. Then, $m = \widetilde{\Omega}(n^2)$.
\end{conjecture}

\paragraph{4. Tight Bounds for Dynamic Streaming} \cref{thm:introStreaming} shows that in dynamic streams with multiplicity $n^{\eps}$, there is an $\tilde{O}(n^{15/8 + 3\eps/8})$ space deterministic algorithm for determining connectivity. We do not believe that this is the best relationship between the multiplicity $n^{\eps}$ and the space complexity of the deterministic streaming algorithm. In particular, in the name of simplicity, our result does not even use the better sketch from \cref{sec:n74}. What, in general, is the best tradeoff between the stream multiplicity and the optimal space complexity of a deterministic streaming algorithm for determining connectivity? As mentioned above, when the stream length reaches a sufficiently large polynomial, this enters the regime of designing \emph{linear sketches}, but shy of $\eps$ being a large constant, there is no known general equivalence.

\paragraph{5. Tight Bounds for Multiple-Round Communication Protocols} \cref{thm:main} pins the complexity of the \emph{one-way} deterministic connectivity problem in the interval $[\tilde{\Omega}(n^{1.5}), \tilde{O}(n^{7/4})]$. How does this bound change when one considers an $r$-round communication protocol?

\paragraph{6. Arbitrary Fault-Regimes} As discussed in the introduction, the central question of the complexity of one-way deterministic connectivity is actually the \emph{same} question as designing space-efficient fault-tolerant connectivity oracles in the regime where $f = n^2$. In this regime, the sketches we build use significantly less than $n \sqrt{f}$ space. Can this be generalized to all values of $f$? I.e., can $f$-fault tolerant connectivity oracles be constructed in just $\widetilde{O}(n f^{1/2 - \Omega(1)})$ bits of space?

\paragraph{7. Designing Distance-Reporting Data Structures} Beyond simply deciding connectivity, the data structure of \cite{khanna2026fault} showed that in $\widetilde{O}(n \sqrt{f})$ bits of space, one can actually \emph{report distances} (up to a factor of $\mathrm{polylog}(n)$) in the graph $G_A - G_B$. When $f = n^2$, this does not outperform $n^2$ bits of space. Is $\Theta(n^2)$ optimal for this task? Or, by using ideas similar to those discussed here, is there an $o(n^2)$ bit data structure / message for this task?

\subsection{Organization}

The remainder of the paper is organized as follows: in \cref{sec:prelim}, we discuss some basic preliminaries. In \cref{sec:n18} we provide a complete proof of the $\widetilde{O}(n^{1.8})$ upper bound. We then develop an even more intricate approach to prove the $\widetilde{O}(n^{1.{75}})$ upper bound in \cref{sec:n74}. We discuss the streaming implementation of our sketch in \cref{sec:streaming}, and then present our lower bounds in \cref{sec:VILS} (for vertex-incidence linear sketches) and \cref{sec:genLB} and \cref{sec:appn32} (for arbitrary sketches).

\section{Preliminaries}\label{sec:prelim}

\subsection{Problem Setting}

Formally, we consider the following problem: Alice is given a graph $G_A = (V, E_A)$, and Bob is given a graph $G_B = (V, E_B)$, with the promise that $E_B \subseteq E_A$. Alice must send a message which is $\leq m$ bits long to Bob, after which Bob must (deterministically, and with no error) report the connectivity (in some cases the connected components) of the graph $G_A - G_B$. The key question is to understand how large $m$ must be. We refer to these problems as the \emph{deterministic one-way connectivity problem} and \emph{deterministic one-way connected components problem} respectively.

Crucially, we can connect the above communication problem to the \emph{chromatic number} of a particularly defined universal graph:

\begin{definition}\label{def:universalGraph}
    We define the \emph{universal graph} $P_n$ to be the graph with $2^{\binom{n}{2}}$ vertices, with one vertex for each possible graph on $n$ vertices. Two graphs $G, H$ have an edge in $P_n$ if the connected components of $G-G \cap H$ and $H - H \cap G$ are different.
\end{definition}

We have the following connection between the deterministic one-way connected components problem and the chromatic number of the universal graph:

\begin{claim}\label{clm:coloringConnection}
    The complexity of the deterministic one-way connected components problem is
    \[
    \Theta(\log(\chi(P_n))).
    \]
\end{claim}

\begin{proof}
    First, we show that the complexity is at least  $\log(\chi(P_n))$. Indeed, suppose that Alice has a protocol which sends $m$ bits to Bob. Then, for any two graphs $G_A \neq G'_A$ which use the same message, it must be the case that $G_A - G_A \cap G'_A$ and $G'_A - G_A \cap G'_A$ have the same connected components. Otherwise, one can consider the setting in which Bob has the graph $G_B = G_A \cap G'_A$. This would lead to a contradiction as the graphs $G_A$ and $G'_A$ receive the same message, yet the resulting response when Bob has graph $G_B$ should be different. Thus for any graphs $G_A, G'_A$ for which $G_A - G_A \cap G'_A$ and $G'_A - G_A \cap G'_A$ have different connected components, $G_A$ and $G'_A$ must get \emph{different messages}. Thus, the chromatic number of $P_n$ lower bounds the number of messages that are required. 
    
    Now we prove that at most $\lceil \log(\chi(P_n)) \rceil$ bits suffice. Indeed, we can consider the protocol where Alice and Bob agree on a canonical coloring of the universal graph $P_n$ which uses $\chi(P_n)$ colors. Then, upon receiving a graph $G_A$, Alice sends the color of the graph $G_A$ in this coloring to Bob. For the graph $G_B$, Bob then finds a graph $G'_A$ which has the same color as $G_A$ for which $G_B \subseteq G'_A$, and reports the connected components of $G'_A - G_B$. Because $G_B \subseteq G_A$ and $G_B \subseteq G'_A$, we know that $G_B \subseteq G'_A \cap G_A$. Further, because $G_A$ and $G'_A$ have the same color in $P_n$, it must be the case that $G_A - G'_A \cap G_A$ and $G'_A -G'_A \cap G_A$ have the same connected components. This then also implies that $G_A - G'_A \cap G_A + (G'_A \cap G_A - G_B) = G_A - G_B$ and $G'_A - G'_A \cap G_A + (G'_A \cap G_A - G_B) = G'_A - G_B$ have the same connected components. This therefore yields a valid protocol.
\end{proof}

\subsection{Concentration Bounds}

We use the following basic Bernoulli concentration result:

\begin{fact}\label{fact:bernoulli}
    Let $Y_1, \dots Y_m$ be independent Bernoulli random variables with expectation $p$. Then, for any $t>0$,
    \[
    \Pr \left [ \left | \frac{\sum_{i \in [m]} Y_i}{m} - p\right |  > t \right ] \leq 2 e^{ - \frac{mt^2}{2(p+t)}}.
    \]
\end{fact}

\subsection{Sparse Recovery}

We also use the following sparse recovery fact (see, for instance, \cite{khanna2026fault}):

\begin{claim}\label{clm:sparseRecoverySketch}
    For any parameters $1 \leq q,k \leq n$, there exists a matrix $P \in \mathbb{Z}^{k \times n}$ (and an efficient algorithm for constructing it) with entries bounded by $\mathrm{poly}(nq)$ such that for any vector $x \in \mathbb{Z}^n$ with $|x_v| \leq q$ and $1 \leq |\mathrm{Supp}(x)| \leq k$, $Px \neq 0$.
\end{claim}

\subsection{Reducing Connected Components to Connectivity}

In this section, we show that the communication complexity of \emph{reporting the connected components} of $G_A - G_B$ is essentially the same as that of \emph{reporting the connectivity} of $G_A - G_B$. We formalize this below:

\begin{claim}\label{clm:componentsToConnectivity}
    Let $n$ denote the number of vertices. Any protocol which solves the \emph{one-way deterministic connectivity problem} on $n+1$ vertices also solves the \emph{one-way deterministic connected component problem} on $n$ vertices.
\end{claim}

\begin{proof}
    Suppose Alice has a graph $G_A$ on $n$ vertices and Bob has a graph $G_B$, with $G_B \subseteq G_A$. Alice wishes to solve the \emph{one-way deterministic connected components problem}, but Alice only has access to a \emph{one-way deterministic connectivity} protocol on $n+1$ vertices. 

    For the new protocol, Alice and Bob  add a new vertex $\star$ to both the graphs $G_A$ and $G_B$. Alice then also adds the edges from $\star$ to every vertex $v \in V$. Alice then sends the message $M$ for the \emph{one-way deterministic connectivity problem} for this new graph $G'_A$.

    Given this message $M$, Bob now wishes to recover the connected components of the graph $G_A-G_B$. Because Bob knows that the edges $S := \{\{\star, v\}: v \in V\}$ are already in the graph $G'_A$, given the message $M$, Bob is able to learn the connectivity of $G'_A - (G_B \cup T) = (G_A - G_B) \cup (S - T)$, for every choice of $T \subseteq \{\{\star, v\}: v \in V\}$. Equivalently, Bob learns the connectivity of 
    \[
    (G_A - G_B) \cup T,
    \]
    for every $T \subseteq S$.

    Given this information, Bob can already learn important information of the graph $G_A - G_B$. Indeed, the number of connected components is exactly the size of the smallest set $T$ such that $(G_A - G_B) \cup T$ is connected. Let $T^*$ denote the smallest such set of edges for which $(G_A - G_B) \cup T^*$ is connected. In particular, for each connected component of $G_A - G_B$, this means there is exactly one edge from each such component to the vertex $\star$ in $T^*$. Now, consider an edge $(v, \star) \in T^*$ and an edge $(v', \star) \notin T^*$. We then consider the set $T^* - \{(v, \star)\} \cup \{(v', \star) \}$. If the query corresponding to this set is still connected, then this \emph{must mean} that $v$ and $v'$ are in the same connected component; otherwise deleting the edge $(v, \star)$ would disconnect that component from the rest of the graph. Conversely, if $v$ and $v'$ are in the same component, this exchange preserves connectivity. By enumerating all of the possible $1$ edge changes, we can then learn exactly all vertices that are in the same component as $v$. Repeating this for all $(v, \star) \in T^*$ then reveals all of the connected components in the graph.
\end{proof}

\section{An $\widetilde{O}(n^{1.8})$ Upper Bound}\label{sec:n18}

In this section, we prove the following theorem:

\begin{theorem}\label{thm:n1.8}
    There is a protocol for the deterministic one-way connectivity problem in which Alice sends $m = \widetilde{O}(n^{1.8})$ bits.
\end{theorem}

\subsection{A Subset-Size Estimation Sketch}

In this section, we prove the following modular lemma which will be crucial in our ultimate communication argument:

\begin{lemma}\label{lem:subsetSizeSketch}
    For any $1 \leq \tau \leq n$, there is a deterministic sketch $\mathcal{S}_{\mathrm{size}}$ such that, when instantiated on any graph $G = (V, E)$, the sketch $\mathcal{S}_{\mathrm{size}}(G)$:
    \begin{enumerate}
        \item Can report $|E_G[T]|$ to within error $\tau \cdot |T|$ for every $T \subseteq V$.
        \item Requires space only $\widetilde{O}\left (\frac{ n^3}{\tau^2} \right )$.
    \end{enumerate}
\end{lemma}

\begin{proof}
We first present the high-level architecture of the proof: we consider a random sample of $m$ edges from the complete edge set $\binom{[n]}{2}$ with replacement, yielding a multiset $E'$. Now, for any fixed graph $G = (V, E)$, we consider the resulting induced sample $E \cap E'$. Our goal will be to show that with some good probability (say, $\geq 1/2$) that $E \cap E'$ provides a good estimate of all of the subset sizes in $G$ (when appropriately scaled). To conclude the proof, we will then use a Newman-esque derandomization to say that there \emph{exists} a set of $O(n^2)$ samples $E'_1, \dots E'_{O(n^2)}$, such that, \emph{for any graph} $G$, at least one of these samples $E'_{j}$ provides a good estimate of the subset sizes in $G$. Our final sketch then consists of (1) the index among the $O(n^2)$ samples, specifying which set of subsampled edges we are considering, and (2) the multiset of edges $E \cap E'$ for this selected subsample. With this intuition in mind, we now provide the formal proof. 

As mentioned above, we start by letting $E'$ denote a uniformly random sample of $m$ edges (with replacement) from $\binom{[n]}{2}$. We denote the edges in $E'$ by $e_1, \dots e_m$, and set $p = m/\binom{n}{2}$. Now, we consider a fixed graph $G = (V, E)$. We define the estimator $f: 2^V \rightarrow \mathbb{R}_{\geq 0}$ which estimates the number of edges $E_G[T]$ for any set $T \subseteq V$ as follows:
\begin{enumerate}
    \item If $|T| \leq \tau$, then $f(T) = 0$.
    \item If $|T| > \tau$, then \[
f(T)=\frac{\binom{n}{2}}{m}\sum_{i=1}^{m}\mathbf 1[e_i\in E_G[T]]
\qquad\text{when }|T|>\tau.
\]
\end{enumerate}
Trivially, for any set $T$ such that $|T| \leq \tau$, the induced error is bounded by $\tau \cdot |T|$. Thus, we focus only on the second case, and seek to show that $|f(T) - |E_G[T]|| \leq \tau \cdot |T|$. So, let us consider any such fixed set $T \subseteq V$, and let $k = |E_G[T]|$.

For each sampled edge $e_i$, we define a random variable $Y_i$, which is $\mathrm{Bern}(\rho)$, where $\rho = \frac{k}{\binom{n}{2}}$. The variable $Y_i$ indicates whether the $i$th sample reveals an edge in the induced subset corresponding to $T$. Our goal is to bound the probability that $|\sum_{i \in [m]} Y_i - m \cdot \rho|>p \cdot \tau \cdot |T|$ (and thus the estimator is within accuracy $\tau \cdot |T|$). Normalizing by $m$, this corresponds to bounding the probability that $\left |\frac{\sum_{i \in [m]} Y_i}{m} - \rho \right |> \frac{p \cdot \tau \cdot |T|}{m} = \frac{\tau |T|}{\binom{n}{2}}$.

By invoking \cref{fact:bernoulli}, we can immediately see that
\[
\Pr\left [ \left | \frac{\sum_{i \in [m]} Y_i}{m} - \rho \right | > \frac{\tau |T|}{\binom{n}{2}} \right ] \leq 2e^{ - \frac{m\left(\frac{\tau |T|}{\binom{n}{2}} \right)^2}{2\left (\rho + \frac{\tau |T|}{\binom{n}{2}} \right)}}.
\]
Using the fact that $\rho = \frac{k}{\binom{n}{2}} \leq \frac{\binom{|T|}{2}}{\binom{n}{2}} \leq \frac{|T|^2}{\binom{n}{2}}$ and that $\frac{\tau |T|}{\binom{n}{2}} \leq \frac{|T|^2}{\binom{n}{2}}$, we then get that $\rho + \frac{\tau |T|}{\binom{n}{2}} \leq \frac{2 |T|^2}{\binom{n}{2}}$. Plugging this in, we can simplify the above expression to 
\[
\Pr\left [ \left | \frac{\sum_{i \in [m]} Y_i}{m} - \rho \right | > \frac{\tau |T|}{\binom{n}{2}} \right ] \leq 2e^{ - \frac{m\left(\frac{\tau |T|}{\binom{n}{2}} \right)^2}{4 \frac{|T|^2}{\binom{n}{2}} }}.
\]

Now, our goal will eventually be to take a union bound over all sets $T$. Thus our goal is to ensure that 
\[
2e^{ - \frac{m\left(\frac{\tau |T|}{\binom{n}{2}} \right)^2}{4 \frac{|T|^2}{\binom{n}{2}} }}  \leq \frac{1}{n^{2|T|}},
\]
which implies that it suffices to choose 
\[
\frac{m\left(\frac{\tau |T|}{\binom{n}{2}} \right)^2}{4 \frac{|T|^2}{\binom{n}{2}} } \geq 3 |T| \log(n).
\]
Re-writing the above, it suffices to choose $m$ such that $m \geq \frac{12 |T| \log(n) \cdot \binom{n}{2}}{\tau^2 }$, or that
\[
m \geq \frac{6 n^3 \log(n)}{\tau^2}.
\]

For such a choice of $m$, by taking a union bound over all sets $T$, we obtain that the estimator $f$ is accurate to within error $\tau \cdot |T|$ for every set $T \subseteq V$ with probability $\geq 1 - \frac{1}{n}$. The space required by the estimator is exactly the space required to store the sampled edges, which is $\tilde{O}(m)$ bits.

Finally, we amplify this into a deterministic sketch for estimating the edge subset densities. To start, we consider sampling $E_1', \dots E_{n^2}'$ (each consisting of $m$ edges). For each multiset of sampled edges $E'_i$ and any graph $G$, we define $f_i: 2^V \rightarrow \mathbb{R}_{\geq 0}$ to be the edge-subset size estimator as defined above. We say that $f_i$ \emph{is good} for $G$ if for every $T \subseteq V$, 
\[
\left||E_G[T]| - f_i(T)\right| \leq \tau \cdot |T|.
\]

We first observe that 
\[
\Pr[\not \exists i \in [n^2]: f_i \text{ is good for }G ] \leq (1/n)^{n^2}.
\]
Thus, by taking a union bound over all graphs $G$, we see that 
\[
\Pr[\exists G: \not \exists i \in [n^2]: f_i \text{ is good for }G ] \leq 2^{n^2} \cdot (1/n)^{n^2} < 1.
\]
In particular, there must then \emph{exist} a collection of subsampled edges $E'_1, \dots E'_{n^2}$ such that \emph{for every} graph $G$, there is some choice of $i \in [n^2]$ such that the estimator corresponding to $E'_i$ is good for $G$. 

This then concludes the lemma: for a graph $G = (V, E)$, the resulting sketch is simply the first index $i \in [n^2]$ which is good for the graph $G$, along with the exact multiset of edges in $E'_i \cap E$.
\end{proof}

\subsection{The Complete Connectivity Sketch}

With the subset-size sketch established, we now give the proof of \cref{thm:n1.8}.

\begin{proof}[Proof of \cref{thm:n1.8}]
Recall that Alice is given the graph $G_A = (V, E_A)$, and must determine the message which she sends to Bob. To specify this message, we store several pieces of information. First, we decompose the vertex set $V$ into $n^{0.8}$ groups of $n^{0.2}$ vertices arbitrarily, which we denote by $C_1, \dots C_{n^{0.8}}$. Now, we store:
\begin{enumerate}
    \item For every vertex $v \in V$ and $i \in [n^{0.8}]$, the number of edges going from $v$ to $C_i$. 
    \item For every vertex $v \in V$, we store a sparse recovery sketch (as specified by \cref{clm:sparseRecoverySketch}) with $k = 2 \cdot n^{0.8}$, $q = n$, applied to the neighborhood of $v$. We denote this sketch as $P \cdot \Gamma(v)$, where $\Gamma(v) \in \zo^n$ is the indicator vector of $v$'s neighborhood. 
    \item The subset-size estimation sketch of \cref{lem:subsetSizeSketch} applied with $\tau = \frac{n^{3/5}}{100}$.
\end{enumerate}

As in \cref{def:universalGraph} and \cref{clm:coloringConnection}, we simply seek to prove that for any two graphs $G, H$ if the sketches of $G$ and $H$ are exactly equal, then it must be the case that $G - G \cap H$ and $H - H \cap G$ have the same connected components. 

So, let us suppose for the sake of contradiction that $G$ and $H$ receive the same sketch, yet $G - G \cap H$ and $H - H \cap G$ have different connected components. We will show that this is not possible.  To start, we let $A_G \in \zo^{n \times n}$ denote the adjacency matrix of $G$, and likewise use $A_H$ for $H$. We then let $D$ denote the difference $D: = A_G - A_H \in \{-1,0,1\}^{n \times n}$. There are now several things we can observe:
\begin{enumerate}
    \item First, it must be the case that for every $C_i$, that $D \cdot \mathbf{1}_{C_i} = 0$. Indeed, this is because for a vertex $v \in V$, $(D \cdot \mathbf{1}_{C_i})_v$ counts the number of edges going from $v$ to $C_i$ in $G$ minus the number of edges going from $v$ to $C_i$ in $H$. But, because we store this information in our sketch, and $G$ and $H$ had the same sketch, it must be that $(D \cdot \mathbf{1}_{C_i})_v = 0$.
    \item Next, we claim that $P \cdot D = 0$ for $P$ as defined in \cref{clm:sparseRecoverySketch}. Indeed, for a vertex $v$, the resulting sparse recovery sketch for $v$'s neighborhood is exactly $P \cdot (A_G)_v$, where $(A_G)_v$ refers to the $v$th column of $A_G$ (likewise, in $H$ the stored sketch for $v$ is $P \cdot (A_H)_v$). But, because we store this sparse recovery sketch at each vertex, and the sketches for $G$ and $H$ are the same, it must be that for every $v$, $P \cdot (A_H)_v = P \cdot (A_G)_v$, or equivalently, that $P \cdot D_v = 0$. Because this holds for every $v$, we then obtain that $P \cdot D = 0$.
\end{enumerate}

Next, because we assumed that $G - G \cap H$ and $H - H \cap G$ have different connected components, this means that there must be (WLOG) some cut $(S, V - S)$ such that no edges in $G - G \cap H$ cross $S$, while $H - H \cap G$ has at least one edge crossing $S$. We denote this number of crossing edges in $H - H \cap G$ by $b$, and call these the \emph{blue} edges. 

Now, for each group $C_j$, we define the vector $d^{(j)} = D \cdot \mathbf{1}_{S \cap C_j}$. We have the following claim:

\begin{claim}
    Let $d^{(j)} = D \cdot \mathbf{1}_{S \cap C_j}$ be defined as above. Then,
    \[
d^{(j)}_u = \begin{cases}
    \#\{\text{blue edges from }u \text{ to } C_j \cap \bar{S}\}: u \in S \\
    -\#\{\text{blue edges from }u \text{ to } C_j \cap S\}: u \notin S
\end{cases}.
\]
\end{claim}

\begin{proof}
We can re-write this as  \[
d^{(j)}_u = (\Gamma_{G - G \cap H}(u) - \Gamma_{H - G \cap H}(u)) \cdot \mathbf{1}_{S \cap C_j} = \Gamma_{G - G \cap H}(u)\cdot \mathbf{1}_{S \cap C_j} - \Gamma_{H - G \cap H}(u)\cdot \mathbf{1}_{S \cap C_j}.
\]
Now, we do a simple case analysis: if $u \notin S$, then it must be the case that there are no edges in $G-G\cap H$ from $u$ to $S \cap C_j$ (otherwise this would violate the property that the cut defined by $S$ is empty). So, if $u \notin S$, then $d^{(j)}_u = - \Gamma_{H - G \cap H}(u)\cdot \mathbf{1}_{S \cap C_j}$ which is exactly the (negative of the) number of blue edges going from $u$ to $C_j \cap S$ (and thus also the number of blue edges going from $u$ to $C_j$).

Otherwise, if $u \in S$, we use the key property that $\Gamma_{G - G \cap H}(u) \cdot \mathbf{1}_{C_j} = \Gamma_{H - G \cap H}(u) \cdot \mathbf{1}_{C_j}$ as proved above. This implies then that 
\[
d^{(j)}_u = (\Gamma_{G - G \cap H}(u) - \Gamma_{H - G \cap H}(u)) \cdot (\mathbf{1}_{S \cap C_j})= (\Gamma_{G - G \cap H}(u) - \Gamma_{H - G \cap H}(u)) \cdot (\mathbf{1}_{S \cap C_j} - \mathbf{1}_{C_j})
\]
\[
= (\Gamma_{H - G \cap H}(u) - \Gamma_{G - G \cap H}(u)) \cdot \mathbf{1}_{\bar{S} \cap C_j}.
\]
Again, it will be the case that $\Gamma_{G - G \cap H}(u) \cdot \mathbf{1}_{\bar{S} \cap C_j} = 0$, and so $d^{(j)}_u = \Gamma_{H - G \cap H}(u) \cdot \mathbf{1}_{\bar{S} \cap C_j}$ which exactly counts the number of blue edges from $u$ to $\bar{S} \cap C_j$ (and thus also the number of blue edges going from $u$ to $C_j$).

To summarize, we get
\[
d^{(j)}_u = \begin{cases}
    \#\{\text{blue edges from }u \text{ to } C_j \cap \bar{S}\}: u \in S \\
    -\#\{\text{blue edges from }u \text{ to } C_j \cap S\}: u \notin S
\end{cases}.
\]
\end{proof}

From the above claim, for the group $C_j$, if we let $B_j$ denote the number of blue-edge incidences at $C_j$ (counting internal edges twice), we obtain that $\Vert d^{(j)}\Vert_1 = B_j$, and moreover, each entry $|d^{(j)}_u| \leq n^{0.2}$ (recall that $C_j$ is of size at most $n^{0.2}$). Furthermore, because $d^{(j)} = D \cdot \mathbf{1}_{S \cap C_j}$ and $P \cdot D = 0$, then it must also be the case that 
\[
P \cdot d^{(j)} = P \cdot D \cdot \mathbf{1}_{S \cap C_j} = 0.
\]

We can now directly invoke \cref{clm:sparseRecoverySketch}. By our choices of $k$ and $q$ in the instantiation of $P$, it must be the case that if $d^{(j)}$ is non-zero, then it has at least $k = 2 \cdot n^{0.8}$ many non-zero entries. We call any group for which $d^{(j)}$ is non-zero an \emph{active} group. The above implies that for any active group $C_j$, it must be that $B_j \geq 2 \cdot n^{0.8}$.
To conclude, we will look at these active groups and show that they imply that there must be some subset $T \subseteq V$ such that the subset size estimation sketches for $H$ and $G$ differ by too much, thus contradicting the fact that $H$ and $G$ receive the same sketch. 

To this end, we let $g$ denote the number of active groups, and let $U$ denote the union of these groups. Because every blue edge has two endpoints, we then see that 
\[
2b = \sum_j B_j \geq 2 \cdot n^{0.8} \cdot g.
\]
At the same time, we also know that $|U| \leq n^{0.2} \cdot g$.

Finally, we look at the sets $X = S \cap U$ and $Y = \bar{S} \cap U$. In the graph $H - G \cap H$, all blue edges cross between $X$ and $Y$, whereas in the graph $G - G \cap H$, there are no edges which cross between $X$ and $Y$. Formally, this means that 
\[
\left | E_{G - G\cap H}[X, Y]\right | = 0,
\]
whereas 
\[
\left | E_{H - G\cap H}[X, Y]\right | \geq n^{0.8} \cdot g,
\]
Re-writing, this means that 
\[
\left | E_{H - G\cap H}[X, Y]\right | - \left | E_{G - G\cap H}[X, Y]\right | \geq n^{0.8} \cdot g,
\]
and thus 
\[
\left | E_{H}[X, Y]\right | - \left | E_{G}[X, Y]\right | \geq  n^{0.8} \cdot g,
\]
(as adding $G \cap H$ to both graphs adds the same number of crossing edges to both).  

Next, we observe that because $U$ is a union of groups $C_j$ and because for every $C_j$ and vertex $u$, the $|E_G[u, C_j]| = |E_H[u, C_j]|$, it must be that 
\[
\sum_{v \in X} |\mathrm{E}_G[v, U]| = \sum_{v \in X} |\mathrm{E}_H[v, U]|.
\]
Rewriting this, we obtain that 
\[
2 \cdot |E_G[X]| + |E_G[X, Y]| = 2 \cdot |E_H[X]| + |E_H[X, Y]|.
\]
Thus, we see that 
\[
2 \cdot \left ( |E_G[X]| -  |E_H[X]|\right ) = |E_H[X, Y]| - |E_G[X, Y]| \geq  n^{0.8} \cdot g,
\]
and therefore 
\[
\left | |E_G[X]| - |E_H[X]|\right | \geq \frac{1}{2}n^{0.8} \cdot g.
\]
This then gives a contradiction; by assumption, we had supposed that $G$ and $H$ received the same subset-size estimation sketch with $\tau = \frac{n^{0.6}}{100}$. On the set $X$ which is of size $\leq n^{0.2} \cdot g$, it would then be the case that the sketch (call this $f$) reports a value $f(X)$ such that 
\[
\left | f(X) - |E_G[X]| \right | \leq \frac{n^{0.6}}{100} \cdot n^{0.2} \cdot g,
\]
and likewise that 
\[
\left | f(X) - |E_H[X]| \right | \leq \frac{n^{0.6}}{100} \cdot n^{0.2} \cdot g,
\]
which would then imply that 
\[
\left | |E_G[X]| - |E_H[X]|\right | \leq \frac{n^{0.8} \cdot g}{50}.
\]
This gives a contradiction with the above lower bound on how edge densities in $X$ differ between $G$ and $H$, and thus it is not possible that $G$ and $H$ received the same sketch.
\end{proof}

\section{An $\widetilde{O}(n^{7/4})$ Bit Upper Bound}\label{sec:n74}

In this section, we prove the following stronger theorem:

\begin{theorem}\label{thm:n74proof}
        There is a protocol for the deterministic one-way connectivity problem in which Alice sends $m = \widetilde{O}(n^{7/4})$ bits.
\end{theorem}

While the overarching intuition is similar to the previous proof, we will require several new ideas. Our key building block will be the following lemma:

\begin{lemma}\label{lem:oneSidedCrossing}
    For any parameter $0 \leq k \leq n^2$, there is a (linear) sketch $\mathcal{S}_{\text{one-sided}}$ which requires space $\widetilde{O}(n \sqrt{k})$ such that, for any two graphs $G$, $H$ on $n$ vertices, if $\mathcal{S}_{\text{one-sided}}(G) = \mathcal{S}_{\text{one-sided}}(H)$, then it must be the case that:
    \begin{enumerate}
        \item For any cut $S \subseteq V$, if $\delta_{G - G \cap H}(S) = 0$, then $\delta_{H - H \cap G}(S) = 0$ or $\delta_{H - H \cap G}(S) > k$.
        \item For any cut $S \subseteq V$, if $\delta_{H - H \cap G}(S) = 0$, then $\delta_{G - G \cap H}(S)= 0$ or $\delta_{G - G \cap H}(S) > k$.
    \end{enumerate}
\end{lemma}

The way to interpret the above sketch is that if two graphs $G$ and $H$ receive the same sketch, then for any cut $S \subseteq V$, if $G$'s crossing edges are a strict subset of $H$'s crossing edges, then it must be the case that $H$ has \emph{many more} (i.e., $k$ many) crossing edges than $G$. In proving \cref{thm:n74proof}, we will use this linear sketch with $k \approx n^{3/2}$, in addition to some other sketches.

We start by showing how \cref{lem:oneSidedCrossing} implies \cref{thm:n74proof}, and then spend the remainder of this section proving \cref{lem:oneSidedCrossing}.

\subsection{Proof of \cref{thm:n74proof}}

Our starting point for the proof of \cref{thm:n74proof} is the following lemma:

\begin{lemma}\label{lem:cutEstimation}
    Consider a fixed parent graph $F$ on $m \geq n^{7/4}$ edges and $n$ vertices. There is a fixed set of subgraphs $M_1, \dots M_{m+1}$, each with $\leq \frac{2m}{3}$ many edges such that for every possible subgraph $G \subseteq F$, there is an index $i \in [m+1]$ such that for every cut $S \subseteq V$,
    \[
    \left | \delta_G(S) - 2 \cdot \delta_{G \cap M_i}(S)\right | \leq 4\sqrt{nm}.
    \]
\end{lemma}

\begin{proof}
    We prove this lemma via random sampling, along with a Newman-style derandomization, mirroring the proof of \cref{lem:subsetSizeSketch}. To start, let us fix an arbitrary graph $G \subseteq F$, and consider a procedure where we subsample the edges randomly at rate $1/2$, and assign weight $2$ to each sampled edge. Clearly, such a procedure produces unbiased estimates of the cut-sizes in $G$. If we let $\hat{G}$ denote the randomly sampled subgraph, we seek to show that for every $S \subseteq V$, 
    \[
    \left |\delta_G(S) - 2 \cdot \delta_{\hat{G}}(S) \right | \leq O(\sqrt{nm}).
    \]
    To do this, we fix a single cut $S$, and observe that the number of edges crossing this cut is naturally at most $m$. Using Hoeffding's inequality, we see that 
    \[
    \Pr \left [\left |\delta_G(S) - 2 \cdot \delta_{\hat{G}}(S) \right | > 4 \sqrt{nm} \right ] \leq 2e^{-8n} \leq 2e^{ -4n}.
    \]
    Taking a union bound over all $2^{n}$ possible cuts, we then see that 
    \[
    \Pr \left [\forall S \subseteq V, \left |\delta_G(S) - 2 \cdot \delta_{\hat{G}}(S) \right | \leq 4 \sqrt{nm} \right ] \geq 1 - 2e^{-3n}.
    \]
    Likewise, we can also observe that with probability $1 - 2^{- \Omega(n)}$ that when we sample $m$ edges at rate $1/2$, the number of surviving edges is $\leq 2m/3$ with probability $1 - 2^{- \Omega(n)}$ (if this event fails, we simply set that subgraph $M_i = \emptyset$). Thus, for large enough, constant $n$, we see that 
    \[
    \Pr[|M_i| \leq 2m/3 \text{ and }\forall S \subseteq V: \left | \delta_G(S) - 2 \cdot \delta_{G \cap M_i}(S)\right | \leq 4\sqrt{nm}] \geq 1/2.
    \]
    To conclude, we again consider an arbitrary graph $G$. We say that a sample of edges $M_i$ is \emph{good} for $G$ if $|M_i| \leq 2m/3$ and $\forall S \subseteq V: \left | \delta_G(S) - 2 \cdot \delta_{G \cap M_i}(S)\right | \leq 4\sqrt{nm}$. The above establishes that such a sample of edges $M_i$ is good for $G$ with probability $\geq 1/2$. Now, let us consider sampling $m+1$ such graphs $M_i$. For a fixed graph $G$, the probability that \emph{none} of the $M_i$ are good is bounded by $\leq (1/2)^{m+1}$. Thus, we obtain that 
    \[
    \Pr_{M_1, \dots M_{m+1}}[\exists G \subseteq F: \not \exists i \in [m+1]: M_i \text{ is good for } G] \leq 2^m \cdot (1/2)^{m+1} < 1. 
    \]
    In particular, this then implies that there \emph{exists} a choice of $M_1, \dots M_{m+1}$ such that \emph{for every graph} $G \subseteq F$, one of $M_1, \dots M_{m+1}$ is good for $G$. This concludes the lemma. 
\end{proof}

Now, we show how to use the above lemma to prove \cref{thm:n74proof}.

\begin{proof}[Proof of \cref{thm:n74proof}]
Our sketch is inductive. To start, we are only promised that our graph $G$ is a subset of the complete graph $F = K_n$. So, we invoke \cref{lem:cutEstimation} with this parent graph, and generate the subgraphs $M_1, \dots M_{m+1}$. We let $i$ denote the first index $i \in [m+1]$ for which $M_i$ is good for $G$. Our sketch for $G$ then consists of this index $i$, the sketch of \cref{lem:oneSidedCrossing} with $k = 8 \sqrt{nm}$ invoked on $G$, and the sketch applied recursively to $M_i \cap G$. Note that whenever the number of edges falls below $n^{7/4}$, we can simply send the entire remaining graph. Formally, if we let $i_F(G)$ denote the first good index for $G$ with parent graph $F$, let $\mathcal{S}_{\mathrm{one-sided}, k}$ denote the sketch of \cref{lem:oneSidedCrossing} with threshold $k$, and let $k_F = 8\sqrt{n |F|}$, then we have that the sketch $\chi_F(G)$ for $G$ when the parent graph is currently $F$ is exactly 
\[
\chi_F(G) = \left (i_F(G), \mathcal{S}_{\mathrm{one-sided}, k_F}(G), \chi_{M_{i_F(G)}}(G \cap M_{i_F(G)}) \right ).
\]
Note that the space bound of $\widetilde{O}(n^{7/4})$ follows immediately, as there are at most $O(\log(n))$ levels in the sketch (the number of edges in $F$ decreases by a $2/3$ factor in each level), and each level stores $O(\log(n))$ bits for the index $i_F(G)$ and at most $\widetilde{O}(n \sqrt{k_F})= \widetilde{O}(n \sqrt{\sqrt{n^3}}) = \widetilde{O}(n^{7/4})$ many bits for the one-sided sketch. All that remains is to prove the correctness of the sketch. For this, we use induction. Indeed, we suppose that for any parent graph $F$ with $< m$ edges that if $\chi_F(G) = \chi_F(H)$ (and that $G \subseteq F, H \subseteq F$) then it must be the case that the connected components of $G - G \cap H$ are the same as the connected components of $H - G \cap H$. Now, we prove the same must be true on $m$ many edges. 

We let $F$ denote a parent graph on $m$ edges and suppose that $\chi_F(G) = \chi_F(H)$. In particular, this implies that $i_F(G) = i_F(H)$ and therefore $M_{i_F(G)} = M_{i_F(H)}$ (we call this graph $M$ going forward). The inductive hypothesis therefore implies that the connected components of $G \cap M - (H \cap G \cap M)$ are the same as the connected components of $H \cap M - (G \cap H \cap M)$. Now, consider a cut $S$ for which $G - G \cap H$ is empty (similar reasoning will work for the case when $H - H \cap G$ is empty), but $H - H \cap G$ is not empty. For this cut $S$, it must also be the case that it is empty in $G \cap M - (H \cap G \cap M)$, as this graph is obtained by intersection with $G$. But, this then also means that the cut $S$ is empty in $H \cap M - (G \cap H \cap M)$. I.e., 
\[
\delta_{G\cap M - G\cap H \cap M}(S) = \delta_{H\cap M - G\cap H \cap M}(S) = 0.
\]
Adding the graph $G \cap H \cap M$ back, we then obtain that 
\[
\delta_{G\cap M}(S) = \delta_{H\cap M}(S).
\]
But, by \cref{lem:cutEstimation}, we know that 
\[
|\delta_G(S) - 2 \delta_{G\cap M}(S)| \leq 4 \sqrt{nm}
\]
and likewise 
\[
|\delta_H(S) - 2 \delta_{H\cap M}(S)| \leq 4 \sqrt{nm}.
\]
Thus, we see that 
\[
|\delta_G(S) - \delta_H(S)| \leq 8 \sqrt{nm}.
\]
This then finally yields a contradiction: indeed, if $\chi_F(G) = \chi_F(H)$, then it must be the case that $\mathrm{S}_{\mathrm{one-sided}, 8 \sqrt{nm}}(H) = \mathrm{S}_{\mathrm{one-sided}, 8 \sqrt{nm}}(G)$. But, if these sketches are equal, it must also be the case that if a cut $S$ in $G-G\cap H$ is empty, then if the same cut $S$ in $H-G\cap H$ is non-empty, it must be of size $> 8 \sqrt{nm}$. This contradicts the above inequality which states that $|\delta_G(S) - \delta_H(S)| \leq 8 \sqrt{nm}$.
\end{proof}

\subsection{Proof of \cref{lem:oneSidedCrossing}}

We now turn our attention towards proving \cref{lem:oneSidedCrossing}. To prove this, we will rely on two specific sketches which we define below.

The first sketch we will use is the following ``vertex cover-forcing'' sketch:

\begin{lemma}\label{lem:vertexCoverSketch}
    For any parameter $k$, there is matrix $P \in \F_q^{n \times \widetilde{O}(\sqrt{k})}$ (with $q = \mathrm{poly}(n)$) such that, for any skew-symmetric matrix $M \in \{-1,0,1\}^{n \times n}$, if $M$ has at most $k$ non-zero entries above its diagonal, and $P^T M P = 0$, then it must be the case that $\Supp(M)$ has a vertex cover of size $\leq 10\sqrt{k}$.
\end{lemma}

Intuitively, the above supplies a sketching matrix $P$ such that whenever $P$ yields the zero sketch, the matrix it is operating on must have all of its edges supported on a small set of vertices. The size of the stored sketch that we end up using will be $\widetilde{O}(n \sqrt{k})$ many bits. Next, we require the following modification of a typical sparse recovery sketch:

\begin{lemma}\label{lem:sparseRecoveryModified}
    For any parameter $t$, there is a matrix $X \in \{\pm 1\}^{n \times \widetilde{O}(t)}$ such that, for any set $U \subseteq [n]$ with $|U| \leq t$:
    \begin{enumerate}
        \item The rows $X|_U$ are linearly independent. 
        \item There exists a vector $h$ such that $(Xh)_v = 0$ when $v \in U$ and $(Xh)_v > 0$ when $v \notin U$.
    \end{enumerate}
\end{lemma}

With these sketches in hand, we can now proceed to the proof of \cref{lem:oneSidedCrossing}.

\begin{proof}[Proof of \cref{lem:oneSidedCrossing}.]
    For $k<1$, the empty sketch suffices, so assume $k\geq1$. To start, we specify the sketches that we will use. For a graph $G$, we let $A_G \in \{0,1\}^{n \times n}$ denote the adjacency matrix of the graph $G$. We then store $A_GP$ and $A_G X$, with $P$ initialized with the desired parameter $k$ in the statement of \cref{lem:oneSidedCrossing}, and $X$ initialized with $t = 10 \sqrt{k}$. Note that these sketches trivially satisfy the desired size bound, yielding an overall complexity of $\widetilde{O}(n \sqrt{k})$ many bits. 

    Now, it remains to show that the sketch provides the guarantee of \cref{lem:oneSidedCrossing}. To this end, we suppose that there is another graph $H$ such that $A_G X = A_H X$ and $A_G P = A_H P$. Equivalently, we can write this as $\Delta X = 0, \Delta P = 0$, where $\Delta = A_G - A_H$. It remains to prove that if there is some cut $S \subseteq V$ for which $\delta_{G - G \cap H}(S) = 0$, then it must be the case that $\delta_{H - G \cap H}(S)$ is either $0$ or $> k$ (note that the other case will follow from the exact same reasoning, so we focus on this case). To prove this, let us suppose that  $\delta_{H - G \cap H}(S)$ is \emph{not} zero. Our goal is then to show that $\delta_{H - G \cap H}(S) > k$. 

    Towards this end, we now define a variation of the matrix $M$ which captures exactly those edges crossing the cut $S$. We let $J_S \in \{0,1\}^{n \times n}$ denote the diagonal matrix with $1$'s exactly in the diagonal coordinates corresponding to $S$. We then let $M = J_S \Delta  - \Delta J_S$. We have the following claim about $M$:

    \begin{claim}\label{clm:Mform}
        Let $M = J_S \Delta  - \Delta J_S$, for $S, \Delta, G, H$ as defined above. Then, it must be the case that:
        \[
        M_{i,j} = \begin{cases}
            0 \text{ if } i,j \in S \\
            0 \text{ if } i, j \notin S \\
            -1 \text{ if } (i,j) \in H - G \cap H \text{ and } i \in S, j \notin S \\
            1 \text{ if } (i,j) \in H - G \cap H \text{ and } i \notin S, j \in S \\
            0 \text{ else}
        \end{cases}.
        \]
        Moreover, $M$ is skew-symmetric.
    \end{claim}
    \begin{proof}
        We first show that if $i,j \in S$ or $i,j \notin S$ then $M_{i,j} = 0$. This follows because $J_S \Delta$ keeps every row of $\Delta$ corresponding to vertices in $S$. $\Delta J_S$ instead captures all columns indexed by vertices in $S$. So, we can observe that 
        \[
        M_{i,j} = \Delta_{i,j} \cdot (\mathbf{1}[i \in S] - \mathbf{1}[j \in S]).
        \]
        When $i, j \in S$ or $i,j \notin S$ this then forces $M_{i,j} = 0$. When $i \in S$ but $j \notin S$, the entry then becomes $M_{i,j} = \Delta_{i,j}$, and when $i \notin S$ but $j \in S$, the entry becomes $M_{i,j} = - \Delta_{i,j}$. Finally, we can observe that the only edges crossing the cut $(S, V -S)$ are those edges which cross the cut in $H - G \cap H$ (recall that the cut is empty in $G - G \cap H$). So, when $i \in S, j \notin S$ and $(i,j) \in H - G \cap H$, we get that $\Delta_{i,j} = -1$, and so $M_{i,j}=-1$. Otherwise, when $i \notin S, j \in S$ and $(i,j) \in H - G \cap H$, we get that $\Delta_{i,j} = -1$, and so $M_{i,j}=1$.

        Note that by the above properties, $M$ will be skew-symmetric. This concludes the claim.
    \end{proof}

    For the matrix $M$ as defined above, we observe that because $\Delta X = 0, \Delta P = 0$, it must also be the case that $P^T M P = 0$ and $X^T M X = 0$. Indeed, this follows because 
    \[
    P^T M P = P^T (J_S \Delta - \Delta J_S) P = P^T J_S \Delta P - P^T \Delta J_S P = 0,
    \]
    where we have used that $P^T \Delta  = (\Delta P)^T = 0$. The same argument proves that $X^T M X = 0$.

    To finish the argument, we will assume that there are $\leq k$ edges crossing the cut $(S, V - S)$ in $H - G \cap H$ and show that this implies that there must be \emph{zero} edges crossing the cut. We will do this by showing that if $M$ has $\leq k$ non-zero entries above the diagonal (which corresponds exactly with $\leq k$ crossing edges over $S$ in $H - G \cap H$) then the combination of $X^T MX = 0$ and $P^T M P = 0$ forces $M = 0$. Thus, if $H - G \cap H$ has $> 0$ edges crossing $S$, this then forces there to be $> k$ many edges crossing $S$, yielding our desired statements. 

    So, to this end, we assume that $M$ has $\leq k$ non-zero entries above the diagonal. We can then immediately invoke \cref{lem:vertexCoverSketch}. Because $P^T M P = 0$ and $M$ has $\leq k$ non-zero entries above the diagonal, this means that the support of $M$ admits a vertex cover of size $\leq 10 \sqrt{k}$. We let $U \subseteq [n]$ denote this vertex cover. Our goal now is to show that the existence of this vertex cover, along with the fact that $X^T MX = 0$ implies that $M = 0$.

    To do this, we let $h$ denote the vector promised by \cref{lem:sparseRecoveryModified} for the set $U$. We then have the property that $(Xh)_v = 0$ for $v \in U$, and $(Xh)_v > 0$ for $v \notin U$. Furthermore, because $X^T M X =0$, it must also be the case that $X^T M X h = 0$. Now, we analyze the vector $M Xh$. We can re-write this as 
    \[
    (MXh)_a = \sum_{b \in [n]} M_{ab} (Xh)_b.
    \]
    Because $(Xh)_b = 0$ when $b \in U$, this means that 
    \[
    (MXh)_a = \sum_{b \notin U} M_{ab} (Xh)_b.
    \]
    However, if $a \notin U$, then by the property of $U$ being a vertex cover, there cannot be any edges of the form $(a, b)$ for $b \notin U$ (all edges must contain at least one vertex from the vertex cover). Hence, for $a \notin U$, $(MXh)_a = 0$. So, this immediately implies that $(MXh)$ is entirely supported on vertices $a \in U$.

    Finally, we analyze $X^T (MXh)$. By the above, we know that this product is $0$, and moreover that $MXh$ is supported entirely on $U$. At the same time however, by \cref{lem:sparseRecoveryModified}, we know that $X|_U$ is linearly independent. Thus, we see that  
    \[
    (X|_U)^T (MXh)|_U = 0,
    \]
    which implies that $(MXh)|_U = 0$, and thus $(MXh) = 0$. As above, we know that for $a \in U$
    \[
    (MXh)_a = \sum_{b \notin U} M_{ab} (Xh)_b = 0.
    \]
    We now crucially use the form of $M$ that we proved in \cref{clm:Mform}. If $a \in S$, then 
    \[
    (MXh)_a = \sum_{b \notin U: M_{ab} \neq 0} (-1) \cdot  (Xh)_b,
    \]
    and if $a \notin S$, then 
    \[
    (MXh)_a = \sum_{b \notin U: M_{ab} \neq 0} (1) \cdot  (Xh)_b.
    \]
    However, as promised by \cref{lem:sparseRecoveryModified}, we know that $(Xh)_b > 0$ for $b \notin U$. Thus, the only way for $(MXh)_a = 0$ is if there are \emph{no} edges $a,b$ for $a \in U$ and $b \notin U$. To summarize, at this point we have thus proved that $\Supp(M) \subseteq U \times U$, and $M$ is $0$ in all other positions. The final piece of our argument seeks to show that $M_{U \times U} = 0$ as well. For this, we recall that because 
    \[
    X^T M X = 0,
    \]
    it must also then be the case that 
    \[
    (X|_U)^T M_{U, U} X|_U = 0,
    \]
    as $M$ is supported only on $U \times U$. But, we previously established that $X|_U$ is linearly independent, and thus full-rank. So, for any $a \in U$, there exists a vector $w_a$ such that $X|_U \cdot w_a = \mathbf{1}_a$. We then see that for any $a, b \in U$
    \[
    M_{a,b} = (w_a)^T(X|_U)^TM_{U, U}X|_U w_b = w_a^T 0 w_b = 0.
    \]
    This then proves that $M = 0$. 

    This concludes the claim, as we showed that for any graphs $G, H$ such that $A_GP = A_H P$ and $A_G X = A_H X$, if there is a cut $S$ such that $S$ is empty in $G - G \cap H$ and non-empty in $H - G \cap H$, then this cut must be of size $> k$ in $H - G \cap H$.
\end{proof}

\subsection{Proof of \cref{lem:vertexCoverSketch}}

It still remains to prove \cref{lem:vertexCoverSketch} and \cref{lem:sparseRecoveryModified}. We start by proving \cref{lem:vertexCoverSketch} in this section. Our proof relies on a \emph{random matrix} $P$, defined over a sufficiently large field.

\begin{proof}[Proof of \cref{lem:vertexCoverSketch}.]
The case $k<1$ is immediate, so assume $k\geq1$. To start, we let the set $U \subseteq [n]$ denote all vertices in $[n]$ which have degree $\geq 10 \sqrt{k}$. Note that because the total number of edges in the support of $M$ is bounded by $k$, and thus the sum of degrees is bounded by $2k$, we have that 
\[
|U| \leq \frac{2k}{10 \sqrt{k}} \leq \frac{\sqrt{k}}{5}.
\]
Outside of the set $U$, we know that all vertices have degree $< 10 \sqrt{k}$.

Going forward, we will work with a matrix $P \in \F_q^{n \times d}$, for $d = \lceil 10 \sqrt{k}\rceil$. For such a matrix $P \in \F_q^{n \times d}$, we let $P_U$ denote the restriction of $P$ to those rows whose indices are in $U$. We next choose a $d \times h$ matrix $Z$ whose columns form a basis of $\ker P_U$. Thus, we have that 
\[
P_U Z = 0.
\]
Moreover, if $|U| \leq \sqrt{k}/5$, then it will also be the case that $h \geq d - |U| \geq \frac{19d}{20}$.

The important property about selecting this $Z$ matrix is that it ensures that any row of $PZ$ corresponding to an index in $U$ is zero. Thus, starting from the assumption that $P^T M P = 0$, we obtain that 
\[
Z^T P^T M P Z = 0,
\]
which now using the structure of $PZ$ implies that 
\[
(PZ)^T M' (PZ) = 0,
\]
where $M' \in \{-1,0,1\}^{n\times n}$ retains only the non-zero entries in $(V-U) \times (V-U)$ (i.e., erasing all entries in the support of $U$).

Thus, our goal now becomes the following: we want to show that there is a matrix $P$ such that, provided that $M$ has at most $k$ edges and $P^T M P = 0$,  \emph{for any} choice of $U$ with $|U|\leq d/20$ such that all vertices outside $U$ have degree $< d$, then there exists a choice of $Z$ such that $(PZ)^T M' (PZ) = 0$ if and only if $M' = 0$. We now prove that such a $P$ exists.

To do so, we choose $P$ uniformly at random from $\F_q^{n \times d}$. Now, for a fixed $U$, we first reveal $P_U$, and then choose a basis $Z$ of the kernel of $P_U$. Note that each row $p_v: v \notin U$ is still chosen completely independently of $P_U$. Thus, when we look at the product $PZ$, each row (for $v \notin U$) is distributed uniformly at random in $\F_q^h$.

As before, we are still assuming that $(PZ)^T M' (PZ) = 0$. This implies that 
\[
(PZ)_{\bar{U}}^T M'_{\bar{U}, \bar{U}} (PZ)_{\bar{U}} = 0.
\]
Note that $(PZ)_{\bar{U}}$ has $h$ many columns. We let $u=|U|$ and $\ell = \lfloor h/2\rfloor$, and split $2\ell$ of the columns of $(PZ)_{\bar{U}}$ into two disjoint groups of $\ell$ columns, which we denote by $A \in \F_q^{n-u \times \ell}$ and $B \in \F_q^{n-u \times \ell}$. By the above identity, we know that it must be the case that 
\[
A^T M'_{\bar{U}, \bar{U}} B = 0.
\]
The key property that we will use is that $A$ and $B$ are still \emph{independent, random} matrices, and thus it is unlikely for $A^T M'_{\bar{U}, \bar{U}} B = 0$ if $M'_{\bar{U}, \bar{U}}$ is not the zero matrix. We summarize this with the below claim:
\begin{claim}\label{clm:MatrixZeroProbability}
    Let $M'_{\bar{U}, \bar{U}}$ be a $\{-1, 0, 1\}$-valued matrix with at most $k$ non-zero entries above the diagonal. Suppose further that every column and every row has at most $10 \sqrt{k}$ non-zero entries. Let $R \geq 1$ denote the rank of $M'_{\bar{U}, \bar{U}}$ over $\F_q$, where $q\geq16$. Then, for independent uniformly random matrices $A, B \in \F_q^{n-u \times \ell}$,
    \[
    \Pr[A^T M'_{\bar{U}, \bar{U}} B = 0] \leq q^{-\ell \cdot \min(\ell, R)/2}.
    \]
\end{claim}

\begin{proof}[Proof of \cref{clm:MatrixZeroProbability}]
We first prove the above claim. For ease of notation, we write  $N=M'_{\bar U,\bar U}$ and $r=\min\{\ell,R\}$. We then factor $N$ as $N = CD$, where $C \in \F_q^{(n-u) \times R}$ has full column rank and $D \in \mathbb{F}_q^{R \times (n-u)}$ has full row rank. We can then see that the matrices $L = A^T C\in\mathbb F_q^{\ell\times R}$ and $H = DB\in\mathbb F_q^{R\times\ell}$ are independent uniform matrices. Moreover, $A^TNB=LH$.

Now, towards proving our claim, let us suppose $L$ has rank $j$. Its kernel has dimension $R-j$, so a uniform vector in $\mathbb F_q^R$ belongs to $\ker L$ with probability $q^{-j}$. The $\ell$ columns of $H$ are independent and uniform, which thus means that
\[
    \Pr[LH=0\mid L]=q^{-j\ell}.
\]
We next must also account for the probability that $L$ has rank $j$. There are at most
\[
    \binom{\ell}{j}q^{Rj+j(\ell-j)}
\]
rank-$j$ matrices in $\mathbb F_q^{\ell\times R}$.
One way to see this is to choose $j$ rows forming a basis of the row space, specify their $Rj$ entries, and specify each remaining row by its $j$ coefficients in this basis.

Dividing by the $q^{R\ell}$ equally likely choices of $L$ yields
\[
    \Pr[\operatorname{rank}(L)=j]
    \le \binom{\ell}{j}q^{-(\ell-j)(R-j)}.
\]
Thus,
\begin{align*}
    \Pr[A^TNB=0]
    &=\sum_{j=0}^{r}
      \Pr[\operatorname{rank}(L)=j]\,q^{-j\ell}\\
    &\le \sum_{j=0}^{r}
      \binom{\ell}{j}q^{-(\ell-j)(R-j)-j\ell}\\
    &=\sum_{j=0}^{r}
      \binom{\ell}{j}q^{-[R(\ell-j)+j^2]}.
\end{align*}

For every $0\le j\le r$, we have $R\ge r$ and $\ell-j\ge 0$.
Therefore,
\begin{align*}
    R(\ell-j)+j^2
    &\ge r(\ell-j)+j^2\\
    &=\ell r-\frac{r^2}{4}
      +\left(j-\frac r2\right)^2\\
    &\ge \ell r-\frac{r^2}{4}
     \ge \frac34\ell r,
\end{align*}
where the last inequality uses $r\le\ell$.
It follows that
\[
    \Pr[A^TNB=0]
    \le 2^\ell q^{-3\ell r/4}
    \le q^{-\ell r/2},
\]
using $q\ge16$ and $r\ge1$ to obtain
$2^\ell\le q^{\ell/4}\le q^{\ell r/4}$.
\end{proof}

Now, with \cref{clm:MatrixZeroProbability} in hand, and using that $\ell \geq 3d/8$, we see that for any $M$ with $\leq k$ non-zero entries above the diagonal, and a rank $R$ remainder $M'$,
\[
\Pr[(PZ)^T M' (PZ) = 0 ] \leq q^{-d \min(d, R)/16}.
\]
Our goal is to take a union bound over all possible rank $R$ remainders $M'$ and all possible choices of the high-degree vertices $U$. Thus, we next bound the number of possible matrices $M'$. This is where we explicitly rely on the low-degreeness of the matrix $M'$:

\begin{claim}\label{clm:matrixNumberBound}
    Let $N \in \{-1, 0, 1\}^{(n-u) \times (n-u)}$ be a matrix such that each row and column has at most $t$ non-zero entries, overall there are at most $k$ non-zero entries, and $N$ has rank $R$ over $\F_q$. The total number of such matrices $N$ is bounded by \[
    \min\left\{(3n)^{2R(t+2)}, (3n)^{2k+2}\right\}.
    \]
\end{claim}

\begin{proof}[Proof of \cref{clm:matrixNumberBound}]
If $R=0$, then $N=0$, so there is exactly one such matrix and the claimed bound holds. Going forward, we assume that $R\ge1$.

We next prove the bound $(3n)^{2R(t+2)}$. Since $N$ has rank $R$, there exist sets $I,J\subseteq[n-u]$, with $|I|=|J|=R$, such that $N_{I,J}$ is invertible over $\mathbb F_q$. We claim that $N$ is determined by these index sets, its rows indexed by $I$, and its columns indexed by $J$. Indeed (over $\F_q$, with odd characteristic),
\[
    N=N_{[n-u],J}(N_{I,J})^{-1}N_{I,[n-u]}.
\]

There are at most $\binom{n-u}{R}^2\le n^{2R}$ choices for $I$ and $J$. Each of the $2R$ recorded rows and columns has at most $t$ nonzero entries, all belonging to $\{-1,1\}$. The number of possibilities for any one of these vectors is at most
\[
    \sum_{s=0}^{\min\{t,n-u\}}\binom{n-u}{s}2^s
    \le \sum_{s=0}^{t}(2n)^s
    \le (3n)^{t+1}.
\]
Thus, we can bound the number of possible matrices by
\[
    n^{2R}\bigl((3n)^{t+1}\bigr)^{2R}
    \le (3n)^{2R(t+2)}.
\]

For the second bound, we simply count matrices by their nonzero positions and signs, without imposing any rank or degree restriction. A matrix with exactly $s$ nonzero entries has at most $\binom{(n-u)^2}{s}2^s$ possible descriptions. Hence the number of matrices with at most $k$ nonzero entries is at most
\[
    \sum_{s=0}^{\min\{k,(n-u)^2\}}\binom{(n-u)^2}{s}2^s
    \le \sum_{s=0}^{k}(2n^2)^s \leq (3n)^{2k+2}.
\]
This finishes the proof. 
\end{proof}

Finally, to conclude we show that there exists a single choice of $P$ such that it ``detects'' any valid non-zero $M_{\bar{U}, \bar{U}}$. To do this, we take a union bound over all possible choices of $U$ and all possible valid choices of $M$. We use the fact that there are at most $n^{u+1} \leq n^{d/20+1}$ sets of size $\leq u$, where $u=\lfloor d/20\rfloor$. Thus, 
\[
\Pr_P[\exists U \subseteq [n], |U| \leq u, \exists \text{ valid }M'\neq0: (PZ)^T M' (PZ) = 0] 
\]
\[
\leq n^{d/20+1} \cdot \left ( \sum_{R = 1}^{d} q^{-dR /16} \cdot (3n)^{2R(d+2)}  + q^{-d^2/16} \cdot (3n)^{4k+2}\right ).
\]
Choosing $q = 3^{\lceil 1000 \log_3(3n) \rceil}$, and letting $E$ be the event $\exists U \subseteq [n], |U| \leq u, \exists \text{ valid }M'\neq0: (PZ)^T M' (PZ) = 0$, we get that 
\[
\Pr_P[E] \leq n^{d/20+1} \cdot \left ( \sum_{R = 1}^{d} (3n)^{-60dR} \cdot (3n)^{2R(d+2)}  + (3n)^{-60d^2} \cdot (3n)^{4k+2}\right )
\]
\[
\leq n^{d/20+1} \cdot \left ( \sum_{R = 1}^{d} (3n)^{-50dR}  + (3n)^{-50d^2} \right ) \leq n^{d/20+1} \cdot (3n)^{-40d} < 1.
\]

In particular, this then guarantees the existence of a matrix $P \in \F_q^{n \times d}$ such that for any skew-symmetric $M$ with at most $k$ non-zero entries above the diagonal, $P^T M P = 0$ implies that $U$ is a vertex cover of $\Supp(M)$. This latter part follows because $P^T M P = 0$ only if, for the choice of $U$ being all vertices of degree $\geq 10 \sqrt{k}$, $(PZ)^T M (PZ) = (PZ)^T M' (PZ) = 0$, which is only possible if $M'$ itself is zero, by our preceding argument. Since $|U|\leq\sqrt{k}/5\leq10\sqrt{k}$, this then finishes the proof of \cref{lem:vertexCoverSketch}.
\end{proof}

\subsection{Proof of \cref{lem:sparseRecoveryModified}}

Finally to conclude this section, we give a formal proof of \cref{lem:sparseRecoveryModified}, as this is the only missing ingredient in the proof of \cref{thm:n74proof}. To construct our desired matrix $X$ for \cref{lem:sparseRecoveryModified}, we will randomly sample such a sign matrix, and argue that it satisfies the properties with probability $> 0$. As part of this argument, we will require the following fact:

\begin{fact}[See, for instance, Corollary 2.5 in \cite{mendelson2005reconstruction}]\label{fact:subgaussian}
There are absolute constants $C_0, c_0 > 0$ such that a random sign matrix $A \in \{-1,1\}^{D \times n}$ with independent uniformly signed entries satisfies, with probability at least $1 - e^{-c_0 D}$, for every $v \in \ker A$:
\[
\Vert v \Vert_2 \leq C_0 \cdot \sqrt{\frac{\log(2n)}{D}} \cdot \Vert v \Vert_1.
\]
\end{fact}

For an exact translation of Corollary 2.5 in \cite{mendelson2005reconstruction} to the above form, we refer the reader to \cref{sec:appendixSubgaussian}.

\begin{proof}[Proof of \cref{lem:sparseRecoveryModified}.]
To construct this matrix $X$, we will show that a random matrix $X$ satisfies these properties with $> 0$ probability. We sample a matrix $A \in \{\pm 1\}^{D \times n}$, where $D = O(t \log(n))$. We write $a_i$ for the $i$th column of $A$, and define $x_i = (1, a_i) \in \{\pm 1\}^{D+1}$, with $x_i^T$ as the $i$th row of $X$. We now prove several properties of these vectors:

\begin{claim}\label{clm:restrictionNormBound}
    Let $A$ be defined as above. Then, for a sufficiently large choice of $D = \Theta(t\log(n))$, with probability $1 - 1 / \mathrm{poly}(n)$ for every $w \in \ker A$ and every $U \subseteq [n]$ with $|U| \leq t$,
    \[
    \Vert w_U \Vert_1 \leq \frac{1}{4} \Vert w \Vert_1.
    \]
\end{claim}

\begin{proof}
    We see that 
    \[
    \Vert w_U \Vert_1 \leq \sqrt{|U|}\Vert w_U \Vert_2
    \]
    by Cauchy-Schwarz. Because $|U| \leq t$, we then obtain that 
    \[
    \sqrt{|U|}\Vert w_U \Vert_2 \leq \sqrt{t} \Vert w \Vert_2.
    \]
    To conclude, we then invoke \cref{fact:subgaussian}, which implies that 
    \[
    \Vert w \Vert_2 \leq C_0 \cdot \sqrt{\frac{\log(2n)}{D}} \cdot \Vert w \Vert_1,
    \]
    which for a sufficiently large choice of $D$ implies that 
    \[
    \Vert w_U \Vert_1 \leq \frac{1}{4} \cdot \Vert w \Vert_1.
    \]
    The probability bound comes from the usage of \cref{fact:subgaussian}.
\end{proof}

Using the above, we can now show that the collections $x_u: u \in U$ (for $U$ sufficiently small) are linearly independent.

\begin{claim}\label{clm:independence}
    Let $x_v: v \in [n]$ be defined as above and satisfy the good event of \cref{clm:restrictionNormBound}. Then, for any $U \subseteq [n], |U| \leq t$, it is the case that $x_u: u \in U$ are linearly independent.
\end{claim}

\begin{proof}
    Suppose for the sake of contradiction that $|U| \leq t$ yet $\sum_{u \in U} c_u x_u = 0$. Looking only at the last $D$ coordinates, we obtain a vector $w_u = c_u$ supported on $U$ such that $w \in \ker A$. \cref{clm:restrictionNormBound} then implies that 
    \[
    \Vert w \Vert_1 = \Vert w_U \Vert_1 \leq \frac{1}{4} \Vert w \Vert_1.
    \]
    This forces every $c_u = 0$.
\end{proof}

We now use these claims to prove the following property of convex combinations of vectors $U \subseteq [n]$:

\begin{claim}\label{clm:convSpanIntersection}
   Suppose $x_v: v \in [n]$ satisfy the good event of \cref{clm:restrictionNormBound}, and let $U \subsetneq [n]$ with $|U|\leq t$. Then,
    \[
    \mathrm{conv}(x_v: v \notin U) \cap \mathrm{span}(x_u: u \in U) = \emptyset.
    \]
\end{claim}

\begin{proof}
    We prove this by contradiction. Suppose that the two sets intersected. This would then imply that there is a selection of $\alpha_u \in \mathbb{R}$ and a selection of $\beta_v \in [0,1]$ such that $\sum_{v: v \notin U}\beta_v = 1$ such that 
    \[
    \sum_{v \notin U} \beta_v x_v = \sum_{u \in U} \alpha_u x_u.
    \]
    Because the first coordinate of every $x$ vector is $1$, this implies that $\sum_{u \in U} \alpha_u = 1$. This already then implies that $U \neq \emptyset$ (as otherwise this sum would be $0$). Thus, we write that $\Vert \alpha \Vert_1 \geq 1$.

    Now, we analyze the remaining $D$ coordinates of the $x$ vectors. Using the coefficients $\alpha, \beta$, we can create a vector $w \in \mathrm{ker} A$, with $w_u =-  \alpha_u: u \in U$, $w_v = \beta_v: v \notin U$.

    Importantly, we see that $\Vert w_U \Vert_1 = \Vert \alpha \Vert_1$, and the total mass is $\Vert w \Vert_1 = \Vert \alpha \Vert_1 + 1$. Because $\Vert \alpha \Vert_1 \geq 1$, this then implies that 
    \[
    \Vert w_U \Vert_1 \geq \frac{1}{2} \Vert w \Vert_1,
    \]
    which contradicts \cref{clm:restrictionNormBound}. Thus, we obtain our desired claim.
\end{proof}

Next, we have the following claim about hyperplanes separating points and convex sets:

\begin{claim}\label{clm:separatingHyperplane}
    Let $U \subsetneq [n]$ with $|U|\leq t$, and let $L = \mathrm{span}(x_u: u \in U)$ and let $\pi: \mathbb{R}^{D+1} \rightarrow L^{\perp}$ be an orthogonal projection. Let $C = \mathrm{conv}(\pi x_v: v \notin U)$. Then, there exists a vector $y \in L^{\perp}$ such that for every $z \in C$,
    \[
    \langle y, z \rangle > 0.
    \]
\end{claim}

\begin{proof}
    This follows from observing that when $U \neq [n]$, $C$ is a non-empty, convex, compact set. Furthermore, as we established above, \cref{clm:convSpanIntersection} implies that 
    \[
    \mathrm{conv}(x_v: v \notin U) \cap \mathrm{span}(x_u: u \in U) = \emptyset.
    \]
    Because $\mathrm{conv}(\pi x_v: v \notin U) = \pi \mathrm{conv}(x_v: v \notin U)$, it must then be the case that $0 \notin C := \mathrm{conv}(\pi x_v: v \notin U)$, as otherwise, some $x\in\mathrm{conv}(x_v:v\notin U)$ would satisfy $\pi x=0$ and hence $x\in L$, contradicting the fact that $\mathrm{conv}(x_v: v \notin U) \cap \mathrm{span}(x_u: u \in U) = \emptyset.$
    
    We can then use that for any point $p$ and compact, closed, convex set $C$ such that $p \notin C$, there is a choice of a vector $y$ such that $\langle z-p, y \rangle > 0$ for every $z\in C$ (see, for instance, \cite{vandenberghe2004convex}, 2.5.1). In our case, we use $p = 0$ and $C = \mathrm{conv}(\pi x_v: v \notin U)$ to conclude the existence of such a $y \in L^{\perp}$.
\end{proof}

Finally, we are ready to conclude. For $U$ as defined above, and $y, \pi$ as in \cref{clm:separatingHyperplane}, we let $h(x) = \langle y, \pi x\rangle$. For $u \in U$, we see that 
\[
\pi x_u = 0,
\]
so $h(x_u) = 0$. For $v \notin U$, we have $\pi x_v \in C$, and so $h(x_v) > 0$. If $U = [n]$, we can simply use $h = 0$. Note that because the same matrix $A$ satisfies \cref{clm:independence} and \cref{clm:restrictionNormBound} for \emph{every} $U \subseteq [n], |U| \leq t$, we obtain the stated claim simultaneously for all such $U$. This concludes the claim.
\end{proof}

\section{A Streaming Implementation}\label{sec:streaming}

In this section, we adapt the aforementioned sketches to show deterministic \emph{dynamic streaming algorithms} for connectivity which use $n^{2 - \Omega(1)}$ bits of space. To start, we have the following definition:

\begin{definition}
    For a number of vertices $n$, we say that a sequence of edge insertions and deletions $(e_i, \Delta_i): i \in [T]$ with $\Delta_i \in \{\pm 1\}$ is a dynamic stream of length $T$. We also use the term \emph{dynamic stream with multiplicity} $n^{\eps}$ to refer to a sequence of edge insertions and edge deletions $(e_i, \Delta_i): i \in [T]$ where any edge slot $(u,v)$ is involved in at most $n^{\eps}$ insertions or deletions. 
\end{definition}

Note that we always assume that a dynamic stream defines a \emph{valid graph} in the sense that for every $(u,v)$ it is the case that 
\[
\sum_{i \in [T]: e_i = (u,v)} \Delta_i \in \zo.
\]

With this terminology, we will show the following:
\begin{theorem}\label{thm:streaming}
    Fix a number of vertices $n$ and a parameter $\eps \geq 0$. There is a deterministic streaming algorithm which, for any dynamic stream with multiplicity $n^{\eps}$, reports the connected components of the resulting graph using space only $\widetilde{O}\left (n^{15/8 + 3\eps/8} \right)$ bits of space. In particular, for $\eps = 1/3 - \Omega(1)$, this requires space $n^{2 - \Omega(1)}$.
\end{theorem}

\begin{proof}
    We modify the procedure of \cref{thm:n1.8}. We denote the dynamic stream of edges by $(e_i, \Delta_i): i \in [T]$ with $\Delta_i \in \{\pm 1\}$. We denote the multi-graph of all insertions (with multiplicity) by $G^{+}$ and the multi-graph of all deletions by $G^{-}$, and let $G = G^+ - G^-$. We denote the vertex set by $V$, and just as in the proof of \cref{thm:n1.8}, we break the vertex set up into smaller groups. For our purposes, we break $V$ into $W$ many groups, each of size $n/ W$. We denote these groups by $C_1, \dots C_W$.

    Now, we initialize $2 \cdot n^{\eps}$ many buffers, which we denote by $B^{+}_1, \dots B^{+}_{n^{\eps}}$ and $B^{-}_1, \dots B^{-}_{n^{\eps}}$. We maintain the invariant that no buffer ever contains more than $n \cdot W$ many edges. Whenever we see an insertion of the form $(e, +1)$, we insert $e$ into the earliest buffer $B^+_i$ which does not already contain that edge (thus ensuring that $B_1^+ \supset B_2^+ \supset \dots \supset B_{n^{\eps}}^+$ and similarly for $B^-$). Likewise, whenever we see a deletion $(e, -1)$, we insert $e$ into the earliest buffer $B^-_i$ which does not already contain that edge. 

    For every vertex $v \in V$, we also store some auxiliary sketches: 
    \begin{enumerate}
        \item We store the number of edges going from $v$ to $C_i: i \in [W]$.
        \item We store a sparse recovery sketch of the neighborhood of $v$, with parameter $k = 2 \cdot W, q = n$. We denote this sketch by $P \cdot \Gamma(v)$ as in the proof of \cref{thm:n1.8}.
    \end{enumerate}

    Now, whenever the buffer $B^{+}_1$ reaches size $n \cdot W$, we apply the sketch of \cref{lem:subsetSizeSketch} with parameter $\tau$.\footnote{Note that the sketch of \cref{lem:subsetSizeSketch} is technically non-uniform, though it is simple to make it uniform. When we wish to compress the buffer $B$ we can simply enumerate all length $m$ multi-sets of edges from the complete graph, with $m \approx 6n^3\log(n) / \tau^2$. For each candidate set of edges, we enumerate all vertex subsets $T$, and check whether the edge density estimate is within $\tau |T|$. We can then \emph{explicitly retain} the first candidate set of edges which passes this check (and its intersection with the buffer) in the same space $\tilde{O}(n^3 / \tau^2)$. So, storing this candidate requires space $\tilde{O}(n^3 / \tau^2)$, and the working space to compute the sketch is only $\tilde{O}(nW +n)$ bits.} We maintain a counter $\mathrm{Count}_+$ which counts the number of times we perform this sketching operation. We store this sketch explicitly, and denote it by $f^{+}_{\mathrm{Count}_+}$, and then delete all of the edges from $B^{+}_1$ (and shift edges from later buffers back into earlier buffers if possible). We likewise perform this operation for $B^-_1$, and denote those sketches by $f^{-}_{\mathrm{Count}_-}$. When the stream terminates, we retain the remaining buffers and let $r(S)$ be their exact net number of edges induced by $S$. Because this sketch is only applied when $B_1$ contains $n \cdot W$ many edges, this operation will be applied a total of 
    \[
    \gamma^+, \gamma^- \leq \frac{n^{2 + \eps}}{nW} = \frac{n^{1 + \eps}}{W}
    \]
    many times. We denote these sketches by $f^{+}_{1}, \dots f^{+}_{\gamma^+}$ and $f^{-}_{1}, \dots f^{-}_{\gamma^-}$.

    At the termination of the stream, we then compute 
    \[
    f = r + \sum_{i \in [\gamma^+]} f^+_i - \sum_{j \in [\gamma^-]} f^-_j.
    \]
    By the properties of \cref{lem:subsetSizeSketch}, we then immediately can see that for every $S \subseteq V$
    \[
    \left|f(S) - |E_G[S]|\right| \leq 2 \cdot \frac{n^{1 + \eps}}{W} \cdot \tau \cdot |S|.
    \]

    It remains to prove that the above function $f$, along with the quantities $E[v, C_i]: i \in [W]$ and the sparse recovery sketches $P \cdot \Gamma(v)$ suffice for determining the connected components of the graph $G$. This proof closely follows the argument from \cref{sec:n18}, with the only difference being that our density estimation has \emph{slightly worse} than the counterpart in that section. For completeness, we still present the full argument below.

    To prove this, let us suppose that there is another graph $H$ for which all of the sketches and function $f$ match, yet there is some cut $S \subseteq V$ for which $S$ is empty in $G$ yet non-empty in $H$ (the reverse case will follow by the same logic). 

    As in the proof of \cref{thm:n1.8}, this implies that if we let $A_G \in \zo^{n \times n}$ and $A_H \in \zo^{n \times n}$ then we can observe the following for the difference matrix $D=A_G-A_H$:

    \begin{enumerate}
    \item First, it must be the case that for every $C_i$, that $D \cdot \mathbf{1}_{C_i} = 0$. Indeed, this is because for a vertex $v \in V$, $(D \cdot \mathbf{1}_{C_i})_v$ counts the number of edges going from $v$ to $C_i$ in $G$ minus the number of edges going from $v$ to $C_i$ in $H$. But, because we store this information in our sketch, and $G$ and $H$ had the same sketch, it must be that $(D \cdot \mathbf{1}_{C_i})_v = 0$.
    \item Next, we claim that $P \cdot D = 0$ for $P$ as defined in \cref{clm:sparseRecoverySketch}. Indeed, for a vertex $v$, the resulting sparse recovery sketch for $v$'s neighborhood is exactly $P \cdot (A_G)_v$, where $(A_G)_v$ refers to the $v$th column of $A_G$ (likewise, in $H$ the stored sketch for $v$ is $P \cdot (A_H)_v$). But, because we store this sparse recovery sketch at each vertex, and the sketches for $G$ and $H$ are the same, it must be that for every $v$, $P \cdot (A_H)_v = P \cdot (A_G)_v$, or equivalently, that $P \cdot D_v = 0$. Because this holds for every $v$, we then obtain that $P \cdot D = 0$.
\end{enumerate}

Next, because we assumed that $G$ and $H$ have different connected components, it must also be the case that $G - G \cap H$ and $H - H \cap G$ have different connected components, as the same cut $(S, V - S)$ satisfies that no edge in $G - G \cap H$ cross $S$, while $H - H \cap G$ has at least one edge crossing $S$. We denote this number of crossing edges in $H - H \cap G$ by $b$, and call these the \emph{blue} edges. 

Now, for each group $C_j$, we define the vector $d^{(j)} = D \cdot \mathbf{1}_{S \cap C_j}$. We have the following claim, exactly as in \cref{thm:n1.8}.

\begin{claim}\label{clm:blueEdgeVector}
    Let $d^{(j)} = D \cdot \mathbf{1}_{S \cap C_j}$ be defined as above. Then,
    \[
d^{(j)}_u = \begin{cases}
    \#\{\text{blue edges from }u \text{ to } C_j \cap \bar{S}\}: u \in S \\
    -\#\{\text{blue edges from }u \text{ to } C_j \cap S\}: u \notin S
\end{cases}.
\]
\end{claim}

From the above claim, for the group $C_j$, if we let $B_j$ denote the number of blue-edge incidences at $C_j$ (counting internal edges twice), we obtain that $\Vert d^{(j)}\Vert_1 = B_j$, and moreover, each entry $|d^{(j)}_u| \leq \frac{n}{W}$ (recall that $C_j$ is of size at most $\frac{n}{W}$). Furthermore, because $d^{(j)} = D \cdot \mathbf{1}_{S \cap C_j}$ and $P \cdot D = 0$, then it must also be the case that 
\[
P \cdot d^{(j)} = P \cdot D \cdot \mathbf{1}_{S \cap C_j} = 0.
\]

We can now directly invoke \cref{clm:sparseRecoverySketch}. By our choices of $k$ and $q$ in the instantiation of $P$, it must be the case that if $d^{(j)}$ is non-zero, then it has at least $k = 2 \cdot W$ many non-zero entries. We call any group for which $d^{(j)}$ is non-zero an \emph{active} group. The above implies that for any active group $C_j$, it must be that $B_j \geq 2 \cdot W$.
To conclude, we will look at these active groups and show that they imply that there must be some subset $T \subseteq V$ such that the subset size estimation sketches for $H$ and $G$ differ by too much, thus contradicting the fact that $H$ and $G$ receive the same sketch. 

To this end, we let $g$ denote the number of active groups, and let $U$ denote the union of these groups. Because every blue edge has two endpoints, we then see that 
\[
2b = \sum_j B_j \geq 2 \cdot W \cdot g.
\]
At the same time, we also know that $|U| \leq \frac{n}{W} \cdot g$.

Finally, we look at the sets $X = S \cap U$ and $Y = \bar{S} \cap U$. In the graph $H - G \cap H$, all blue edges cross between $X$ and $Y$, whereas in the graph $G - G \cap H$, there are no edges which cross between $X$ and $Y$. Formally, this means that 
\[
\left | E_{G - G\cap H}[X, Y]\right | = 0,
\]
whereas 
\[
\left | E_{H - G\cap H}[X, Y]\right | \geq  W \cdot g,
\]
Re-writing, this means that 
\[
\left | E_{H - G\cap H}[X, Y]\right | - \left | E_{G - G\cap H}[X, Y]\right | \geq W \cdot g,
\]
and thus 
\[
\left | E_{H}[X, Y]\right | - \left | E_{G}[X, Y]\right | \geq  W \cdot g,
\]
(as adding $G \cap H$ to both graphs adds the same number of crossing edges to both).  

Next, we observe that because $U$ is a union of groups $C_j$ and because for every $C_j$ and vertex $u$, the $|E_G[u, C_j]| = |E_H[u, C_j]|$, it must be that 
\[
\sum_{v \in X} |\mathrm{E}_G[v, U]| = \sum_{v \in X} |\mathrm{E}_H[v, U]|.
\]
Rewriting this, we obtain that 
\[
2 \cdot |E_G[X]| + |E_G[X, Y]| = 2 \cdot |E_H[X]| + |E_H[X, Y]|.
\]
Thus, we see that 
\[
2 \cdot \left ( |E_G[X]| -  |E_H[X]|\right ) = |E_H[X, Y]| - |E_G[X, Y]| \geq  W \cdot g,
\]
and therefore 
\[
\left | |E_G[X]| - |E_H[X]|\right | \geq W \cdot g/2.
\]
This then gives a contradiction; 
by assumption, we had supposed that $G$ and $H$ received the same subset-size estimation sketch with accuracy $2 \cdot \frac{n^{1+\eps}\tau}{W}$. On the set $X$ which is of size $\leq \frac{n}{W} \cdot g$, it would then be the case that the sketch (call this $f$) reports a value $f(X)$ such that 
\[
\left | f(X) - |E_G[X]| \right | \leq 2 \cdot \frac{n^{1+\eps}\tau}{W} \cdot \frac{n}{W} \cdot g,
\]
and likewise that 
\[
\left | f(X) - |E_H[X]| \right | \leq 2 \cdot \frac{n^{1+\eps}\tau}{W} \cdot \frac{n}{W} \cdot g,
\]
which would then imply that 
\[
\left | |E_G[X]| - |E_H[X]|\right | \leq 4 \cdot \frac{n^{1+\eps}\tau}{W} \cdot \frac{n}{W} \cdot g.
\]
This gives a contradiction as long as 
\[
4 \cdot \frac{n^{1+\eps}\tau}{W} \cdot \frac{n}{W} \cdot g < W \cdot g/2,
\]
or equivalently, that 
\[
W^3 > 8 n^{2 + \eps} \tau.
\]
We therefore set $W:= 3 n^{2/3 + \eps/3} \tau^{1/3}$. If $W>n/2$, storing the graph directly takes $\widetilde{O}(n^2)=\widetilde{O}(nW)$ bits, so we may assume $W\leq n/2$.

All that remains is to compute the space complexity of the streaming algorithm. At any point in time, storing the buffers $B^{\pm}_i$ takes $\widetilde{O}(nW)$ space (recall that $B_1^+ \supset B_2^+ \supset \dots \supset B_{n^{\eps}}^+$ and similarly for $B^-$, thus we can simply encode them by storing all present edges $e$ along with the \emph{number} of positive and negative occurrences of $e$ in the buffers). Storing the edge counts $v$ to $C_i$ takes space $\widetilde{O}(nW)$, and likewise the sparse recovery sketches take space $\widetilde{O}(nW)$. Storing the sketches defining $f$ requires 
\[
\widetilde{O}(n^{1 + \eps}/ W) \cdot \widetilde{O}(n^3/\tau^2) = \widetilde{O} \left ( \frac{n^{4 + \eps}}{W \tau^2}\right )
\]
bits of space. Plugging in for $W$, the space complexity is bounded by 
\[
\widetilde{O}\left (n^{5/3 + \eps/3} \tau^{1/3} + \frac{n^{10/3 + 2\eps/3}}{ \tau^{7/3}} \right)
\]
many bits. Setting $\tau = n^{5/8 + \eps/8}$, this works out to 
\[
\widetilde{O}\left (n^{15/8 + 3\eps/8} \right)
\]
bits of space. Note that in this proof we showed that there is a streaming algorithm using space $\tilde{O}(n^{15/8 + 3\eps/8})$ bits of space \emph{maintains a sketch} which suffice for determining connectivity. We can also show that the algorithm can \emph{uniformly} decode this sketch to determine the connected components of the resulting graph. We detail this claim in \cref{sec:uniformDecoder}.
\end{proof}

\section{A Lower Bound Against Vertex Incidence Sketches}\label{sec:VILS}

In this section, we present a warm-up for our general lower bound results for deterministic connectivity by first presenting a lower bound for \emph{deterministic connectivity via vertex incidence linear sketches}.

\subsection{Preliminaries}

\subsubsection{Linear Sketches}

We start by defining linear sketches:

\begin{definition}
    For vectors in $\mathbb{R}^U$, a \emph{linear sketch} of size $m$ is specified by a matrix $A \in \mathbb{Z}^{m \times U}$. Importantly, we require that every entry in $A$ has its magnitude bounded by $\mathrm{poly}(U)$. For a vector $v \in \mathbb{R}^U$, the resulting sketch that is stored is exactly $A \cdot v$.
\end{definition}

We will focus particularly on the case of \emph{vertex incidence linear sketches}:

\begin{definition}
    When the universe $U = \binom{n}{2}$ corresponds with the edge set of the graph $K_n = (V, E)$, we say that a matrix $A \in \Z^{m \times U}$ is a \emph{vertex incidence linear sketch} if, for every $i \in [m]$, the corresponding row $A_i$ satisfies that there exists a vertex $j \in [n]$ such that $\Supp(A_i) \subseteq \Gamma_{K_n}(j)$. In words, this requires that each linear measurement of $A$ is defined only on the neighborhood of a single vertex. 
\end{definition}

\begin{definition}
    For a vertex-incidence linear sketch $A \in \Z^{m \times \binom{n}{2}}$, and a vertex $j \in [n]$, we define the \emph{multiplicity} of $j$ to be the number of rows $i \in [m]$ for which $j$ is the assigned vertex satisfying $\Supp(A_i) \subseteq \Gamma_{K_n}(j)$. In the event that a row is a single edge and is therefore in the support of two vertices, we assign it to whichever vertex comes lexicographically first. We assign a zero row to the first vertex. We call this the \emph{base vertex} of a row $A_i$.
\end{definition}

We will frequently use the notation $A \cdot G$ to denote the resulting linear sketch that results from multiplying $A$ by the indicator vector of the graph $G$.

In this section, we will prove the following theorem, which shows that any vertex-incidence linear sketch which suffices for reporting connectivity \emph{must} use at least $\widetilde{\Omega}(n^{3/2})$ bits of space. 

\begin{theorem}\label{thm:VILSlowerbound}
    Let $A \in \Z^{m \times \binom{n}{2}}$ be a vertex-incidence linear sketch whose entries are bounded by $\mathrm{poly}(n)$. Suppose that, for any connected graph $G$ and any disconnected graph $G'$ it is the case that $A \mathbf{1}_G \neq A \mathbf{1}_{G'}$. Then, $m = \widetilde{\Omega}(n^{3/2})$.
\end{theorem}

\subsubsection{Graph Lifts}

\begin{definition}
    Let $G = K_n$ with vertex set $V = [n]$. A \emph{$2$-lift} of $G$ is specified by a bit string $x \in \zo^{E(K_n)}$ (note that we will often drop the notation $E(K_n)$ and simply use $K_n$ to denote its edge set). The resulting lifted graph $G_x$ is defined on $2n$ vertices, $V_0, V_1$. For each edge $(u,v) \in G$, we add the edges $(u_0, v_0)$ and $(u_1, v_1)$  if $x_{(u,v)} = 0$, and add the edges $(u_0, v_1)$ and $(u_1, v_0)$ if $x_{(u,v)} = 1$.
\end{definition}

Next, we define the \emph{cut-code} of $K_n$:

\begin{definition}
    We let $\mathcal{C}_n \subseteq \zo^{K_n}$ denote the \emph{cut-code} of $K_n$. This is formed by taking $B \in \F_2^{K_n \times n}$ to be the incidence matrix of $K_n$ over $\F_2$, and setting $\mathcal{C}_n = \mathrm{Im}(B)$. In this way, $\mathcal{C}_n$ is a linear space of dimension $n-1$.
\end{definition}

Importantly, we have the following claim

\begin{claim}\label{clm:cutCodeConnection}
Let $G = K_n$. Then, for $x \in \zo^{K_n}$ the graph $G_x$ is connected if and only if $x \notin \mathcal{C}_n$.
\end{claim}

Lastly, we can also make the following observation:

\begin{fact}\label{fact:liftedExchange}
    Let $A \in \Z^{m \times \binom{2n}{2}}$ be a vertex-incidence linear sketch for graphs on $2n$ vertices. Then, we can define a new vertex-incidence linear sketch $C \in \Z^{m \times \binom{n}{2}}$ along with an offset vector $b \in \Z^{m}$ such that for any $x \in \zo^{K_n}$, 
    \[
    A \cdot G_x = b + C \cdot x.
    \]
\end{fact}

\begin{proof}
    Observe that \[
    A \cdot G_x = A \cdot G_0 + \sum_{e = (u,v) \in E} \mathbf{1}[x_e] \cdot A \cdot (e_{u_0, v_1} + e_{u_1, v_0} - e_{u_0, v_0} - e_{u_1, v_1} ) = A \cdot G_0 + A \cdot D \cdot x,
    \]
    where $D \in \Z^{\binom{2n}{2} \times \binom{n}{2}}$, and the $(u,v)$th column of $D$ is exactly  $(e_{u_0, v_1} + e_{u_1, v_0} - e_{u_0, v_0} - e_{u_1, v_1} )$.
    Thus, we set $b = A \cdot G_0$. For each $e=(u,v) \in E(K_n)$, the corresponding column in $C$ has entries 
    \[
    C_e = A\cdot (e_{u_0, v_1} + e_{u_1, v_0} - e_{u_0, v_0} - e_{u_1, v_1} ).
    \]
    The bound on the magnitude of the entries of $C$ is directly inherited from the bound on the magnitude of entries in $A$.

    Now, consider a row $j \in [m]$ of the original sketching matrix $A$ which was supported in the neighborhood of some vertex $i_a$ for $a \in \zo$. We claim that row $j$ of $C$ is still supported in the neighborhood of the corresponding vertex $i$ in the base graph. This follows because the $j$th row of $C$ is exactly
    \[
    C_j = A_j \cdot D.
    \]
    For each column of $D$, indexed by an edge $(u,v)$, if $i \neq u$ and $i \neq v$, then it must be the case that the $(u,v)$th column of $D$ is disjoint from the support of the row $A_j$ (as by definition, all edges in the support of $A_j$ are incident to $i_a$). This concludes the fact. 
\end{proof}

\subsection{Finding Collisions}

Our goal now is to find, under certain conditions, an $x \in \zo^{K_n}$ such that $G_x$ is not connected yet the sketch for $x$ collides with some other $x'$. I.e., there exists $x' \in \zo^{K_n}$ such that $Cx = Cx'$. Note that it is possible that $G_{x'}$ is also disconnected, and so this may not imply any useful condition for answering our above question. Nevertheless, it is a \emph{necessary condition} for us to show that such $x$'s exist. 

\subsubsection{Preprocessing the Sketches}

We now start by making some basic observations regarding the space complexity of our sketches.

\begin{claim}\label{clm:highDegreeVertices}
    Suppose that $C \in \Z^{m \times \binom{n}{2}}$ is a vertex-incidence linear sketch. Then, there must exist at least $\frac{n}{2}$ vertices whose multiplicity is bounded by $\frac{2m}{n}$.
\end{claim}

\begin{proof}
    Indeed, otherwise there must have been $> \frac{n}{2}$ vertices whose multiplicity is at least $\frac{2m}{n}$. Because each row is assigned to exactly one vertex for multiplicity calculations, this implies a contradiction, as this would require $> m $ many rows. 
\end{proof}

Thus, going forward, we let $V_{\mathrm{low}} \subseteq [n]$ denote a set of $\lfloor n/2\rfloor$ many vertices such that their multiplicity is bounded by $\frac{2m}{n}$.

Now, we also observe that many sketches collide:

\begin{claim}\label{clm:boundSizeImage}
    Suppose that $C \in \Z^{m \times \binom{n}{2}}$ is a vertex-incidence linear sketch. Let $u \in V_{\mathrm{low}}$, and let $C|_u$ denote only those rows which are assigned base vertex $u$. Then, letting
    \[
    \mathrm{Im}(C|_u) = \{(C|_u) x: x \in \zo^{K_n} \}
    \]
    denote all possible sketches of the neighborhood of $u$, we have that 
    \[
    |\mathrm{Im}(C|_u)| \leq \left (\mathrm{poly}(n) \right )^{\frac{2m}{n}}.
    \]
\end{claim}

\begin{proof}
    This follows because $u \in V_{\mathrm{low}}$, and hence there are at most $\frac{2m}{n}$ rows with base vertex $u$. For each such row, because $C$ is a vertex-incidence linear sketch, each entry is $\mathrm{poly}(n)$-bounded. This $\mathrm{poly}(n)$ bound still holds after multiplying $C|_u$ by the indicator vector $x \in \zo^{K_n}$. Thus, there are $\leq \mathrm{poly}(n)$ possible choices for each entry in the resulting vector $(C|_u) x$. Using our bound on the number of rows in $(C|_u)$ this then gives a bound of $\left (\mathrm{poly}(n) \right )^{\frac{2m}{n}}$ possible sketches.
\end{proof}

\subsubsection{Finding Replaceable Neighborhoods}

Starting with $V_{\mathrm{low}}$ from above, we now choose two disjoint subsets $L,R \subseteq V_{\mathrm{low}}$. We leave their sizes undefined for now. We next want to show that with high probability over a \emph{random choice} of disconnected, lifted graph, it is the case that each vertex in $L$ has a \emph{replaceable} neighborhood. We formalize this below:

\begin{claim}\label{clm:singletonFiberBound}
    Let $c = B y \in \mathcal{C}_n$ for a randomly chosen $y \in \zo^n$. 
    Fix any vertex $u \in L$, and let $c|_{E[u, R]} \in \zo^{K_n}$ be the result of only keeping non-zero entries corresponding to edges between $u$ and $R$. Then, with probability at least $1 - \frac{\left (\mathrm{poly}(n) \right )^{\frac{2m}{n}}}{2^{|R|}}$, it is the case that 
    \[
    \left | \left \{ S \subseteq E[u, R]: (C|_u) \mathbf{1}_S = (C|_u) c|_{E[u, R]}  \right \} \right | \geq 2.
    \]
\end{claim}

\begin{proof}
    By \cref{clm:boundSizeImage}, we know that $|\mathrm{Im}(C|_u)| \leq \left (\mathrm{poly}(n) \right )^{\frac{2m}{n}}.$ This immediately implies that 
    \[
    \left | \left \{ S \subseteq E[u, R]: \forall S'\subseteq E[u, R], S' \neq S, (C|_u) \mathbf{1}_S \neq  (C|_u) \mathbf{1}_{S'}  \right \} \right | \leq \left (\mathrm{poly}(n) \right )^{\frac{2m}{n}}.
    \]
    In words, this is stating that the number of \emph{singleton fibers} of the map $C|_u$ operating on the edges $E[u, R]$ is bounded by $\left (\mathrm{poly}(n) \right )^{\frac{2m}{n}}$, as each such element in a singleton fiber requires a different sketch, and the number of sketches is bounded by $\left (\mathrm{poly}(n) \right )^{\frac{2m}{n}}$.
    
    Moreover, there are $2^{|R|}$ possible choices of the vectors $\mathbf{1}_S$ for $S \subseteq E[u, R]$. Because $y$ is chosen uniformly at random, this means that $c|_{E[u, R]} \in \zo^{K_n}$ is uniformly random on the edges going from $u$ to $R$; thus, we see that 
    \[
    \Pr[c|_{E[u, R]} \text{ is in a singleton fiber}] \leq \frac{\left (\mathrm{poly}(n) \right )^{\frac{2m}{n}}}{2^{|R|}}.
    \]
\end{proof}

As an immediate corollary, we have the following:

\begin{corollary}\label{cor:leftNeighborChoice}
Suppose that $|R|$ is sufficiently large such that \[
\frac{\left (\mathrm{poly}(n) \right )^{\frac{2m}{n}}}{2^{|R|}} \leq \frac{1}{n^2}.
\]
Then, with probability at least $1 - \frac{1}{n}$ over the choice of $c = By$, it is the case that for every $u \in L$, 
\[
\left | \left \{ S \subseteq E[u, R]: (C|_u) \mathbf{1}_S = (C|_u) c|_{E[u, R]}  \right \} \right | \geq 2.
\]
More importantly, provided $R$ satisfies the size bound, such a $c$ is \emph{guaranteed to exist}.
\end{corollary}

\begin{proof}
    This follows from a simple union bound over the events defined in \cref{clm:singletonFiberBound}.
\end{proof}

The key fact from the above is the following: letting $c = By$ for a random $y \in \zo^n$ (as was done in \cref{clm:singletonFiberBound}), for every vertex $u$ in $L$, we can \emph{replace} the ``neighborhood'' of $u$ (i.e., the bit string $\zo^{E[u, R]}$) \emph{without} altering $u$'s sketches.  

Note however, that this does not leave the \emph{entire} sketch unchanged. Indeed, from the perspective of the \emph{right-hand vertices} $R$, replacing the neighborhood of $u$ can alter their sketches. This is where the next key observation comes into play: at every left-vertex $u \in L$, conditioned on \cref{cor:leftNeighborChoice}, we have two options for edge set $E[u, R]$ that share the same sketch: one choice is the original edge set which agrees with $c = By$, and the other is an arbitrary set $S \subseteq E[u, R]$ as guaranteed by \cref{cor:leftNeighborChoice}. We let $S_u$ denote this set for each vertex $u \in L$. We now introduce one piece of notation:

\begin{definition}\label{def:replacedCollisionVector}
    For a set $A \subseteq L$, we let $x_A \in \zo^{K_n}$ denote the lifting vector such that:
    \[
    (x_A)_e = \begin{cases}
        c_e \text{ if } e \notin E[L, R] \\
        c_e \text{ if } e \in E[u, R] \text{ and } u \in L-A \\ 
        \mathbf{1}[e \in S_u] \text{ if } e \in E[u, R] \text{ and } u \in A
    \end{cases}.
    \]
\end{definition}

In words, this simply captures that for $u \in A$, we replace the edges between $u$ and $R$ with $S_u$. In this way, we do not alter the sketches stored at $u$. We now have the following claim:

\begin{claim}\label{clm:collidingCSketch}
    Suppose $c = By$ is chosen in such a way that for every $u \in L$,
    \[
\left | \left \{ S \subseteq E[u, R]: (C|_u) \mathbf{1}_S = (C|_u) c|_{E[u, R]}  \right \} \right | \geq 2.
\]
If 
\[
2^{|L|} > \left (\mathrm{poly}(n) \right )^{\frac{2m |R|}{n}},
\]
there exist two choices of $A \neq B \subseteq L$ such that for the vectors $x_A, x_B \in \zo^{K_n}$ defined as in \cref{def:replacedCollisionVector}, 
\[
C (x_A) = C (x_B).
\]
\end{claim}

\begin{proof}
    To start, observe that for every $A \neq B \subseteq L$, it must be the case that for every $u \in L$ that $(C|_u) x_A = (C|_u) x_B = (C|_u)c$. This is because \cref{cor:leftNeighborChoice} guarantees that $(C|_u) c|_{E[u, R]} = (C|_u) \mathbf{1}_{S_u}$, where $S_u$ is the other set of edges in the fiber, and all other edges incident to $u$ remain unchanged.

    Moreover, observe that for vertices outside of $L, R$, their neighborhoods are unchanged, and thus their vertex-incidence sketches are also unchanged. All that remains to show is that the sketches of $C$ corresponding to base vertices from $R$ are \emph{also} unchanged. For this, we use \cref{clm:boundSizeImage}: for each $v \in R$, 
    \[
    |\mathrm{Im}(C|_v)| \leq \left (\mathrm{poly}(n) \right )^{\frac{2m}{n}}.
    \]
    Thus, in total, 
    \[
    |\mathrm{Im}(C|_R)| \leq \left (\mathrm{poly}(n) \right )^{\frac{2m|R|}{n}}.
    \]
    Provided that 
    \[
    2^{|L|} > \left (\mathrm{poly}(n) \right )^{\frac{2m |R|}{n}},
    \]
    this then implies that there must be two choices of $A \neq B  \subseteq L$ for which $(C|_R) x_A = (C|_R) x_B$. This then concludes the claim. 
\end{proof}

\begin{corollary}\label{cor:differenceZero}
    Under the conditions of \cref{clm:collidingCSketch}, there exists $A \neq B \subseteq L$ such that $C(x_A - x_B) = 0$.
\end{corollary}

\subsubsection{Finding Disconnected Graphs with Colliding Sketches}

The vectors $x_A, x_B$ as guaranteed by \cref{clm:collidingCSketch} unfortunately still do not satisfy the conditions we set out to guarantee, as it may be that $x_A, x_B \notin \mathcal{C}_n$. However, as mentioned in \cref{cor:differenceZero}, we know that $C(x_A - x_B) = 0$. Thus, we now focus exactly on this vector $x_A - x_B$. We can observe that (using the set-up from \cref{def:replacedCollisionVector}):
\[
    (x_A - x_B)_e = \begin{cases}
        0 \text{ if } e \notin E[L, R] \\
        0 \text{ if } e \in E[u, R] \text{ and } u \in A \cap B \\ 
        0 \text{ if } e \in E[u, R] \text{ and } u \notin A \cup B \\ 
        \mathbf{1}[e \in S_u] - c_e \text{ if } e \in E[u, R] \text{ and } u \in A-B \\
        c_e - \mathbf{1}[e \in S_u]  \text{ if } e \in E[u, R] \text{ and } u \in B-A
    \end{cases}.
    \]

Our goal is now to find some $y' \in \zo^n$ such that for $c' = By'$, it is the case that $c' + (x_A - x_B) \in \zo^{K_n}$. In this way, it will be the case that $c' + (x_A - x_B)$ still corresponds with some valid lift, and yet the sketches for $c'$ and $c' + (x_A - x_B)$ will be the same. To ensure that $c' + (x_A - x_B) \in \zo^{K_n}$, the entries corresponding to edges in any of the first three cases can be arbitrary, i.e., when (1) $e \notin E[L, R]$, (2) $e \in E[u, R] \text{ and } u \in A \cap B$, or (3) $e \in E[u, R] \text{ and } u \notin A \cup B$. However, for the final two cases we need to be more careful. 

For an edge $e \in E[u, R] \text{ and } u \in A-B$, the value in the corresponding edge slot is $\mathbf{1}[e \in S_u] - c_e$. When $c_e = 1$, this means the value can be either $0$ or $-1$, and when $c_e =0$, the value can be either $1$ or $0$. Thus, when we construct the new vector $c'$, we will ensure that $(c')_e = c_e$; in this way, $(c')_e + (x_A - x_B)_e \in \zo$.

This leaves the final case, when $e \in E[u, R] \text{ and } u \in B-A$. In this case, the value in the corresponding edge slot is $c_e - \mathbf{1}[e \in S_u]$. When $c_e = 1$, this means the value can be either $1$ or $0$, and when $c_e =0$, the value can be either $0$ or $-1$. Thus, when we construct the new vector $c'$, we will ensure that $(c')_e = (1 - c_e)$; in this way, $(c')_e + (x_A - x_B)_e \in \zo$.

In light of the above, we now have the following claim:

\begin{claim}\label{clm:legitCollision}
    Let $x_A, x_B$ be defined as above, with respect to the vector $c = By$. Define $y' \in \zo^n$ such that 
    \[
    y'_u = \begin{cases}
        (1 - y_u) \text{ if } u \in B - A \\
        y_u \text{ else }
    \end{cases}.
    \]
    Then, for $c' = By'$, it is the case that $c' + (x_A - x_B) \in \zo^{K_n}$.
\end{claim}

\begin{proof}
    As discussed above, for any $e \notin E[L, R]$, or $e \in E[u, R] \text{ and } u \in A \cap B$, or $e \in E[u, R] \text{ and } u \notin A \cup B $, it is the case that $(c' + (x_A - x_B))_e = (c')_e \in \zo$.

    For any $e \in E[u, R] \text{ and } u \in A-B$, $(x_A - x_B)_e = \mathbf{1}[e \in S_u] - c_e$. Thus, when $c_e = 1$, $(x_A - x_B)_e \in \{0, -1\}$ and when $c_e = 0$,  $(x_A - x_B)_e \in \zo$. Crucially, we can observe that for such $e$, it is the case that $c'_e = c_e$, as for $e = (u,v)$, $y'_u = y_u, y'_v= y_v$. Thus, it is still the case that  $(c' + (x_A - x_B))_e \in \zo$.

    Finally, we consider any $e \in E[u, R] \text{ and } u \in B-A$. In this case, $(x_A - x_B)_e =  c_e -\mathbf{1}[e \in S_u]$. Thus, when $c_e = 1$, $(x_A - x_B)_e \in \{0, 1\}$ and when $c_e = 0$,  $(x_A - x_B)_e \in \{0,-1\}$. Crucially, we can observe that for such $e$, it is the case that $c'_e = 1-c_e$, as for $e = (u,v)$ (with $u \in L, v \in R$), $y'_u = 1 - y_u$ while $y'_v = y_v$. Thus, it is still the case that  $(c' + (x_A - x_B))_e \in \zo$. This concludes the claim. 
\end{proof}

\subsection{Violating the Sketch Guarantees}

\cref{clm:legitCollision} shows that the graphs $G_{c'}$ and $G_{c' + (x_A - x_B)}$ have the same sketch, as 
\[
C(c') = C(c' + (x_A - x_B)).
\]
We also already know (by construction) that $G_{c'}$ is disconnected, as $c' = By'$ and thus is a codeword in the cut code. All that remains to be shown then is that $G_{c' + (x_A - x_B)}$ is connected:

\begin{claim}\label{clm:connected}
    Let $x_A, x_B, y, y', c, c'$ be defined as in \cref{clm:legitCollision}. Then, $G_{c' + (x_A - x_B)}$ is connected.
\end{claim}

\begin{proof}
    To prove this, it suffices to show that $c' + (x_A - x_B) \notin \mathcal{C}_n$. For this, we now make a few observations: first, we can re-write $c' + (x_A - x_B)$ using field operations over $\F_2$. Because $c' \in \zo^{K_n}$ and $(x_A - x_B) \in \{1,0,-1\}^{K_n}$, if we define $F = \Supp(x_A - x_B)$, we see that 
    \[
    c' + (x_A - x_B) = c' \oplus \mathbf{1}_F.
    \]
    This is because whenever $(x_A - x_B)$ is non-zero in some entry, it must be the case that it is flipping the corresponding bit in $c'$.

    Now, in order to show that $c' \oplus \mathbf{1}_F \notin \mathcal{C}_n$, it suffices to show that $\mathbf{1}_F \notin \mathcal{C}_n$, as the code $\mathcal{C}_n$ is closed under addition of codewords over $\F_2$ (and $c'$ is a codeword). Finally, we observe that codewords in $\mathcal{C}_n$ correspond only with indicators of edges in cuts of the complete graph $K_n$. Note that in any such \emph{non-empty} cut, it must be the case that \emph{every vertex} has some edge which crosses the cut. But, $F$ is only non-zero in edges between $L$ and $R$; in particular, because we choose $|L \cup R| \leq n/2$, there must be some vertex $v$ in $[n] - L - R$ such that $v$ has none of its neighborhood in $F$. This proves that $\mathbf{1}_F \notin \mathcal{C}_n$, as we know that $F \neq \emptyset$.
\end{proof}

We now conclude the proof of \cref{thm:VILSlowerbound}:

\begin{proof}[Proof of \cref{thm:VILSlowerbound}.]
We first fix any vertex-incidence linear-sketching matrix $A \in \Z^{m \times \binom{2n}{2}}$ with polynomially-bounded entries and $m \leq \frac{n^{3/2}}{100 \log^4(n)}$. We assume that $A$ decides connectivity; this means that for any connected graph $G$ and any disconnected graph $G'$ it is the case that $A \mathbf{1}_G \neq A \mathbf{1}_{G'}$.

By \cref{fact:liftedExchange}, we know that $A$ admits a transformation into a vertex-incidence linear sketch $C$ of graphs on $n$ vertices, such that for any lifted graph $G_x$ (with $x \in \zo^{K_{n}}$), $C x \neq C x'$ if exactly one of $G_x, G_{x'}$ is connected.

Next, we invoke \cref{clm:highDegreeVertices}. This implies there is a set $V_{\mathrm{low}} \subseteq [n]$ of size $\lfloor n/2\rfloor$ such that each vertex in $V_{\mathrm{low}}$ has multiplicity $\leq \frac{2m}{n} \leq \frac{\sqrt{n}}{50 \log^4(n)}$ in the vertex-incidence linear sketch. We then choose disjoint subsets $L,R \subseteq V_{\mathrm{low}}$, where $|L| = \lfloor n/4\rfloor$ and $|R| = \left\lfloor\frac{\sqrt{n}}{10 \log^2(n)}\right\rfloor$. Importantly, for sufficiently large $n$, this ensures that 
 \[
\frac{\left (\mathrm{poly}(n) \right )^{\frac{2m}{n}}}{2^{|R|}} \leq \frac{1}{n^2},
\]
and likewise 
\[
2^{|L|} > \left (\mathrm{poly}(n) \right )^{\frac{2m |R|}{n}}.
\]
By \cref{clm:boundSizeImage}, \cref{clm:singletonFiberBound}, \cref{cor:leftNeighborChoice}, and \cref{clm:collidingCSketch}, this then implies that there exists a vector $c = By$, (for $B$ being the incidence matrix of $K_n$, and $y \in \zo^n$), along with $x_A, x_B$ as defined in \cref{def:replacedCollisionVector}, such that 
\[
C(x_A) = C(x_B),
\]
and thus $C(x_A - x_B) =0$. By \cref{clm:legitCollision}, we can then find a $y' \in \zo^n$ such that for $c' = By'$, it is the case that $c' + (x_A - x_B) \in \zo^{K_n}$, which furthermore by \cref{clm:connected}, we know corresponds to a connected graph. To summarize then, we know that $G_{c' + (x_A - x_B) }$ is connected and $G_{c'}$ is \emph{not connected} (as this is in the image of the cut-code, see \cref{clm:cutCodeConnection}). Moreover, $c' + (x_A - x_B) \in \zo^{K_n}$ and $c' \in \zo^{K_n}$, so these correspond to valid lifting strings. Finally, because $C(x_A - x_B) = 0$, we additionally know that 
\[
C(c' + (x_A - x_B)) = C(c'),
\]
and therefore, 
\[
A \cdot \mathbf{1}_{G_{c' + (x_A - x_B)}} = A \cdot \mathbf{1}_{G_{c'}} .
\]
This is finally a contradiction, as we have two graphs, only one of which is connected, but for which the sketching matrix $A$ reports the same vector. The same lower bound for an odd number of vertices follows by adding one fixed leaf to each graph.
\end{proof}

\section{A General Lower Bound}\label{sec:genLB}

In this section, we now present an \emph{unconditional} lower bound on the chromatic number of $P_n$:

\begin{theorem}\label{thm:n54LB}
    Fix a parameter $n$ denoting the number of vertices. Then, $\log(\chi(P_n)) = \Omega(n^{5/4})$.
\end{theorem}

As an immediate corollary, using \cref{clm:componentsToConnectivity}, we obtain the following lower bound on the complexity of the one-way deterministic connectivity problem:

\begin{corollary}
    Any communication protocol which solves the one-way deterministic connectivity problem on graphs with $n$ vertices requires messages of length $\Omega(n^{5/4})$.
\end{corollary}
Towards proving this theorem, we first introduce several basic preliminaries. 

\subsection{Chromatic Lower Bound Preliminaries}

We use the following chromatic number lower bound due to Lov\'asz \cite{lovasz1979shannon}:

\begin{fact}[\cite{lovasz1979shannon}, Theorem 4]\label{fact:chromaticLB}
    Let $H$ be a graph on $N$ vertices and let $M \in \R^{N \times N}$ be such that $M \succeq 0, M_{xx} = 1$ for every $x \in [N]$, and $M_{xy} = 0$ whenever $x\neq y$ and $(x,y) \notin H$. Then,
    \[
    \chi(H) \geq \frac{\mathbf{1}^T M \mathbf{1}}{N}.
    \]
\end{fact}

Going forward, we will define a matrix whose off-diagonal support is contained in the edges of the \emph{universal graph} $P_n$.

\subsection{Defining the Matrix}

We now define the exact matrix supported on $P_n$ that we will analyze. Recall that the vertices of $P_n$ are all the possible graphs on $n$ vertices. For even $n$, rather than considering all possible base graphs, we break the $n$ vertices into two groups of $k = n/2$ vertices, and consider only all the possible subgraphs of $K_{k \times k}$. We index these graphs by $\mathcal{X}_k = \zo^{k \times k}$.

For $x \in \mathcal{X}_k$ we let $G_x$ denote the graph indexed by $x$, and we let $H_k$ be the restriction of the universal graph $P_n$ to the graphs $G_x:x \in \mathcal{X}_k$.

As mentioned, the off-diagonal support of our ultimate choice of matrix $M$ will be contained in $H_k$, but we will additionally require a parameterization in terms of $\theta \in (0, \pi/2)$. Letting $s = \sin \theta$ and writing 
\[
A_+ = \begin{bmatrix}
    1 & 2s \\
    0 & 1
\end{bmatrix},
\]
and 
\[
A_- = \begin{bmatrix}
    1 & 0 \\
    2s & 1
\end{bmatrix},
\]
our final choice of $M$ is then exactly defined as
\[
M_{k, \theta} =\E_{\xi \sim \{\pm 1\}^k, \zeta\sim \{\pm 1\}^k} \left [\bigotimes_{i,j \in [k]} A_{\xi_i \cdot  \zeta_j}\right ].
\]
The way to understand this matrix is that if we denote $L$ and $R$ to be the left and right sides of the bipartite graphs we consider, we then sample uniformly random signs to all vertices in $L$ and $R$, and then take a tensor product over all edges between $L$ and $R$, where the assigned matrix for each edge slot $(i,j)$ is $A_{\xi_i \cdot  \zeta_j}$. $M_{k, \theta}$ is the \emph{expectation} of this entire operation. 

For this matrix, it is not hard to see that we have the following properties:

\begin{lemma}\label{lem:MeasyProp}
    For every $k$ and $\theta \in (0, \pi/2)$, the matrix $M_{k, \theta}$ satisfies:
    \begin{enumerate}
        \item $M_{k, \theta}$ has all real entries. 
        \item All diagonal entries are $1$. 
        \item $M_{k, \theta}$ is symmetric. 
        \item $(M_{k, \theta})_{x,y} = 0$ if $x \neq y$ and the connected components of $G_x - G_y\cap G_x $ equal the connected components of $G_y - G_x \cap G_y$.
        \item $\mathbf{1}^T M_{k, \theta} \mathbf{1} = (2 + 2 \sin \theta)^{k^2}$.
    \end{enumerate}
\end{lemma}

\begin{proof}
We prove these items in order. Item $1$ follows because each of the matrices $A_+, A_-$ is real, and thus their tensor product and expectation must also be real. Likewise, because $A_+, A_-$ have $1$'s on the diagonal, it must also be the case that their tensor product and expectation has $1$'s on the diagonal. Symmetry follows from the fact that for every $\xi$, if we consider the \emph{flipped vector} $\bar{\xi}$ such that $\bar{\xi}_i = (-1) \cdot \xi_i$, then it is the case that for any choice of $\zeta$,
\[
\bigotimes_{i,j \in [k]} A_{\xi_i \cdot  \zeta_j} = \left ( \bigotimes_{i,j \in [k]} A_{\bar{\xi}_i \cdot  \zeta_j} \right )^T.
\]
We can thus re-write 
\[
M_{k, \theta} = \E_{\xi \sim \{\pm 1\}^k, \zeta\sim \{\pm 1\}^k} \left [\bigotimes_{i,j \in [k]} A_{\xi_i \cdot  \zeta_j}\right ] \]
\[
= \frac{1}{2} \cdot \E_{\xi \sim \{\pm 1\}^k, \zeta\sim \{\pm 1\}^k} \left [\bigotimes_{i,j \in [k]} A_{\xi_i \cdot  \zeta_j} + \left ( \bigotimes_{i,j \in [k]} A_{\xi_i \cdot  \zeta_j} \right )^T\right ],
\]
which is then trivially symmetric. 

We now prove item $4$. Consider any entry $(x,y)$ with $x\neq y$ such that $(M_{k, \theta})_{xy} > 0$. We will prove that this implies that the connected components satisfy $\mathrm{CC}(G_x - G_x \cap G_y) \neq \mathrm{CC}(G_y - G_x \cap G_y)$. (Note that if $G_x-G_x\cap G_y$ is empty, the other difference is nonempty because $x\neq y$, so the conclusion follows immediately.) Otherwise, there is some choice of $\xi, \zeta$ for which 
\[
\left (\bigotimes_{i,j \in [k]} A_{\xi_i \cdot  \zeta_j}\right )_{xy} > 0,
\]
and because all entries are nonnegative in the $A$ matrices, the only way this is possible is if every factor in the product contributing to entry $(x, y)$ is positive. Now, consider an edge $e = (i,j)$. There are two cases:
\begin{enumerate}
    \item If $e \in G_x - G_x \cap G_y$, then $x_{ij} = 1$ and $y_{ij} = 0$. In order for entry $x,y$ to be positive, this forces the product to contain $A_-$, and thus $\xi_i \zeta_j = -1$.
    \item If $e \in G_y - G_x \cap G_y$, then $x_{ij} = 0$ and $y_{ij} = 1$. In order for entry $x,y$ to be positive, this forces the product to contain $A_+$, and thus $\xi_i \zeta_j = 1$.
\end{enumerate}

Now, let us consider all vertices $S$ which received sign $+1$. For any edge $(u,v) \in G_x - G_x \cap G_y$, it must be the case that $\xi_u \zeta_v = -1$; equivalently, $u, v$ are on opposite side of the cut defined by $S$. On the other hand,  for any edge $(u,v) \in G_y - G_x \cap G_y$, it must be the case that $\xi_u \zeta_v = 1$; equivalently, $u, v$ are on \emph{same} side of the cut defined by $S$. This is to say that \emph{every edge} in $G_x - G_x \cap G_y$ crosses $(S, \bar{S})$ while \emph{no} edge in $G_y - G_x \cap G_y$ crosses the cut. This then forces $G_x - G_x \cap G_y$ and $G_y - G_x \cap G_y$ to have different connected components. 

Finally, we prove item $5$. For this, we observe that the entry-wise sum of a tensor product is the same as the product of the entry-wise sums. Thus, 
\[
\mathbf{1}^T M_{k, \theta} \mathbf{1} = \prod_{i,j \in [k]} (2 + 2 \sin \theta) = (2 + 2 \sin \theta)^{k^2}.
\]
This concludes the lemma.
\end{proof}

Ultimately, we will take $\theta \approx 1 / k^{3/4}$, but before we can invoke \cref{fact:chromaticLB}, we must prove that $M_{k, \theta}$ is \emph{positive semi-definite}. It is exactly this which we aim to show in the following section.

\subsection{Proving Positive Semi-Definiteness}

Towards proving that $M_{k, \theta}$ is PSD, we define the following intermediate quantity, which takes as input a sign matrix $\eta \in \{\pm 1 \}^{k \times k}$:
\[
F_{\eta}(\theta) = \E_{\xi \sim \{\pm 1 \}^k } \left [\prod_{j = 1}^k \cos\left (\theta \sum_{i = 1}^k  \eta_{i,j} \xi_i \right ) \right ].
\]

\subsubsection{Characterizing PSD-ness in Terms of $F$}

Crucially, these functions $F_{\eta}$ enable us to make an exact characterization of when $M_{k, \theta}$ is PSD:

\begin{lemma}\label{lem:PSDreduction}
    Fix an integer $k\geq1$ and $\theta \in (0, \pi/2)$. Then, $M_{k, \theta}$ is PSD if and only if $F_{\eta}(\theta) \geq 0$ for every $\eta \in \{ \pm 1\}^{k \times k}$.
\end{lemma}

\begin{proof}
    As before, we let $s = \sin \theta, c = \cos \theta$ and let $m =k^2$. Establishing the above characterization relies on the following fact:

    \begin{fact}\label{fact:matrixReWrites}
        The matrices $T = \begin{bmatrix}
            c & 0 \\
            -s & 1 
        \end{bmatrix}$ and $J = \begin{bmatrix}
            0 & 1 \\
            -1 & 0 
        \end{bmatrix}$ satisfy the following conditions:
        \begin{enumerate}
            \item $\det T =c > 0$.
            \item For $\sigma \in \{ \pm 1 \}$, we have $T A_{\sigma} T^T = c (cI + \sigma s J)$.
        \end{enumerate}
    \end{fact}

    \begin{proof}
        The first item follows trivially from the determinant formula. The second item follows from the fact that 
        \[
        T A_{\sigma} T^T = \begin{bmatrix}
            c^2 & \sigma cs \\
            -\sigma c s & c^2
        \end{bmatrix},
        \]
        where we have used the fact that $c^2 + s^2  = 1$.
    \end{proof}

    Now, we define an additional intermediate matrix 
    \[
    C =T^{\otimes m} M_{k, \theta} (T^T)^{\otimes m}.
    \]
    For this matrix, we have the following claim:

    \begin{claim}\label{clm:eigenvaluesOfC}
        The eigenvalues of $C$ (counting multiplicity) are exactly the values of $c^m F_{\eta}(\theta)$ across all possible $\eta \in \{\pm 1\}^{k \times k}$.
    \end{claim}

\begin{proof}[Proof of \cref{clm:eigenvaluesOfC}]
    To see this, we observe that by using \cref{fact:matrixReWrites}, we have 
    \[
    C =T^{\otimes m} M_{k, \theta} (T^T)^{\otimes m} = \E_{\xi \sim \{\pm 1\}^k, \zeta\sim \{\pm 1\}^k} \left [T^{\otimes m}\bigotimes_{i,j \in [k]} A_{\xi_i \cdot  \zeta_j} (T^T)^{\otimes m}\right ]
    \]
    \[
    = c^m \E_{\xi \sim \{\pm 1\}^k, \zeta\sim \{\pm 1\}^k} \left [\bigotimes_{i,j \in [k]} (cI + s \xi_i \zeta_j  J) \right ].
    \]

    Importantly, for this matrix the first term $cI$ is just a multiple of the identity. The second term is a scalar multiple of $J$; for which we can explicitly compute its eigenvectors to be 
    \[
    \frac{1}{\sqrt{2}}\begin{bmatrix}
        1 \\
        \mathbf{i} \cdot 1
    \end{bmatrix} \quad \text{and} \quad   \frac{1}{\sqrt{2}}\begin{bmatrix}
        1 \\
        \mathbf{i} \cdot (-1)
    \end{bmatrix},
    \]
    which have eigenvalues $\mathbf{i} \cdot (\pm1)$ respectively.
    As mentioned, because $cI$ is just a multiple of the identity, this means we can write an explicit, shared orthonormal basis of eigenvectors of all matrices in the final expression. Indeed, these eigenvectors are exactly 
    \[
    v_{\eta} = \bigotimes_{i,j \in [k]} \frac{1}{\sqrt{2}}\begin{bmatrix}
        1 \\
        \mathbf{i} \cdot (\eta_{i,j})
    \end{bmatrix}: \eta \in \{ \pm 1 \}^{k \times k}.
    \]
    Because the eigenvalues of the tensor product are simply the product of the eigenvalues, for the eigenvector $v_{\eta}$, we can then exactly compute its eigenvalue of $C$ to be 
    \[
    \lambda_{\eta} = c^m \E_{\xi \sim \{\pm 1\}^k, \zeta\sim \{\pm 1\}^k} \left [\prod_{i,j} (c + \mathbf{i}s\eta_{i,j}\xi_i \zeta_j) \right ]
    \]
    \[
    = c^m \E_{\xi \sim \{\pm 1\}^k, \zeta\sim \{\pm 1\}^k} \left [\exp\left (\mathbf{i}\theta \cdot \sum_{ij} \eta_{ij}\xi_i \zeta_j \right )\right ]
    \]
    \[
    = c^m \E_{\xi \sim \{\pm 1\}^k} \left [\prod_{j = 1}^k\cos\left (\theta \cdot \sum_{i \in [k]} \eta_{ij}\xi_i  \right )\right ] = c^m F_{\eta}(\theta).
    \]
    Note that in this last line, we have used the fact that 
    \[
    \E_{\zeta \sim \{ \pm 1 \}} e^{\mathbf{i}t \zeta}= \frac{e^{\mathbf{i}t} + e^{-\mathbf{i}t}}{2} = \cos t.
    \]
    In general starting with $\E_{\xi \sim \{\pm 1\}^k, \zeta\sim \{\pm 1\}^k} \left [\exp\left (\mathbf{i}\theta \cdot \sum_{ij} \eta_{ij}\xi_i \zeta_j \right )\right ]$, we let $L_j = \sum_{i \in [k]}\eta_{ij} \xi_i$. We then have that 
    \[
    \E_{\xi \sim \{\pm 1\}^k, \zeta\sim \{\pm 1\}^k} \left [\exp\left (\mathbf{i}\theta \cdot \sum_{ij} \eta_{ij}\xi_i \zeta_j \right )\right ]
    \]
    \[
 = \E_{\xi, \zeta} \left [  \exp(\mathbf{i}\theta \sum_{j \in [k]} L_j\zeta_j ) \right]
    \]
    \[
    = \E_{\xi, \zeta} \left [  \prod_{j = 1}^k e^{\mathbf{i}\theta L_j \zeta_j} \right]
    \]
    \[
     =\E_{\xi} \left [  \prod_{j = 1}^k \E_{\zeta_j} [e^{\mathbf{i}\theta L_j \zeta_j}] \right]
    \]
    \[
     = \E_{\xi \sim \{\pm 1\}^k} \left [\prod_{j = 1}^k\cos\left (\theta \cdot \sum_{i \in [k]} \eta_{ij}\xi_i  \right )\right ].
    \]
\end{proof}

Finally, we can conclude the proof of \cref{lem:PSDreduction}. Indeed, because $T$ is invertible by \cref{fact:matrixReWrites}, we then have that $M_{k, \theta}$ is PSD if and only if $C$ is PSD. We can then apply \cref{clm:eigenvaluesOfC} to conclude that  $M_{k, \theta}$ is PSD if and only if $F_{\eta}(\theta) \geq 0$ for every $\eta \in \{ \pm 1\}^{k \times k}$.

\end{proof}

\subsubsection{Proving $F$ is Non-Negative}

We now turn to proving that $F_{\eta}(\theta) \geq 0$ for every $\eta \in \{ \pm 1\}^{k \times k}$ provided that our $\theta$ is sufficiently small.

Our key lemma in this subsection is the following:

\begin{lemma}\label{lem:FlowerBound}
    For every integer $k \geq 1$, every $\theta > 0$, and every $\eta \in \{ \pm 1 \}^{k \times k}$,
    \[
    F_{\eta}(\theta) \geq e^{-\theta^2k^2} - 4k \exp \left ( - \frac{1}{2k\theta^2}\right ).
    \]
\end{lemma}

\begin{proof}
    As in the definition of $F$, we let $\xi \in \{\pm 1\}^k$ be a vector of uniformly random signs. We let $L_j = \sum_{i \in [k]} \eta_{ij} \xi_i$, we let $Q = \sum_{j \in [k]} L_j^2$, and we let $B = \max_j \mathbf{1}[|\theta L_j| > 1]$.

    With this, we then have the following intermediate claim:

    \begin{claim}\label{clm:QandBbounds}
        For $Q$ and $B$ as defined above, we have that 
        \[
        \E[Q] = k^2
        \]
        and 
        \[
        \Pr[B] \leq 2k \exp \left ( - \frac{1}{2k\theta^2}\right ).
        \]
    \end{claim}

    \begin{proof}[Proof of \cref{clm:QandBbounds}.]
        The bound on $Q$ follows from the fact that each $L_j$ is the sum of independent, uniformly random signs, and thus $\E[L_j^2] = k$. We can then obtain the bound on $Q$ by summing over all $k$ choices of $L_j$.

        For the second point, Hoeffding's inequality implies that for any $u\geq0$, 
        \[
        \Pr[|L_j| > u] \leq 2e^{-u^2 / 2k}.
        \]
        For our purposes, we set $u = \frac{1}{\theta}$ and take a union bound over the $k$ choices of $j$. 
    \end{proof}

    We also have the following claim which bounds a product of cosines:
    \begin{claim}\label{clm:cosineProductBound}
        For any real numbers $z_1, \dots z_k$,
        \[
        \prod_{j = 1}^k \cos(z_j) \geq \exp \left ( -\sum_{j = 1}^k z_j^2\right ) - 2 \mathbf{1}_{\max_j |z_j| > 1}.
        \]
    \end{claim}
    \begin{proof}[Proof of \cref{clm:cosineProductBound}.]
        If $|z_j| \leq 1$ for every $j$, then we use the trivial fact that 
        \[
        \cos(z_j) \geq e^{-z_j^2}.
        \]
        Otherwise, the left side is at least $-1$, and the right side is at most $-1$.
    \end{proof}

    We are now ready to conclude \cref{lem:FlowerBound}. Indeed, 
    \[
    F_{\eta}(\theta) = \E_{\xi \sim \{\pm 1 \}^k } \left [\prod_{j = 1}^k \cos\left (\theta \sum_{i = 1}^k  \eta_{i,j} \xi_i \right ) \right ],
    \]
    and so applying \cref{clm:cosineProductBound} with $z_j = \theta L_j$ yields that 
    \[
    F_{\eta}(\theta) \geq \E_{\xi \sim \{\pm 1 \}^k } \left [\exp \left ( -\sum_{j = 1}^k z_j^2\right ) - 2 \mathbf{1}_{\max_j |z_j| > 1} \right ]
    \]
    \[
    = \E_{\xi \sim \{\pm 1 \}^k} \left [ e^{-\theta^2 Q}\right ] - 2 \Pr[B]
    \]
    \[
    \geq  e^{-\theta^2 \E[Q]} - 2 \Pr[B]
    \]
    \[
    \geq e^{-\theta^2k^2} - 4k \exp \left (- \frac{1}{2k \theta^2} \right ),
    \]
    where in the second to last-line we have used Jensen's inequality on the function $u \rightarrow e^{-\theta^2 u}$. This concludes the lemma. 
\end{proof}

Using \cref{lem:FlowerBound}, we can now prove the following corollary:

\begin{corollary}\label{cor:MPSD}
    For any integer $k \geq 1$, define $\theta_k = \frac{1}{4k^{3/4}}$. Then, for every $\eta \in \{\pm 1 \}^{k \times k}$ it is the case that 
    \[
    F_{\eta}(\theta_k) > 0,
    \]
    and thus $M_{k, \theta_k} \succ 0$.
\end{corollary}

\begin{proof}
    Indeed, for our choice of $\theta_k$, we obtain that 
    \[
    F_{\eta}(\theta_k) \geq e^{-\theta_k^2k^2} - 4k \exp \left (- \frac{1}{2k \theta_k^2} \right )
    \]
    \[
    \geq e^{- \sqrt{k}/16} - 4ke^{-8 \sqrt{k}}.
    \]
    For $k \geq 1$, we can see that $4k < e^{2\sqrt{k}}$, and thus $e^{- \sqrt{k}/16} - 4ke^{-8 \sqrt{k}} > 0$. By using \cref{clm:eigenvaluesOfC} and the invertibility of $T$, we then obtain that $M_{k, \theta_k} \succ0$.
\end{proof}

We are now ready to conclude the main theorem.

\begin{proof}[Proof of \cref{thm:n54LB}.]
    We know that $\chi(P_n) \geq \chi(H_k)$. For the matrix $M_{k, \theta_k}$ as defined above, we know that $M_{k, \theta_k}$ is PSD by \cref{cor:MPSD}, and by \cref{lem:MeasyProp}, it is symmetric, with $1$'s on the diagonal, and the off-diagonal support of $M_{k, \theta_k}$ is contained in the edges of $H_k$. Thus, 
    \[
    \chi(H_k) \geq \frac{\mathbf{1}^T M_{k, \theta_k} \mathbf{1}}{2^{k^2}} \geq \frac{(2 + 2\sin\theta_k)^{k^2}}{2^{k^2}} \geq \left (1 + \sin(1 / (4k^{3/4}))\right )^{k^2}
    \]
    \[
    \geq \left (1 + \frac{1} {8k^{3/4}}\right )^{k^2} = 2^{\Omega(n^{5/4})}.
    \]
    Taking logarithms concludes the theorem.
\end{proof}

\section{An $\widetilde{\Omega}(n^{3/2})$ Lower Bound}\label{sec:appn32}

As before, we use the \emph{universal graph} $P_n$ to denote the graph on $2^{\binom{n}{2}}$ many vertices, where each vertex corresponds with one of the possible graphs on $n$ vertices.
Recall that two distinct vertices $G,H$ are
adjacent exactly when
\begin{equation}\label{eq:conflict-definition}
 \cc(G\setminus H)\ne\cc(H\setminus G).
\end{equation}

In this section, we prove an even stronger lower bound on the chromatic number of the universal graph. The key theorem we show in this section is the following:

\begin{theorem}\label{thm:mainAppendix}
There exists a constant $n_0$ such that for every integer $n\ge n_0$, the universal graph satisfies
\begin{equation}\label{eq:main-bound}
 \chi(P_n)
 \ge 2^{\Omega( n^{3/2}/\log n)}
\end{equation}
\end{theorem}

As alluded to before, this coloring lower bound relies on an explicit inequality in complex analysis. We first mention this statement that we will require in the next section, and then subsequently use it to prove the lower bound. We then prove this inequality in \cref{sec:keyInequalityApp}.

\subsection{The Key Inequality}

To introduce the key inequality, we first require a few basic preliminaries. A \emph{Rademacher random variable} is a random sign that takes each
of $-1$ and $1$ with probability $1/2$. Let
$\sigma,\tau\in\{-1,1\}^k\subset\R^k$ be independent random column
vectors, with all $2k$ coordinates being these independent Rademacher variables.
For a fixed real sign matrix $S\in\{-1,1\}^{k\times k}$, we define
\[
 \phi_S:\R\longrightarrow\mathbb C,\qquad
 \phi_S(t)=\E_{\sigma,\tau}\exp(i t\sigma^T S\tau),
\]
where $i^2=-1$. The product $\sigma^T S\tau$ is a real scalar as we are multiplying matrices of dimension $(1\times k)(k\times k)(k\times1)$.
The function $\phi_S$ is the so-called ``characteristic function'' of that scalar
random variable. If we pair up $\sigma$ with $-\sigma$, we can see that its
imaginary part is zero, so in fact $\phi_S(t)\in[-1,1]$ for real $t$. However, it is \emph{not} clear whether $\phi_S(t)$ should be negative or positive, and it is proving this which constitutes the vast majority of the effort in our lower bound. Ultimately, we seek to show the following: 

\begin{theorem}[Uniform positivity of the characteristic function]
\label{thm:uniform}
For $a=10^{-7}$, every positive integer $k$, and every
$S\in\{-1,1\}^{k\times k}$, put
\[
 \theta=\frac{a}{\sqrt{k}\log(4k)}.
\]
Then
\begin{equation}\label{eq:analytic-target}
 \phi_S(\theta)
 \ge\frac14\exp\left(-\frac{64a^2k}{\log(4k)}\right)>0.
\end{equation}
\end{theorem}

The word ``uniform'' means that the same bound holds for every choice
of the $k^2$ signs in $S$. Simply averaging over random matrices $S$ would not
suffice, as it turns out that each choice of $S$ corresponds with an \emph{eigenvalue} of the eventual matrix that we create. Because we seek to apply the framework of \cite{lovasz1979shannon}, we then require that \emph{all} eigenvalues are positive.

As mentioned, we omit the proof of \cref{thm:uniform} from the main body of this paper, as its techniques are largely disjoint from the rest of the paper (and from the TCS community more generally). The complete argument is detailed in \cref{sec:keyInequalityApp}, and a lean proof may be found at \url{https://github.com/Louiealbp/ChromaticInequality}.

\subsection{Connecting the Characteristic Function to Graph Coloring}
\label{sec:graph}

With \cref{thm:uniform} in hand, we are now ready to explain how it implies \cref{thm:mainAppendix}. As in \cref{sec:genLB}, we will choose a large family of graphs (a subset of all possible graphs on $n$ vertices), and define a matrix whose rows and columns are indexed by these graphs. We will ultimately then ensure that this matrix satisfies the requirements of \cite{lovasz1979shannon}'s framework for coloring lower bounds, while simultaneously ensuring that the entry-wise sum is as large as possible.

We now present the reduction more carefully. To start, we use \emph{sign matrices}, where a \emph{sign matrix} is a real matrix whose
entries belong to $\{-1,1\}$. For a $k$-by-$k$ sign matrix $S$, we use the same $\phi$-function as above:
\[
\phi_S:\R\longrightarrow\R,
\qquad
\phi_S(\theta)=\E_{\sigma,\tau}
  e^{i\theta\sigma^TS\tau},
\]
where $i^2=-1$ and $\sigma,\tau\in\{-1,1\}^k$ are independent uniform
vectors. Although the definition uses complex numbers, its value is real:
replacing $\sigma$ by $-\sigma$ preserves its distribution, but conjugates the
value. Thus, one may equivalently see that $\phi_S(\theta)=\E\cos(\theta\sigma^TS\tau)$.

It should not be obvious at all a priori why this function will appear. In the remainder of this section, we will exactly detail the construction of our family of graphs, and define a matrix supported on its edges (as in \cref{fact:chromaticLB}).  Ultimately, the function $\phi_S(\theta)$ will be revealed to control the eigenvalues of this matrix and bounding its value will be necessary for ensuring that the conditions of \cref{fact:chromaticLB} are satisfied.

\subsubsection{The Graph Family}

We now define the family of graphs which we use to prove the coloring lower bound. For our parameter $k$, we let $H=K_{k,k}$ have left vertices $u_1,\ldots,u_k$ and right vertices
$v_1,\ldots,v_k$. Its $m=k^2$ edges are $u_iv_j$, with $1\le i,j\le k$.
Now, we fix an ordering of these edges. We use $\F_2=\{0,1\}$ for the field in which
addition and multiplication are taken modulo two. 
The space $\F_2^m$ consists of one binary coordinate per edge, and for
$x\in\F_2^m$, we write $x_{ij}$ for its coordinate on the edge from $u_i $ to $v_j$. 

We will use two kinds of vectors with different purposes. The binary
vectors $x,b,\ell$ have $m$ coordinates and describe edge subsets or edge
labels. Later, a real vector indexed by all $x\in\F_2^m$ has $2^m$
coordinates, one for each graph in our family. Binary addition applies to
the edge labels, but we will use ordinary real addition for the entries of those
large matrices and real vectors. In particular, an expression such as
$x+b\le y+b$ first adds the labels modulo two and then compares their
resulting zero-one values as integers.

We now recall the notion of the \emph{lift of a graph} which we used in \cref{sec:VILS}. For each vertex $w\in V(H)$, we replace it by two labeled vertices $(w,0)$ and
$(w,1)$. For an edge $u_iv_j$, the bit $x_{ij}$ specifies which of two
matchings to use:
\[
\begin{array}{c|c}
x_{ij}=0 & \{(u_i,0),(v_j,0)\},\quad
             \{(u_i,1),(v_j,1)\}\\[2pt]
x_{ij}=1 & \{(u_i,0),(v_j,1)\},\quad
             \{(u_i,1),(v_j,0)\}.
\end{array}
\]
The resulting graph is denoted $L_x$. More compactly, it has $4k$ many vertices, and has the edges
\[
\{(u_i,t),(v_j,t+x_{ij})\},
\qquad 1\le i,j\le k,\quad t\in\F_2.
\]

Every $L_x$ is a \emph{simple graph} on the same $4k$ labeled vertices. For the remainder of the proof, we focus our attention only on these $2^m$ distinct graphs which are obtained via the two lift operation.

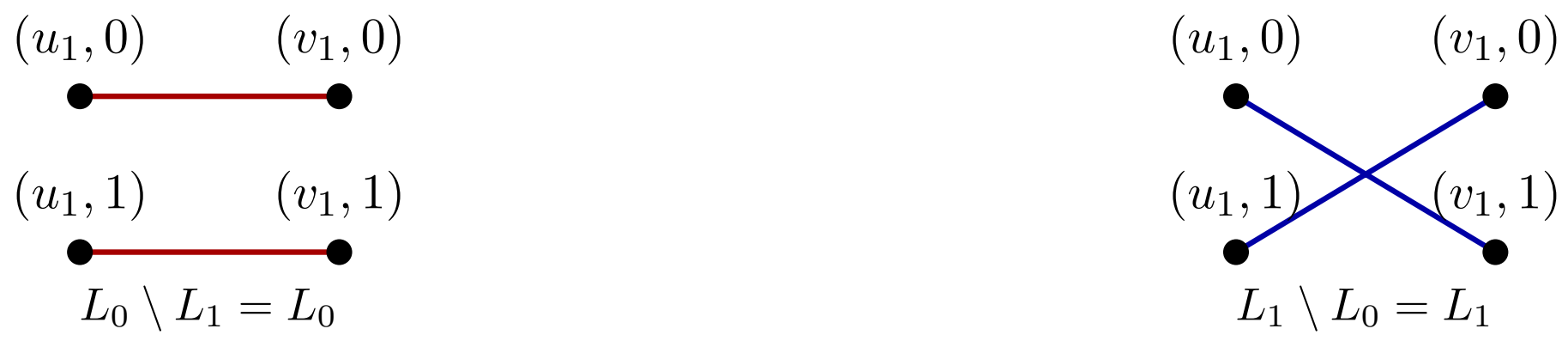
\begin{figure}[tb]
\centering
\begin{minipage}{.46\textwidth}
\centering
\begin{tikzpicture}[x=2cm,y=1.2cm]
\draw[very thick,red!65!black] (0,1)--(1,1) (0,0)--(1,0);
\foreach \name/\xx/\yy/\lab in
 {u0/0/1/{(u_1,0)},v0/1/1/{(v_1,0)},u1/0/0/{(u_1,1)},v1/1/0/{(v_1,1)}}
 {\node[circle,fill=black,inner sep=2pt,label=above:{$\lab$}]
  (\name) at (\xx,\yy) {};}
\end{tikzpicture}
\par\small $L_0\setminus L_1=L_0$
\end{minipage}
\hfill
\begin{minipage}{.46\textwidth}
\centering
\begin{tikzpicture}[x=2cm,y=1.2cm]
\draw[very thick,blue!65!black] (0,1)--(1,0) (0,0)--(1,1);
\foreach \name/\xx/\yy/\lab in
 {u0/0/1/{(u_1,0)},v0/1/1/{(v_1,0)},u1/0/0/{(u_1,1)},v1/1/0/{(v_1,1)}}
 {\node[circle,fill=black,inner sep=2pt,label=above:{$\lab$}]
  (\name) at (\xx,\yy) {};}
\end{tikzpicture}
\par\small $L_1\setminus L_0=L_1$
\end{minipage}
\caption{The two-lift example with one base edge. The crossing in the right
picture is not a vertex. The vertices with labels $(u_1,0)$ and $(v_1,1)$
are connected on the right and disconnected on the left.}
\label{fig:two-lift-example}
\end{figure}

\subsubsection{The Switching Operation}

We now define a new operation on these lifted graphs which we call \emph{switching}, which will be a necessary ingredient for defining our eventual matrix (in particular, for proving that it is supported on the edges of the universal graph $P_n$). To specify this operation, we choose bits $a_w\in\F_2$ at the $2k$ base vertices of the graph. For vertices which have $a_w = 1$, we then perform a \emph{relabeling} of its vertices, where we map
\[
(w,t)\longmapsto(w,t+a_w).
\]
More clearly, for lifted vertices of the form $(u, 0), (u,1)$ whenever $a_u = 1$, we re-label $(u,0)$ with $(u,1)$ and re-label $(u,1)$ with $(u, 0)$. 
Under this permutation, $L_x$ becomes $L_{x+b}$, where
$b_{ij}=a_{u_i}+a_{v_j}$.

With this operation defined, we next define the \emph{binary cut space}:
\[
\mathcal B
=\bigl\{(a_{u_i}+a_{v_j})_{i,j}:a\in\F_2^{V(H)}\bigr\}
\subseteq\F_2^m.
\]
Intuitively, this is \emph{all possible sets of cut edges} which can be obtained via a $\zo$-assignment to the vertices of $H$.
It is a linear subspace because the map from vertex bits to edge bits is
linear over $\F_2$ (simply the XOR of the endpoints). Likewise, because the graph is bipartite, the all-ones vector belongs to $\mathcal B$: take
$a_{u_i}=0$ on the left and $a_{v_j}=1$ on the right. Lastly, because $H$ is connected, the kernel of this map consists exactly of the two constant vertex
labelings. Consequently $\dim\mathcal B=2k-1$ and
$|\mathcal B|=2^{2k-1}$.

\begin{lemma}[A sufficient condition for conflict]\label{lem:lift-comparison}
Suppose $x,y\in\F_2^m$ are distinct and there is $b\in\mathcal B$ such that
$x+b\le y+b$ coordinatewise, when the coordinates are viewed as ordinary
integers in $\{0,1\}$. Then $L_x$ and $L_y$ are adjacent in $P_{4k}$.
\end{lemma}

\begin{proof}
We start by applying the same switching permutation to both graphs. In more detail, starting with $L_x$ and $L_y$, we apply the switching permutation corresponding to $b$. If we denote this permutation of vertex labels by $\pi$, we then have that $\pi(L_x) = L_{x+b}$ and $\pi(L_y) = L_{y + b}$. Moreover, for any single permutation $\pi$ and any graphs $G, H$ on the same
vertex set, it will always be the case that 
\[
\pi(G - G \cap H) = \pi(G) - \pi(G) \cap \pi(H),
\]
and thus $H - G \cap H$ and $G - G \cap H$ will have the same connected components if and only if $\pi(G) - \pi(G) \cap \pi(H)$ and $\pi(H) - \pi(G) \cap \pi(H)$ have the same connected components. 

So, if we then consider the graphs $L_{x+b}$ and $L_{y+b}$, we can observe that $L_{x+b} - L_{x + b} \cap L_{y + b}$ and $L_{y+b} - L_{x + b} \cap L_{y + b}$ have the same connected components if and only if $L_x - L_x \cap L_y$ and $L_y - L_x \cap L_y$ have the same connected components. For the rest of the proof, relabel the switched vectors $x+b,y+b$ as $x,y$, so that $x\leq y$. 

Now, we consider all of the edges $D=\{e:x_e\ne y_e\}$ where $x$ and $y$ disagree. Note that for all \emph{agreement} edges, (i.e., $e = (u,v): x_e = y_e$), when we apply the lift we obtain the same lifted edges in $L_x$ and $L_y$ corresponding to $e$. Thus, when we ultimately consider the difference $L_x - L_x \cap L_y$ and $L_y - L_x \cap L_y$, all of these edges will cancel out. Whenever an edge $e$ is \emph{in} $D$, this means that the lifted edges corresponding to $e$ are \emph{disjoint} between $L_x$ and $L_y$. More explicitly, because we now have $x \leq y$, we know that for $e \in D$, $x_e = 0$ and $y_e = 1$. In the graph $L_x$, this then implies that all lifted edges corresponding to $e = (u,v)$ are of the form 
\[
\{(u,0), (v,0) \} \quad \{(u,1), (v,1) \}
\]
and in $L_y$, the lifted edges are of the form 
\[
\{(u,0), (v,1) \} \quad \{(u,1), (v,0) \}.
\]

To conclude the argument, we then observe that for every edge $e = (u,v) \in L_x - L_x \cap L_y$, it is always of the form $\{(u,0), (v, 0) \}$ or $\{(u,1), (v, 1) \}$. Thus, there is never any crossing edge between the lifted ``$1$-copies'' and the lifted ``$0$-copies''. Meanwhile, in $L_y - L_x \cap L_y$, there is at least one such edge, forcing $L_y - L_x \cap L_y$ and $L_x - L_x \cap L_y$ to have different connected components. 
\end{proof}

\subsubsection{The Lower Bound Matrix}

With the above established, we are now ready to define our final matrix $M$ which witnesses the coloring lower bound. To do this, we fix $p\in(0,1)$, and define two real $2$-by-$2$ matrices, with rows and
columns indexed by $0,1$:
\[
Q=\begin{pmatrix}1&2p\\0&1\end{pmatrix},
\qquad
X=\begin{pmatrix}0&1\\1&0\end{pmatrix}.
\]
The tensor power $Q^{\otimes m}$ is the real $2^m$-by-$2^m$ matrix indexed
by $x,y\in\F_2^m$ whose entries are
\[
(Q^{\otimes m})_{xy}=\prod_{e\in E(H)}Q_{x_e,y_e}.
\]
The $xy$th entry of $Q^{\otimes m}$ is nonzero \emph{exactly} when $x\le y$ coordinatewise. For
$b\in\F_2^m$, let $X^b$ be the $2^m$-by-$2^m$ permutation matrix sending
the coordinate vector at $x$ to the coordinate vector at $x+b$. Thus
$X^b=\bigotimes_e X^{b_e}$, where $X^0$ is the $2$-by-$2$ identity and
$X^1=X$.

To define our witnessing matrix $M$, we then take the average of this matrix over the entire cut-space. I.e.,
\begin{equation}\label{eq:graph-fitting-matrix}
M=\frac1{|\mathcal B|}\sum_{b\in\mathcal B}X^bQ^{\otimes m}X^b
  \in\R^{2^m\times2^m}.
\end{equation}
It is not hard to see that all entries of each summand are nonnegative, and every diagonal entry is
one. Likewise, if $x\ne y$ and $M_{xy}\ne0$, at least one summand is nonzero at
$(x,y)$, so $x+b\le y+b$ for some $b\in\mathcal B$. By our previous argument (see Lemma~\ref{lem:lift-comparison})
this then forces the connected components of $L_x - L_x \cap L_y$ and $L_y - L_x \cap L_y$ to be different.

Furthermore, the matrix $M$ is real and symmetric, although $Q$ itself is not. Indeed
$XQX=Q^T$. Taking the transpose of a summand in
\eqref{eq:graph-fitting-matrix} therefore replaces $b$ by
$b+\one$, where $\one\in\F_2^m$ is the all-one bit vector. Since
$\one\in\mathcal B$, this replacement merely permutes the summands. The remaining piece of the argument is to prove that $M$ is positive definite.

\subsubsection{Some Useful Transformations}

As mentioned above, with the above matrix established, our primary goal is to prove that it is PSD, and doing so requires an understanding of the \emph{eigenvalues} of the matrix. In this section, we provide some useful characterizations towards this end.

For $u,x\in\F_2^m$, write $u\cdot x=\sum_eu_ex_e\in\F_2$. Define
\[
\mathcal C=\mathcal B^\perp
=\{c\in\F_2^m:c\cdot b=0\text{ for every }b\in\mathcal B\}.
\]
Note that the elements are exactly the edge subsets with even degree at every base
vertex. To see this, for any vertex labeling $a$ we have, in $\F_2$,
\[
\sum_{u_iv_j\in E(H)}c_{ij}(a_{u_i}+a_{v_j})
=\sum_{w\in V(H)}a_w\deg_c(w).
\]
This vanishes for every $a$ exactly when all the degrees are even. Thus
$\mathcal C$ is the binary cycle space, and
$\dim\mathcal C=m-(2k-1)=(k-1)^2$. Every $c\in\mathcal C$ also has even
Hamming weight $|c|$, meaning an even number of coordinates equal to one:
its number of edges is the sum of its even degrees over the left vertices.
Here, we are crucially using the bipartiteness to ensure even cardinality. 

The \emph{Walsh transform} is the real $2^m$-by-$2^m$ matrix
\[
W_{ux}=2^{-m/2}(-1)^{u\cdot x},\qquad u,x\in\F_2^m.
\]
Its rows are the functions from $\F_2^m$ to $\{-1,1\}$ given by
$x\mapsto(-1)^{u\cdot x}$, normalized to have Euclidean norm one
when viewed as vectors in $\R^{2^m}$. We now have the following lemma:

\begin{lemma}\label{lem:walsh-orthogonality}
For every $a\in\F_2^m$,
\[
\sum_{x\in\F_2^m}(-1)^{a\cdot x}
=\begin{cases}2^m,&a=0,\\0,&a\ne0.\end{cases}
\]
Consequently $W^T=W$ and $W^2=I_{2^m}$.
\end{lemma}

\begin{proof}
The case $a=0$ is immediate. Otherwise choose a coordinate $e$ with
$a_e=1$ and pair $x$ with $x+e_e$, where $e_e\in\F_2^m$ has a one only in that
coordinate. The two summands have opposite signs. Finally, the inner
product of the rows indexed by $u,v$ is $2^{-m}\sum_x(-1)^{(u+v)\cdot x}$,
which equals one when $u=v$ and zero otherwise.
\end{proof}

In Walsh coordinates, adding $b$ to the input multiplies character $u$ by
$(-1)^{u\cdot b}$. Therefore $WX^bW$ is diagonal, with that entry at $u$. By directly multiplying these matrices in a single coordinate, we get the following
$2\times2$ matrices $W_1$ and $J$:
\[
W_1QW_1=\begin{pmatrix}1+p&-p\\p&1-p\end{pmatrix}=:J,
\qquad
W_1=2^{-1/2}\begin{pmatrix}1&1\\1&-1\end{pmatrix}.
\]
The $(u,v)$th entry of $WMW$ is consequently
\begin{equation}\label{eq:graph-cut-average}
(WMW)_{uv}
=\left(\prod_eJ_{u_e,v_e}\right)
 \frac1{|\mathcal B|}\sum_{b\in\mathcal B}(-1)^{(u+v)\cdot b}.
\end{equation}
Note that in the last average, we obtain $1$ if $u+v\in\mathcal C$. Otherwise, if $u + v \notin \mathcal{C}$, we choose
$b_0\in\mathcal B$ with $(u+v)\cdot b_0=1$ and pair $b$ with $b+b_0$, obtaining an average of $0$. This ``cut averaging'' operation has therefore exactly deleted those entries outside $u+v\in\mathcal C$.

\subsubsection{Calculating the Eigenvalues}

We are now ready to prove the positive definiteness of the matrix $M$. For a bit vector $u$, let $|u|$ denote its Hamming weight. We use the
positive diagonal matrix $D\in\R^{2^m\times2^m}$ and the diagonal sign
matrix $R\in\R^{2^m\times2^m}$ defined as
\[
D_{uu}=d_u=(1+p)^{m-|u|}(1-p)^{|u|},\qquad
R_{uu}=(-1)^{\lfloor|u|/2\rfloor},
\qquad z=\frac p{\sqrt{1-p^2}}.
\]
All $d_u$ are positive because $0<p<1$. Next, we define
\[
A=RD^{-1/2}WMWD^{-1/2}R\in\R^{2^m\times2^m}.
\]
This is a \emph{congruence}: $A=PMP^T$ for the invertible matrix
$P=RD^{-1/2}W\in\R^{2^m\times2^m}$. A congruence preserves positive definiteness, since
$v^TAv=(P^Tv)^TM(P^Tv)$, but it does not generally preserve eigenvalues.
We will compute the eigenvalues of $A$ and use their positivity to deduce
positive definiteness of $M$.

Towards this end, we now seek to exactly understand the entries of the matrix $A$. Recall that in the previous section, we showed that 
\[
(WMW)_{uv} = \mathbf{1}[u + v \in \mathcal{C}] \prod_{e = 1}^m J_{u_e, v_e},
\]
where $\mathcal{C}$ is the binary cycle space of $K_{k,k}$. The intuition behind the creation of the matrix $A$ is that we want to modify $WMW$ so that entry $(u,v)$ depends only on $|u+v|$, rather than our choices of $u$ and $v$ separately, as they do now (in the $J_{u_e, v_e}$ term). The first step towards this simplification is to modify our matrix $J = \begin{bmatrix}
    1+p & -p \\ 
    p & 1-p 
\end{bmatrix}$ to a matrix where the diagonal entries are all $1$. This is exactly what we achieve with the matrix $D$. For instance, $D_1 = \begin{bmatrix}
    1+p & 0 \\ 0 & 1-p
\end{bmatrix}$, and thus one can see that 
\[
D_1^{-1/2} J D_1^{-1/2} = \begin{bmatrix}
    1 & -z \\ 
    z & 1
\end{bmatrix}.
\]

So, for the aforementioned definition of $D$, we then obtain that for
\[
K := D^{-1/2} WMW D^{-1/2},
\]
that 
\[
K_{uv} = \mathbf{1}[u + v \in \mathcal{C}] \prod_{e = 1}^m \begin{bmatrix}
    1 & -z \\ 
    z & 1
\end{bmatrix}_{u_e, v_e}.
\]

For a supported entry $(u,v)$, we can see that the magnitude of the entry is precisely $z^{|u+v|}$, as we pick up a factor of $1$ whenever $u_e =v_e$, and a factor of $\pm z$ whenever $u_e \neq v_e$. Likewise, the sign of entry $u,v$ is exactly $(-1)^{|v\setminus u|}$, where
$v\setminus u$ denotes coordinates with $v_e=1,u_e=0$. Finally, we now consider 
\[
(RKR)_{uv} = (RD^{-1/2} WMW D^{-1/2}R)_{uv} = \mathbf{1}[u+v\in\mathcal C] R_{uu} \cdot R_{vv} \cdot z^{|u+v|} \cdot (-1)^{|v\setminus u|}
\]
\[
=\mathbf{1}[u+v\in\mathcal C] z^{|u + v|} \cdot (-1)^{\lfloor |v| /2\rfloor +\lfloor |u| /2\rfloor + |v\setminus u|}.
\]
On any supported entry, we know that $u + v \in \mathcal{C}$, so $|u+v|$ is even, so then $|u|$ and $|v|$ have the same
parity, say $\epsilon\in\{0,1\}$. 
The identity
\[
\lfloor|u|/2\rfloor+\lfloor|v|/2\rfloor+|v\setminus u|
-\frac{|u+v|}{2}=|v|-\epsilon
\]
then implies that 
\[
\lfloor|u|/2\rfloor+\lfloor|v|/2\rfloor+|v\setminus u| = \frac{|u+v|}{2} + |v|-\epsilon,
\]
and since $|v|$ and $\epsilon$ have the same parity, we can then write the sign purely as $(-1)^{\frac{|u+v|}{2}}$ (the final $|v|$ and $\eps$ have the same parity and cancel out).

So, all this then implies that
\begin{equation}\label{eq:graph-convolution}
A_{uv}=\begin{cases}
(-1)^{|u+v|/2}z^{|u+v|},&u+v\in\mathcal C,\\
0,&u+v\notin\mathcal C.
\end{cases}
\end{equation}

As desired, the entry now depends only on $u+v$, and therefore enables an \emph{explicit} calculation of its eigenvectors. 
For $\ell\in\F_2^m$, multiply $A$ by the character column in
$\R^{2^m}$ with coordinate
$(-1)^{\ell\cdot v}$ and substitute $a=u+v$ in the sum over $v$.
The result at coordinate $u$ is $(-1)^{\ell\cdot u}$ times
\begin{equation}\label{eq:graph-eigenvalues}
\lambda_\ell(z)
=\sum_{a\in\mathcal C}(-1)^{|a|/2+\ell\cdot a}z^{|a|}.
\end{equation}
For fixed $\ell$, this formula defines a polynomial
$\lambda_\ell:\R\to\R$; the argument above uses it at the positive
value $z=p/\sqrt{1-p^2}$.
Because the characters are orthogonal, this gives an explicit basis of $2^m$ independent eigenvectors. This therefore computes every eigenvalue, with multiplicity, as our matrix is $2^m \times 2^m$. 

We do note that different $\ell$ can give the same character on $\mathcal{C}$, but this is not an issue for us. To be more concrete, two labels
$\ell,\ell'$ define the same character on $\mathcal C$ if and only if
$(\ell+\ell')\cdot a=0$ for every $a\in\mathcal C$. Since the binary dot
product is nondegenerate and $\mathcal C=\mathcal B^\perp$, this says
$\ell+\ell'\in\mathcal B$. The equal characters therefore form cosets
of $\mathcal B$, each of size $2^{2k-1}$. There are
$2^{m-(2k-1)}=2^{(k-1)^2}$ different restricted characters, which accounts
for all $2^m$ eigenvector positions. 

\subsubsection{Re-Writing Eigenvalues as Characteristic Functions}

Above, we established that $\lambda_\ell(z)
=\sum_{a\in\mathcal C}(-1)^{|a|/2+\ell\cdot a}z^{|a|}$ exactly captures the eigenvalue corresponding to the $\ell$th character. We now explain how to connect this to the aforementioned \cref{thm:uniform}.

We set $p=\sin\theta$, so $z=\tan\theta$. For
$\ell\in\F_2^m$, let $S_\ell\in\{-1,1\}^{k\times k}$ have entries
$(S_\ell)_{ij}=(-1)^{\ell_{ij}}$. We claim that
\begin{equation}\label{eq:graph-characteristic}
\lambda_\ell(\tan\theta)
=(\cos\theta)^{-m}\phi_{S_\ell}(\theta).
\end{equation}

Indeed, recall that 
\[
\phi_S(\theta) = \E_{\sigma, \tau}[\exp(i \theta \sigma^T S \tau)],
\]
where $\sigma, \tau \in \{ \pm 1\}^k$ are independent uniform signs. Thus, we can rewrite
\[
\phi_S(\theta) = \E_{\sigma, \tau} \left [ \prod_{i = 1}^k \prod_{j = 1}^k e^{i \theta S_{i,j}\sigma_i \tau_j}\right ].
\]
There are $m = k^2$ terms in this product, one for each possible choice of base edge in the graph $K_{k, k}$.

Now, because $(S_\ell)_{ij}\sigma_i\tau_j\in\{-1,1\}$, Euler's formula implies that
\[
e^{i\theta(S_\ell)_{ij}\sigma_i\tau_j}
=\cos\theta\bigl(1+i\tan\theta\,(S_\ell)_{ij}\sigma_i\tau_j\bigr).
\]
Thus, 
\[
\phi_S(\theta) = (\cos \theta)^m \E_{\sigma, \tau} \left [ \prod_{i,j} (1  + i (\tan \theta ) S_{i,j}\sigma_i \tau_j)\right ].
\]

We next multiply out and expand this product. Letting $z = \tan \theta$, the product is 
\[
\prod_{i,j} (1  + i (\tan \theta ) S_{i,j}\sigma_i \tau_j).
\]
For each $(i,j)$, we have one of two choices: either (1) we can choose $1$, or (2) we can choose $i (\tan \theta ) S_{i,j}\sigma_i \tau_j$. Thus, we can index each term via an edge subset $a \subseteq K_{k,k}$, indicating for which edges we select the term $i (\tan \theta ) S_{i,j}\sigma_i \tau_j$. By collecting these terms, we can then exactly observe that 
\[
\prod_{e = (i,j) \in a} (i (\tan \theta ) S_{i,j}\sigma_i \tau_j) = 
(iz)^{|a|}
\left(\prod_{(i,j)\in a}S_{ij}\right)
\underbrace{
\left(\prod_i\sigma_i^{\deg_a(u_i)}\right)
\left(\prod_j\tau_j^{\deg_a(v_j)}\right)
}_{\text{random spin factor}}.
\]

Because of the independence and symmetry of the ``spin factor'', we see that its expectation is zero \emph{unless}
every degree is even, in which case its expectation is one. In this way, the surviving
subsets are exactly those in $\mathcal C$. For such a subset, $|a|$ is even,
$i^{|a|}=(-1)^{|a|/2}$, and the product of its edge signs is
$(-1)^{\ell\cdot a}$. Substituting both identities gives
\[
\phi_{S_\ell}(\theta)
=
(\cos\theta)^m
\sum_{a\in\mathcal C}
(-1)^{|a|/2+\ell\cdot a}
(\tan\theta)^{|a|}.
\]

But recall the eigenvalue formula derived earlier:
\[
\lambda_\ell(z)
=
\sum_{a\in\mathcal C}
(-1)^{|a|/2+\ell\cdot a}z^{|a|}.
\]
This proves \eqref{eq:graph-characteristic} term by term.

Now, because of \cref{thm:uniform}, the assumed positivity of every $\phi_S(\theta)$ now makes every eigenvalue
of $A$ positive. Thus $A$ and, by congruence, $M$ are positive definite.

\subsubsection{Finishing the Lower Bound}

Finally, we prove \cref{thm:mainAppendix}.

\begin{proof}[Proof of \cref{thm:mainAppendix}.]
We define $M$ as above, i.e., 
\[
M=\frac1{|\mathcal B|}\sum_{b\in\mathcal B}X^bQ^{\otimes m}X^b
  \in\R^{2^m\times2^m},
\]
where 
\[
Q=\begin{pmatrix}1&2p\\0&1\end{pmatrix},
\qquad
X=\begin{pmatrix}0&1\\1&0\end{pmatrix}.
\]

As we already established, $M$ is real, symmetric, with $1$'s on the diagonal, and its non-zero off-diagonal entries are completely supported on the edges of $P_{4k}$.

Moreover, we know that $M$ is PSD if and only if $A$ is PSD, where $A$ is the matrix defined such that 
\[
A_{uv}=\begin{cases}
(-1)^{|u+v|/2}z^{|u+v|},&u+v\in\mathcal C,\\
0,&u+v\notin\mathcal C.
\end{cases}
\]

We next use our formula for the eigenvalues of $A$. We set $p=\sin\theta$ and $z=\tan\theta$. We likewise let
$\ell\in\F_2^m$, and let $S_\ell\in\{-1,1\}^{k\times k}$ have entries
$(S_\ell)_{ij}=(-1)^{\ell_{ij}}$.
As computed in \cref{eq:graph-characteristic}, the eigenvalues of $A$ are exactly
\begin{equation}
\lambda_\ell(\tan\theta)
=(\cos\theta)^{-m}\phi_{S_\ell}(\theta).
\end{equation}

By invoking \cref{thm:uniform}, we then know that for $a=10^{-7}$, every positive integer $k$, and every
$S\in\{-1,1\}^{k\times k}$, if
\[
 \theta=\frac{a}{\sqrt{k}\log(4k)},
\]
then
\begin{equation}
 \phi_S(\theta)
 \ge\frac14\exp\left(-\frac{64a^2k}{\log(4k)}\right)>0.
\end{equation}

So, we are permitted to select $\theta=\frac{a}{\sqrt{k}\log(4k)}$, and for this choice of $\theta$ (and $p = \sin(\theta)$) it is the case that the matrix $M$ is PSD. For this same choice of $\theta$, we can see that 
\[
\frac{\mathbf{1}^T M \mathbf{1}}{2^{k^2}} = \frac{(2 + 2 \sin \theta)^{k^2}}{2^{k^2}} = (1 + \sin \theta)^{k^2} \geq \left (1 + \frac{a}{2\sqrt{k}\log(4k)}\right )^{k^2}
\]
\[
\geq 2^{\Omega(k^{3/2} / \log(k))}.
\]

Because $M$ is supported on $2^{k^2}$ many subgraphs (vertices) of $P_{4k}$, we then obtain via \cref{fact:chromaticLB} that 
\[
\chi(P_{4k}) \geq 2^{\Omega(k^{3/2} / \log(k))}.
\]
For general $n$, take $k=\lfloor n/4\rfloor$ and add $n-4k$ isolated vertices to every graph in the construction. This preserves conflicts, so $\chi(P_n)\geq\chi(P_{4k})$, concluding the theorem.
\end{proof}

\subsection{Proving \cref{cor:ftcoLB}}

For completeness, we also include here a proof of \cref{cor:ftcoLB}.

\begin{proof}[Proof of \cref{cor:ftcoLB}.]
The intuition for the proof is quite simple. We have already proved a lower bound of the form $\tilde{\Omega}(n f^{1/4})$ when $f = n^2$. To extend this to smaller values of $f$, we break up the vertex set of size $n$ into $\frac{n}{\sqrt{f}}$ blocks of size $\sqrt{f}$. Each one of these blocks is completely disjoint from the others. Moreover, because our global budget in the number of faults is $f$, if all edge faults target a single one of these blocks, there can be as many as $f \geq \binom{\sqrt{f}}{2}$ many faults, meaning that there is essentially no bound on the number of edge faults (i.e., all edges can be deleted). Intuitively, we should then expect an $\tilde{\Omega}((\sqrt{f})^{3/2})$ lower bound for the message length for \emph{each} of these blocks, of which there are $\frac{n}{\sqrt{f}}$ many. Summing across these blocks would then yield a lower bound of 
\[
\tilde{\Omega}((\sqrt{f})^{3/2}) \cdot \frac{n}{\sqrt{f}} = \tilde{\Omega}(n f^{1/4}).
\]

Formally, for $f\geq64$, we let $k=\lfloor\sqrt f/8\rfloor$ and $t=\lfloor(n-1)/(4k)\rfloor$. We then consider graphs consisting of $t$ disjoint $4k$-vertex blocks, with block $i$ equal to the lifted graph $L_{x_i}$ from the construction above. We let $M$ be the matrix from the preceding proof, with $\theta=a/(\sqrt{k}\log(4k))$, and use the matrix $M^{\otimes t}$ indexed by these graphs. It is PSD and has diagonal entries equal to $1$.

We now claim that every nonzero off-diagonal entry corresponds to two graphs distinguishable by at most $f$ common edge deletions. Indeed, consider strings $x,y \in \zo^{tk^2}$ such that $(M^{\otimes t})_{xy} \neq 0$. We then focus our attention on a single block $i$ with $x_i\neq y_i$, where as above, $x_i$ refers to the subset of the lifting string $x$ for block $i$. Since $(M)_{x_i,y_i}\neq0$, if we set $F_i=E(L_{x_i})\cap E(L_{y_i})$ it must be the case that $L_{x_i} - F_i$ and $L_{y_i} - F_i$ have different connected components, using \cref{lem:lift-comparison}. In particular, because $F_i$ is completely contained in block $i$, the number of edges that have been deleted is at most $2k^2 \leq f$. Thus, for the two graphs indexed by $x$ and $y$, it is the case that a \emph{single set} of $f$ deletions suffices for distinguishing the connected components of $x$ and $y$.

Crucially, this then means that the messages of any $f$-fault tolerant connectivity oracle properly color this conflict graph. By \cref{fact:chromaticLB}, their bit length is at least
\[
\log_2\frac{\mathbf1^T M^{\otimes t}\mathbf1}{2^{tk^2}}
=t k^2\log_2(1+\sin\theta)
=\Omega\!\left(\frac{t k^{3/2}}{\log(4k)}\right)
=\widetilde\Omega(n f^{1/4}),
\]
where $k=\Theta(\sqrt f)$ and $t=\Theta(n/k)$ since $f\leq n^2$. Note that whenever $f$ is (say) $\leq 1000$, we simply use the fact that $1$-fault tolerant connectivity oracles require space $\Omega(n)$. This concludes the claim.
\end{proof}

\section{AI Acknowledgment}

Many of the proofs and bounds presented in this paper were generated entirely by AI. The authors first began studying this problem in late 2023, and were able to establish bounds of the form $\tilde{O}(n \sqrt{f})$ for the one-way communication complexity whenever Bob's graph contains fewer than $f$ edges (see \cite{khanna2026fault}, where this bound is established even for approximate \emph{distance reporting}). The authors then revisited this question in Summer 2026, and were able to prove (without the use of AI) the $\tilde{\Omega}(n^{3/2})$ lower bound for connectivity when restricted to vertex-incidence linear sketches. Subsequently, the authors provided the chromatic number formulation of the connectivity problem to ChatGPT 6-Astra Ultra, and made extensive use of this model to  push the lower bound beyond vertex-incidence linear sketches, yielding the $\Omega(n^{1.25})$ lower bound of \cref{thm:main} and the $\tilde\Omega(n^{1.5})$ lower bound found in \cref{thm:mainAppendix}. The authors initially believed that the problem should have complexity $\Omega(n^2)$, but nevertheless were successful in prompting Astra to \emph{upper-bound} the chromatic number, resulting in both the $\widetilde{O}(n^{1.8})$ and $\widetilde{O}(n^{1.75})$ upper bounds. The extension to the streaming setting was proposed and done solely by the authors, as was all of the writing in the main body of the paper. The appendix has only been lightly refined, and is primarily GPT-written. We include it as we believe it will be a promising starting point for future work on this problem, and also include a lean verification (\url{https://github.com/Louiealbp/ChromaticInequality}) of the inequality proved therein.

\bibliographystyle{alpha}
\bibliography{ref}

\appendix

\section{Proving the Key Inequality in \cref{sec:appn32}}\label{sec:keyInequalityApp}

\subsection{Organization and prerequisites}

Section~\ref{sec:preliminaries} defines the probabilistic and matrix
language used later. It also proves the specific general facts needed
in the argument. In particular, the Brascamp--Lieb variance inequality
is attributed to its classical sources and proved here by one-dimensional
integration by parts followed by induction. The elementary entropy
identities are also proved, so familiarity with information theory is
not required. The reader is assumed to know ordinary multivariable
calculus, basic finite-dimensional linear algebra, and the usual rules
for absolutely convergent sums and integrals. When a limiting argument
is used, we specify the dominating bound or monotonicity that justifies it.

The rest of the proof is organized by purpose:
\begin{enumerate}
 \item Section~\ref{sec:graph} constructs the graphs and proves a
 conditional graph-coloring bound from the positivity of $\phi_S$.
 This identifies the precise analytic statement that remains to be proved.
 \item Section~\ref{sec:lattice} controls mass in convex sets and
 sums over periodic cells. An explicit noisy decoder makes the lattice
 estimate independent of any independence between the cell coordinates.
 \item Sections~\ref{sec:zero-field} and~\ref{sec:all-field} control
 covariances of random sign vectors under negative quadratic weights.
 The second section adds arbitrary linear terms to the exponent.
 \item Sections~\ref{sec:cosine-reference} and~\ref{sec:joint}
 replace the oscillating cosine by a positive comparison function and
 control the resulting joint distribution of signs and real variables.
 \item Section~\ref{sec:uniform} compares the central cell with all
 other cells and proves Theorem~\ref{thm:uniform}.
\end{enumerate}

\subsection{Preliminaries: matrices, probability, and convexity}
\label{sec:preliminaries}

This section fixes conventions and proves the reusable facts that appear
throughout the paper. Several later estimates are short once their
matrix or probabilistic form is recognized. The purpose here is to make
those forms available without leaving implicit assumptions about
dimensions, normalization, or the meaning of an expectation.

\subsubsection{Vectors, matrices, and positive-definite order}

For a positive integer $d$, $\R^d$ is the set of real column vectors
$x=(x_1,\ldots,x_d)^T$. The transpose $x^T$ is a $1\times d$ row
vector. For $x,y\in\R^d$, their inner product and Euclidean norm are
\[
 x^Ty=\sum_{j=1}^d x_jy_j,\qquad
 \norm{x}=(x^Tx)^{1/2}.
\]
The notation $\R^{r\times s}$ denotes real matrices with $r$ rows and
$s$ columns. Such a matrix $B$ defines a linear map $B:\R^s\to\R^r$.
Its transpose $B^T\in\R^{s\times r}$ maps $\R^r$ to $\R^s$.
The identity matrix $I_d\in\R^{d\times d}$ leaves vectors in $\R^d$
unchanged. For $B\in\R^{r\times s}$, its squared Frobenius norm is
$\norm{B}_F^2=\sum_{i=1}^r\sum_{j=1}^s B_{ij}^2$.
For a square matrix $M\in\R^{d\times d}$, its trace is
$\tr M=\sum_{j=1}^d M_{jj}$. The notation
$\diag(t_1,\ldots,t_d)$ means the $d\times d$ diagonal matrix with
the displayed diagonal entries.
The operator norm of $B\in\R^{r\times s}$ is
$\norm{B}_{\mathrm{op}}=\max_{x\in\R^s,\,\norm{x}=1}\norm{Bx}$;
it measures the largest factor by which $B$ can stretch a unit vector.

A symmetric matrix $M\in\R^{d\times d}$ is positive semidefinite (PSD),
written $M\succeq0$, if $x^TMx\ge0$ for every $x\in\R^d$.
It is positive definite, written $M\succ0$, if that inequality is
strict whenever $x\ne0$. For symmetric matrices $M,N$ of the same
size, the notation $M\preceq N$ means $N-M\succeq0$. It compares
quadratic forms in every direction; it does not compare entries.

The spectral theorem says that a real symmetric matrix can be written
$M=U\Lambda U^T$, where $U\in\R^{d\times d}$ is orthogonal
($U^TU=I_d$) and $\Lambda$ is diagonal with real entries, the
eigenvalues. In particular, if $M\succ0$, all its eigenvalues are
positive. Define $M^{1/2}=U\Lambda^{1/2}U^T$ and
$M^{-1/2}=U\Lambda^{-1/2}U^T$, both $d\times d$ matrices.
These definitions will let us turn weighted quadratic forms into
ordinary squared norms.

\begin{lemma}[A variational formula for the inverse]\label{pre:variational}
Let $M\in\R^{d\times d}$ be symmetric positive definite and
$u\in\R^d$. Then
\begin{equation}\label{eq:pre-variational}
 u^TM^{-1}u=\sup_{x\in\R^d}\{2u^Tx-x^TMx\}.
\end{equation}
The supremum is attained at $x=M^{-1}u$.
\end{lemma}
\begin{proof}
Complete the square:
\[
 2u^Tx-x^TMx
 =u^TM^{-1}u-(x-M^{-1}u)^TM(x-M^{-1}u).
\]
The last term is nonnegative and is zero exactly at the claimed
maximizer. Every term in the identity is a real scalar.
\end{proof}

\begin{lemma}[Inverse order and weighted Cauchy--Schwarz]
\label{pre:inverse-order}
If $M,N\in\R^{d\times d}$ are symmetric and
$0\prec M\preceq N$, then $N^{-1}\preceq M^{-1}$.
Also, for $M\succ0$ and $u,v\in\R^d$,
\begin{equation}\label{eq:pre-weighted-cs}
 (u^Tv)^2\le(u^TMu)(v^TM^{-1}v).
\end{equation}
\end{lemma}
\begin{proof}
For each fixed $u\in\R^d$, replacing $M$ by the larger matrix $N$
can only decrease the expression inside the supremum in
\eqref{eq:pre-variational}. This proves inverse order by its definition
in terms of quadratic forms. For the second assertion, apply the usual
Cauchy--Schwarz inequality to $M^{1/2}u\in\R^d$ and
$M^{-1/2}v\in\R^d$. Their inner product is $u^Tv$.
\end{proof}

Two other consequences will be used repeatedly. If $M\succeq0$ and
$B\in\R^{d\times r}$, then $B^TMB\succeq0$, because its quadratic
form at $x\in\R^r$ equals $(Bx)^TM(Bx)$. If in addition $B$ is
square and invertible, $B^TMB\succ0$ if and only if $M\succ0$.
This operation is called a congruence. It preserves positivity, though
it generally does not preserve individual eigenvalues.

\begin{lemma}[Block matrices and Schur complements]\label{pre:schur}
Let $P\in\R^{r\times r}$ and $H\in\R^{s\times s}$ be symmetric,
let $B\in\R^{r\times s}$, and suppose $P\succ0$. Define
\[
 \Pi=\begin{pmatrix}P&-B\\-B^T&H\end{pmatrix}
       \in\R^{(r+s)\times(r+s)},\qquad
 K=H-B^TP^{-1}B\in\R^{s\times s}.
\]
Then $\Pi\succ0$ if and only if $K\succ0$. When they are positive
definite, for $u\in\R^r$ and $v\in\R^s$,
\begin{equation}\label{eq:pre-block-inverse}
 \begin{pmatrix}u\\v\end{pmatrix}^{\!T}\Pi^{-1}
 \begin{pmatrix}u\\v\end{pmatrix}
 =u^TP^{-1}u+(v+B^TP^{-1}u)^TK^{-1}(v+B^TP^{-1}u).
\end{equation}
If also $H\succ0$, the upper-left $r\times r$ block of $\Pi^{-1}$
is $(P-BH^{-1}B^T)^{-1}$.
\end{lemma}
\begin{proof}
For $x\in\R^r$ and $t\in\R^s$, completing the square gives
\[
 x^TPx-2x^TBt+t^THt
 =(x-P^{-1}Bt)^TP(x-P^{-1}Bt)+t^TKt.
\]
This proves the positivity criterion: if $K$ has a nonpositive direction,
choose $x=P^{-1}Bt$; if $K\succ0$, the displayed sum is positive
for every nonzero pair $(x,t)$.

Apply \eqref{eq:pre-variational} to $\Pi$. For each fixed $t$, first
maximize in $x$. That maximum is
\[
 (u+Bt)^TP^{-1}(u+Bt)+2v^Tt-t^THt.
\]
Now complete the square in $t$, obtaining
\eqref{eq:pre-block-inverse}. Finally, maximizing in the reverse order
when $v=0$ gives $u^T(P-BH^{-1}B^T)^{-1}u$ for every $u$.
Equality of these quadratic forms identifies the upper-left block.
\end{proof}

\subsubsection{Probability masses, probability densities, and expectations}

A probability mass function on a finite or countable set $\Omega$ is
a function $p:\Omega\to[0,1]$ with $\sum_{x\in\Omega}p(x)=1$.
It assigns probability $p(x)$ to the outcome $x$. If
$f:\Omega\to\R$ is a scalar function with
$\sum_x p(x)\abs{f(x)}<\infty$, its expectation is
\[
 \E f(X)=\sum_{x\in\Omega}p(x)f(x).
\]
For a finite set, $\Unif(\Omega)$ is the law giving each point mass
$1/\abs{\Omega}$. Thus a uniform sign vector
$\sigma\in\{-1,1\}^d\subset\R^d$ gives probability $2^{-d}$
to each of the $2^d$ possible sign columns.

A probability density on $\R^d$ is an integrable function
$p:\R^d\to[0,\infty)$ with
\[
 \int_{\R^d}p(x)\,dx=1.
\]
Here $dx=dx_1\cdots dx_d$ denotes ordinary $d$-dimensional volume.
The probability of a measurable set $A\subseteq\R^d$ is
$\Pp(X\in A)=\int_A p(x)\,dx$. A density value $p(x)$ is not the
probability of the single point $x$; for a continuous law, each single
point has probability zero. In particular, density values may exceed
one. The notation $d\mu(x)=p(x)\,dx$ is another way of specifying
the probability law $\mu$.

For a scalar function $f:\R^d\to\R$ with
$\int \abs{f(x)}p(x)\,dx<\infty$, define
$\E_\mu f(X)=\int f(x)p(x)\,dx$. Vector and matrix expectations
are taken entry by entry. For example, if $X\in\R^d$ has integrable
coordinates, $\E X\in\R^d$. The indicator of a set
$A\subseteq\Omega$ is the function $\one_A:\Omega\to\{0,1\}$
equal to one on $A$ and zero elsewhere. Consequently
$\E\one_A(X)=\Pp(X\in A)$.

Frequently we start from a nonnegative weight $w$ rather than a
normalized law. If $0<Z=\int w(x)\,dx<\infty$, then $w/Z$ is a
density. The scalar $Z$ is its normalizing constant, also called a
partition function. On a finite set the corresponding definition uses
$Z=\sum_x w(x)$. Writing a density as ``proportional to $w$'' always
means division by this positive constant.

We will also use mixed distributions: a sign column
$\sigma\in\{-1,1\}^r$ and a real column $T\in\R^s$ can have a
joint law specified by a nonnegative function
$p:\{-1,1\}^r\times\R^s\to[0,\infty)$ satisfying
\[
 \sum_{\sigma\in\{-1,1\}^r}\int_{\R^s}p(\sigma,t)\,dt=1.
\]
The marginal density of $T$ is $p_T(t)=\sum_\sigma p(\sigma,t)$.
Where $p_T(t)>0$, the conditional mass function of $\sigma$ given
$T=t$ is $p(\sigma\mid t)=p(\sigma,t)/p_T(t)$.
Every conditional expectation used below is a sum or integral against
one of these conditional laws. Independence means that the joint law
is the product of the marginal laws.

\begin{definition}[Variance and covariance]\label{pre:covariance}
For a random column $X\in\R^d$ with finite second moments, put
$m=\E X\in\R^d$ and define
\[
 \Cov(X)=\E[(X-m)(X-m)^T]\in\R^{d\times d}.
\]
For scalar random variables $U,V$ with finite second moments, define
$\Var(U)=\E[(U-\E U)^2]$ and
$\Cov(U,V)=\E[(U-\E U)(V-\E V)]$.
\end{definition}

For every deterministic $u\in\R^d$,
\begin{equation}\label{eq:pre-directional-variance}
 u^T\Cov(X)u=\Var(u^TX)\ge0.
\end{equation}
Thus covariance matrices are positive semidefinite. If $B\in\R^{r\times d}$,
expanding the definition gives $\Cov(BX)=B\Cov(X)B^T\in\R^{r\times r}$.
The elementary Cauchy--Schwarz inequality for random variables,
$(\E UV)^2\le\E U^2\E V^2$, follows by minimizing
$\E(U-tV)^2\ge0$ over real $t$. Applying it after centering gives
$\Cov(U,V)^2\le\Var(U)\Var(V)$.

\begin{lemma}[The law of total variance]\label{pre:total-variance}
If $U$ is a scalar random variable with finite second moment and $T$
is another random variable or vector, then
\begin{equation}\label{eq:pre-total-variance}
 \Var(U)=\E[\Var(U\mid T)]+\Var(\E[U\mid T]).
\end{equation}
For $X\in\R^d$ the analogous matrix identity is
$\Cov(X)=\E[\Cov(X\mid T)]+\Cov(\E[X\mid T])$.
\end{lemma}
\begin{proof}
Let $m(T)=\E[U\mid T]$, a real-valued function of the conditioning
variable. Decompose
$U-\E U=(U-m(T))+(m(T)-\E U)$ and square. The cross term has
expectation zero, since $\E[U-m(T)\mid T]=0$. The two remaining
terms give \eqref{eq:pre-total-variance}. Apply that scalar identity
to $u^TX$ for every $u\in\R^d$ to obtain the matrix identity.
\end{proof}

\subsubsection{Exponential tilts and moment-generating functions}

Let $X$ have a base probability law $\mu$ on a subset of $\R^d$.
Suppose all exponential moments in question are finite. For a column
$h\in\R^d$ define the partition function and its logarithm by
\[
 Z:\R^d\to(0,\infty),\quad Z(h)=\E_\mu e^{h^TX},
 \qquad \psi=\log Z:\R^d\to\R.
\]
The tilted law $\mu_h$ assigns to each outcome its original weight
multiplied by $e^{h^TX}/Z(h)$. This favors outcomes with larger values
of $h^TX$, while keeping the total probability equal to one.

\begin{lemma}[Derivatives of a log partition function]\label{pre:tilt}
If differentiation under the expectation is valid, then
\begin{equation}\label{eq:pre-tilt-derivatives}
 \nabla\psi(h)=\E_{\mu_h}X\in\R^d,\qquad
 \nabla^2\psi(h)=\Cov_{\mu_h}(X)\in\R^{d\times d}.
\end{equation}
Here $\nabla$ denotes the column of first partial derivatives and
$\nabla^2$ denotes the matrix of second partial derivatives.
\end{lemma}
\begin{proof}
The $j$th derivative of $Z$ is $\E[X_j e^{h^TX}]$.
Dividing by $Z$ gives the $j$th coordinate of the first formula.
Differentiating it with respect to $h_i$ gives
$\E_{\mu_h}[X_iX_j]-\E_{\mu_h}X_i\E_{\mu_h}X_j$,
the $(i,j)$ covariance entry. In this paper these differentiations
involve either finite sums, bounded domains, or densities with Gaussian
decay. On a compact set of values of $h$, multiplying the density by
$e^{h^Tx}$ and by a fixed polynomial still leaves an integrable bound;
this justifies differentiation under the integral.
\end{proof}

\begin{lemma}[From covariance to exponential moments and tails]
\label{pre:mgf-tail}
Suppose that for every $h\in\R^d$ the tilted covariance satisfies
$\Cov_{\mu_h}(X)\preceq D$, where $D\in\R^{d\times d}$ is
fixed and positive semidefinite. With $m=\E_\mu X\in\R^d$,
\begin{equation}\label{eq:pre-mgf}
 \E_\mu e^{u^T(X-m)}\le e^{u^TDu/2}
 \qquad(u\in\R^d).
\end{equation}
If $b\in\R^d$, $v=b^TDb>0$, and $t>0$, then
\begin{equation}\label{eq:pre-tail}
 \Pp\bigl(\abs{b^T(X-m)}>t\bigr)
 \le2e^{-t^2/(2v)}.
\end{equation}
\end{lemma}
\begin{proof}
For fixed $u\in\R^d$, set $g(s)=\psi(su)$ for real $s$.
The preceding lemma gives $g''(s)\le u^TDu$.
Twice integrating between $0$ and $1$ yields
$\psi(u)-\psi(0)-u^Tm\le u^TDu/2$. Exponentiating gives
\eqref{eq:pre-mgf}.

For the tail, the elementary Markov bound says that
$\Pp(W\ge a)\le\E W/a$ for $W\ge0$ and $a>0$, because
$W\ge a\one_{\{W\ge a\}}$. Apply it to
$W=e^{s b^T(X-m)}$ with $s>0$. Equation~\eqref{eq:pre-mgf}
gives a bound $e^{-st+s^2v/2}$. Its minimum is at $s=t/v$ and is
$e^{-t^2/(2v)}$. Apply the same argument to $-b$ and add the two
one-sided bounds. If $v=0$, the variance bound already implies
$b^T(X-m)=0$ with probability one, and no division by $v$ is needed.
\end{proof}

A centered random column satisfying \eqref{eq:pre-mgf} is called
sub-Gaussian with covariance proxy $D$. A proxy is an upper bound
that controls exponential moments; it need not equal the actual
covariance matrix. This definition does not assert that the coordinates
are independent.

\subsubsection{Gaussian densities and integration by parts}

\begin{definition}[A Gaussian law]\label{pre:gaussian}
For a mean column $m\in\R^d$ and a symmetric positive-definite
matrix $T\in\R^{d\times d}$, the law $N(m,T)$ has density
\begin{equation}\label{eq:pre-gaussian-density}
 p_{m,T}:\R^d\to(0,\infty),\qquad
 p_{m,T}(x)=\frac{\exp[-(x-m)^TT^{-1}(x-m)/2]}
 {(2\pi)^{d/2}(\det T)^{1/2}}.
\end{equation}
We abbreviate $p_{0,T}$ to $p_T$. In dimension one we also write
$p_v(t)=(2\pi v)^{-1/2}e^{-t^2/(2v)}$ for $v>0$.
\end{definition}

To check normalization, the elementary Gaussian integral is
$\int_\R e^{-t^2/2}\,dt=\sqrt{2\pi}$. One proof squares the
integral and uses polar coordinates in $\R^2$; the result is
$2\pi\int_0^\infty r e^{-r^2/2}\,dr=2\pi$.
The change of variables $x=m+T^{1/2}z$ reduces
\eqref{eq:pre-gaussian-density} to the product of $d$ such normalized
one-dimensional densities. Symmetry gives $\E z_j=0$; integration
by parts gives $\E z_j^2=1$; independence gives
$\E z_i z_j=0$ when $i\ne j$. Therefore a random column
$G\sim N(m,T)$ has $\E G=m$ and $\Cov(G)=T$.

Completing the square in the exponent gives, for $u\in\R^d$,
\begin{equation}\label{eq:pre-gaussian-mgf}
 \E e^{u^TG}=\exp(u^Tm+u^TTu/2).
\end{equation}
Every scalar linear projection $b^T(G-m)$, $b\in\R^d$, has the
one-dimensional law $N(0,b^TTb)$. To see this directly, write it as
$(T^{1/2}b)^TZ$. An orthogonal change of coordinates preserves the
standard Gaussian density, because that density depends only on
$\norm{z}^2$. Choose such a change sending $T^{1/2}b$ to
$\sqrt{b^TTb}\,e_1$, where $e_1\in\R^d$ is the first coordinate
column. The projection is therefore $\sqrt{b^TTb}$ times a standard
normal variable. If its variance is zero, it is the constant zero
random variable.

\begin{lemma}[Gaussian integration by parts]\label{pre:gaussian-ibp}
Let $G\sim N(0,T)$ with $T\in\R^{d\times d}$ positive definite.
For a differentiable function $f:\R^d\to\R$ with at most
polynomial growth, and similarly controlled derivatives,
\begin{equation}\label{eq:pre-gaussian-ibp}
 \E[G_i f(G)]=\sum_{j=1}^d T_{ij}\E[\partial_j f(G)]
 \qquad(1\le i\le d).
\end{equation}
For a twice differentiable such function,
\begin{equation}\label{eq:pre-gaussian-second-ibp}
 \E[G_iG_j f(G)]
 =T_{ij}\E f(G)+\sum_{a,b=1}^d T_{ia}T_{jb}
                                  \E[\partial_a\partial_b f(G)].
\end{equation}
The same identities apply separately to the real and imaginary parts
of a complex-valued function.
\end{lemma}
\begin{proof}
Differentiating the density gives
$\partial_j p_T(x)=-(T^{-1}x)_j p_T(x)$. In vector form this says
$x p_T(x)=-T\nabla p_T(x)$. Multiply its $i$th coordinate by
$f(x)$ and integrate by parts in each coordinate. Boundary terms
vanish because Gaussian decay dominates the stated polynomial growth.
This gives \eqref{eq:pre-gaussian-ibp}. Apply that identity first
to $x\mapsto x_j f(x)$, using
$\partial_a(x_j f)=\one_{\{a=j\}}f+x_j\partial_a f$, and then
apply it again to the remaining term. This gives
\eqref{eq:pre-gaussian-second-ibp} with the displayed indices.
\end{proof}

\begin{lemma}[The Gaussian characteristic function]\label{pre:fourier}
For $G\sim N(0,T)$ and a deterministic column $u\in\R^d$,
\begin{equation}\label{eq:pre-gaussian-fourier}
 \E e^{iu^TG}=e^{-u^TTu/2}.
\end{equation}
\end{lemma}
\begin{proof}
First let $Z\sim N(0,1)$ and define $F:\R\to\mathbb C$ by
$F(t)=\E e^{itZ}$. Differentiation and one-dimensional integration
by parts give $F'(t)=i\E[Ze^{itZ}]=-tF(t)$. Since $F(0)=1$,
solving this differential equation gives $F(t)=e^{-t^2/2}$.
For the general case write $G=T^{1/2}Z$, where now
$Z\in\R^d$ has independent standard normal coordinates. Then
$u^TG=(T^{1/2}u)^TZ$. Independence multiplies the one-dimensional
formulas, giving $\exp(-\norm{T^{1/2}u}^2/2)$, as claimed.
This argument uses real integrations only.
\end{proof}

\subsubsection{Entropy, decoding, and a Gaussian information bound}

The lattice argument later uses entropy to count how many cell labels
can be distinguished after adding Gaussian noise. The following
definitions and identities are standard in information theory; see,
for example, Chapter~2 of Cover and Thomas~\cite{CoverThomas2006}.
We give the needed derivations, including the case of a finite label
observed through a continuous vector. No differential entropy is needed.

If a random variable $K$ takes values in a finite set $\mathcal K$
with masses $p_k$, its entropy in natural-logarithm units, or nats, is
\[
 H(K)=-\sum_{k\in\mathcal K}p_k\log p_k,
\]
where $0\log0$ is defined as zero. This is the expected value of the
nonnegative surprise $-\log p_K$. A distribution concentrated at one
point has entropy zero; the uniform distribution on $N$ points has
entropy $\log N$.

For mass functions $p,q:\mathcal K\to[0,1]$, with $q_k>0$ whenever
$p_k>0$, define the relative entropy
\[
 D(p\|q)=\sum_k p_k\log\frac{p_k}{q_k}\in[0,\infty].
\]
For densities $p,q:\R^r\to[0,\infty)$, use instead
$D(p\|q)=\int p(v)\log(p(v)/q(v))\,dv$, with the same convention
on zero terms. If the set where $q=0$ has positive probability under
$p$, the value is $+\infty$. A discrepancy on a set of probability zero
does not change an integral. Nonnegativity follows from
$\log t\le t-1$ for $t>0$:
\[
 -D(p\|q)=\E_p\log\frac q p
 \le\E_p\left(\frac q p-1\right)\le0.
\]
The last inequality allows $q$ to put some mass outside the support
of $p$; the sum or integral of $q$ on that support is at most one.

Suppose that, conditional on $K=k$, an observation $V\in\R^r$
has density $p_k^V:\R^r\to[0,\infty)$. Its marginal density is
$p_V(v)=\sum_k p_k p_k^V(v)$. Where $p_V(v)>0$, the posterior
probabilities are
\[
 r_k(v)=\Pp(K=k\mid V=v)=\frac{p_k p_k^V(v)}{p_V(v)}.
\]
For each fixed $k$, $r_k$ is a function on the set where $p_V>0$
with values in $[0,1]$. More generally, for any observation $W$, whether
continuous, discrete, or constant, define
\[
 H(K\mid W)=\E\left[-\sum_k\Pp(K=k\mid W)
                                  \log\Pp(K=k\mid W)\right].
\]
This is the average entropy of the conditional mass function of $K$.
For the observation $V$ with density above, it becomes
\[
 H(K\mid V)=\int p_V(v)
       \left[-\sum_k r_k(v)\log r_k(v)\right]\,dv.
\]
The mutual information is the scalar
$I(K;V)=H(K)-H(K\mid V)$. Direct substitution of the posterior
formula gives
\begin{equation}\label{eq:pre-mutual-information}
 I(K;V)=\sum_k p_k\int p_k^V(v)
                      \log\frac{p_k^V(v)}{p_V(v)}\,dv\ge0.
\end{equation}
The nonnegativity is a sum of relative entropies.

\begin{proposition}[The entropy facts used in the proof]\label{pre:entropy}
The following statements hold whenever the displayed quantities are
defined and the finite labels have the indicated conditional laws.
\begin{enumerate}
 \item For any reference density $q:\R^r\to[0,\infty)$,
 \begin{equation}\label{eq:pre-information-reference}
 \sum_k p_k D(p_k^V\|q)
   = I(K;V)+D(p_V\|q)\ge I(K;V).
 \end{equation}
 \item If $K=(K_1,\ldots,K_d)$ is a finite-valued vector, $V\in\R^r$
 is an observation, and $W_j=g_j(V)$ for functions
 $g_j:\R^r\to\R$, then
 \begin{equation}\label{eq:pre-conditional-subadditivity}
 H(K\mid V)\le\sum_{j=1}^d H(K_j\mid V)
              \le\sum_{j=1}^d H(K_j\mid W_j).
 \end{equation}
 \item If $q(j\mid w)$ is any probability distribution on a finite
 or countable label set for each observation $w$, then
 \begin{equation}\label{eq:pre-cross-entropy}
 H(K_j\mid W_j)\le\E[-\log q(K_j\mid W_j)].
 \end{equation}
 The upper bound is allowed to be infinite.
 \item If $\beta>0$, $z_k\in\R^r$ are deterministic columns, and
 $V=z_K+Z$ for independent $Z\sim N(0,I_r/\beta)$, then
 \begin{equation}\label{eq:pre-gaussian-information}
 I(K;V)\le\frac\beta2\E\norm{z_K}^2.
 \end{equation}
 \item If $e_k\in\R$ are finitely many energies and
 $p_k=e^{-\beta e_k}/\sum_j e^{-\beta e_j}$, then
 \begin{equation}\label{eq:pre-gibbs-entropy}
 \log\sum_k e^{-\beta e_k}=H(K)-\beta\E e_K.
 \end{equation}
\end{enumerate}
\end{proposition}
\begin{proof}
For the first identity split
$\log(p_k^V/q)=\log(p_k^V/p_V)+\log(p_V/q)$, integrate, and
sum over $k$. The first part is
\eqref{eq:pre-mutual-information}; in the second part the mixture
identity $\sum_k p_kp_k^V=p_V$ gives $D(p_V\|q)$.

For subadditivity, fix a value $v$ of the observation and denote the
posterior mass function of the entire label by $r(k\mid v)$.
Let $r_j(k_j\mid v)$ be its coordinate marginals. Nonnegativity of
\[
 D\left(r(\cdot\mid v)\,\middle\|\,
              \prod_{j=1}^d r_j(\cdot\mid v)\right)
\]
is precisely the assertion that the entropy of the joint posterior is
at most the sum of the entropies of its marginals. Averaging in $v$
proves the first inequality of
\eqref{eq:pre-conditional-subadditivity}. The product of the marginals
is a comparison distribution; the actual coordinates need not be
independent.

For the second inequality, consider one label $K_j$ and put
$W=g_j(V)$. Given $V$, the value of $W$ is already known. The
posterior law of $K_j$ given $W$ is the conditional average, over
values of $V$ producing that $W$, of its posterior law given $V$.
Expanding relative entropy therefore gives
\[
 H(K_j\mid W)-H(K_j\mid V)
 =\E_V D\bigl(\mathcal L(K_j\mid V)\,
                     \|\,\mathcal L(K_j\mid W)\bigr)\ge0.
\]
Here $\mathcal L$ means the probability law, and each relative entropy
is between two mass functions on the same finite label set. This also
shows directly why discarding part of an observation cannot improve
the information about a label.

For the cross-entropy bound, at each observation $w$ let $r(\cdot\mid w)$
be the true posterior. The identity
\[
 \sum_j r(j\mid w)[-\log q(j\mid w)]
 =-\sum_j r(j\mid w)\log r(j\mid w)
   +D(r(\cdot\mid w)\|q(\cdot\mid w))
\]
and nonnegativity of the last term prove
\eqref{eq:pre-cross-entropy} after averaging. A countable comparison
distribution is permitted even when the true label set is finite.

For the Gaussian bound choose $q$ to be the density of
$N(0,I_r/\beta)$. The conditional density is that of
$N(z_k,I_r/\beta)$. Their log ratio at $v\in\R^r$ is
\[
 \log\frac{p_k^V(v)}{q(v)}
 =\beta z_k^Tv-\frac\beta2\norm{z_k}^2.
\]
Its expectation under the conditional law, whose mean is $z_k$, is
$\beta\norm{z_k}^2/2$. Substitute this into
\eqref{eq:pre-information-reference}. Finally, taking the logarithm
of the mass function in the last assertion gives
$-\log p_k=\beta e_k+\log\sum_j e^{-\beta e_j}$.
Average this identity to obtain \eqref{eq:pre-gibbs-entropy}.
\end{proof}

\subsubsection{Convexity and the Brascamp--Lieb variance inequality}

We write $C^1$ for continuously differentiable functions and $C^2$ for
twice continuously differentiable functions. A set $K\subseteq\R^d$ is convex if
$(1-t)x+ty\in K$ for $x,y\in K$ and $0\le t\le1$.
A function $V:K\to\R$ is convex if
$V((1-t)x+ty)\le(1-t)V(x)+tV(y)$ under the same conditions.
For a twice continuously differentiable function, convexity is
equivalent to positive semidefiniteness of its Hessian. If the stronger
inequality $\nabla^2V(x)\succeq H\succ0$ holds with a fixed
$d\times d$ matrix $H$, then integrating the second derivative along
the segment from $x$ to $y$ gives
\begin{equation}\label{eq:pre-strong-convexity}
 V(y)\ge V(x)+\nabla V(x)^T(y-x)
                    +\tfrac12(y-x)^TH(y-x).
\end{equation}
Such a potential is called uniformly convex. On all of $\R^d$,
the bound shows that $e^{-V}$ is integrable and decays at least as a
Gaussian after a translation. In particular, multiplying it by any
fixed polynomial or by $e^{h^Tx}$ leaves an integrable function.

A continuous function $F:\R^d\to\R$ is coercive if
$F(x)\to+\infty$ as $\norm{x}\to\infty$. A coercive continuous
function attains its minimum on every nonempty closed set: a minimizing
sequence is bounded, has a convergent subsequence in finite dimensions,
and its limit remains in the set. We will use this elementary fact to
justify the minimizers in the lattice energy.

The next theorem is the variance form of the Brascamp--Lieb inequality,
originally proved in~\cite[Theorem~4.1]{BL1976}. It is stated as
equation~(1.3) in~\cite{CCL2013}, whose appendix also discusses the
inductive proof. We include the proof because this inequality connects
the curvature of a density to the fluctuations of a random vector.

\begin{theorem}[Brascamp--Lieb variance inequality]\label{pre:BL}
Let $\Omega$ be either $\R^d$ or a closed rectangular box
$\prod_{j=1}^d[a_j,b_j]$ with $a_j<b_j$. Let
$V:\Omega\to\R$ be twice continuously differentiable, with an
extension to a neighborhood in the closed-box case. Suppose
$\nabla^2V(x)\succeq H\succ0$ for a fixed symmetric
$H\in\R^{d\times d}$. Define
\[
 Z_V=\int_\Omega e^{-V(x)}\,dx,\qquad
 p(x)=\begin{cases}Z_V^{-1}e^{-V(x)},&x\in\Omega,\\
                    0,&x\notin\Omega.\end{cases}
\]
Then $0<Z_V<\infty$, and $p:\R^d\to[0,\infty)$ is a density.
For a continuously differentiable function $f:\Omega\to\R$
with the indicated finite second moments and integrals,
\begin{equation}\label{eq:pre-BL}
 \Var_p(f(X))
 \le\E_p\!\left[\nabla f(X)^T(\nabla^2V(X))^{-1}\nabla f(X)\right]
 \le\E_p[\nabla f(X)^TH^{-1}\nabla f(X)].
\end{equation}
The gradient is a column in $\R^d$, and both matrices inside these
quadratic forms are $d\times d$.
\end{theorem}

\begin{proof}
We first prove the inequality on compact boxes, where every
differentiation and integration below concerns continuous functions
on a bounded set. We then pass to the whole space by taking larger
boxes. The second inequality in \eqref{eq:pre-BL} follows at once
from inverse order, so it is enough to prove the first.

\paragraph{Step 1: one dimension.}
Let the interval be $[a,b]$, let $p=e^{-V}/Z_V$, and let
$f_0=f-\E_p f$. Define
\[
 U:[a,b]\to\R,\qquad
 U(t)=\frac1{p(t)}\int_a^t f_0(s)p(s)\,ds.
\]
The centering of $f_0$ implies $U(a)=U(b)=0$.
Differentiating $U(t)p(t)$ gives
$f_0=U'-V'U$. Therefore integration by parts gives
\begin{equation}\label{eq:pre-BL-1d-first}
 A:=\E_p f_0^2
 =\int_a^b f_0(t)(U(t)p(t))'\,dt
 =-\int_a^b f'(t)U(t)p(t)\,dt.
\end{equation}
There is no boundary term, since $U$ vanishes at both endpoints.
Expanding the square in a second expression for $A$ gives
\begin{align*}
 A&=\int_a^b (U'-V'U)^2p\,dt\\
  &=\int_a^b (U')^2p\,dt
       -\int_a^b V'(U^2)'p\,dt
       +\int_a^b(V')^2U^2p\,dt\\
  &=\int_a^b (U')^2p\,dt+\int_a^b V''U^2p\,dt.
\end{align*}
For the last equality, integrate the middle term by parts and use
$p'=-V'p$. Its contribution is
$\int(V''-(V')^2)U^2p$, canceling the final $(V')^2$ term.
Thus $\E_p[V''U^2]\le A$. Weighted Cauchy--Schwarz in
\eqref{eq:pre-BL-1d-first} now yields
\[
 A^2\le\E_p\left[\frac{(f')^2}{V''}\right]
               \E_p[V''U^2]
       \le A\E_p\left[\frac{(f')^2}{V''}\right].
\]
If $A=0$ the assertion is immediate; otherwise divide by $A$.
This proves the one-dimensional inequality on a closed interval.

\paragraph{Step 2: separate one coordinate from the others.}
Suppose the result has been proved in dimension $d-1$, and write
$x=(y,z)\in\R\times\R^{d-1}$ for a point of a $d$-dimensional
box. For each $y$, let $\E_y$ denote expectation over the conditional
density of $z$ proportional to $e^{-V(y,z)}$. Define the marginal
potential and the conditional mean by
\[
 W(y)=-\log\int e^{-V(y,z)}\,dz,
 \qquad g(y)=\E_y f(y,z).
\]
Both are real-valued functions on the interval of possible $y$ values.
The marginal density of $y$ is proportional to $e^{-W(y)}$.
Differentiating a quotient of integrals gives
\begin{equation}\label{eq:pre-BL-marginal-derivatives}
 W''(y)=\E_y V_{yy}-\Var_y(V_y),
 \qquad g'(y)=\E_y f_y-\Cov_y(f,V_y).
\end{equation}
Subscripts $y$ on $V$ and $f$ mean derivatives; subscripts on
$\E_y$, $\Var_y$, and $\Cov_y$ mean conditional averages over $z$.

At a point $(y,z)$, write the Hessian blocks as
\[
 \nabla^2V=
 \begin{pmatrix}V_{yy}&b^T\\b&C\end{pmatrix},\quad
 b=\nabla_z V_y\in\R^{d-1},\quad
 C=\nabla_z^2V\in\R^{(d-1)\times(d-1)}.
\]
The matrix $C$ is positive definite. Define its scalar Schur complement
$s=V_{yy}-b^TC^{-1}b>0$.
The inductive hypothesis applied to the scalar function $z\mapsto V_y(y,z)$
shows that
$\Var_y(V_y)\le\E_y[b^TC^{-1}b]$.
Consequently \eqref{eq:pre-BL-marginal-derivatives} gives
$W''(y)\ge\E_y s>0$.
All these functions are continuous on the compact box, so this is a
strictly positive lower bound on the entire $y$ interval. Thus the
one-dimensional result applies to the marginal density and the function
$g$.

The law of total variance and that one-dimensional result imply
\begin{equation}\label{eq:pre-BL-conditional-bound}
 \Var(f)\le\E\left[
   \Var_y(f)+
   \frac{(\E_y f_y-\Cov_y(f,V_y))^2}
        {\E_y V_{yy}-\Var_y(V_y)}\right],
\end{equation}
where the outside expectation is over $y$.

\paragraph{Step 3: keep the covariance terms together.}
Fix $y$ for this step. Put
$A_0=\Var_y(f)$, $B_0=\Cov_y(f,V_y)$,
$C_0=\Var_y(V_y)$, $r=\E_y f_y$, and $q=\E_y V_{yy}$;
these are real scalars. Because $q-C_0=W''(y)>0$,
completing a scalar square gives
\begin{align*}
 A_0+\frac{(r-B_0)^2}{q-C_0}
 &=\sup_{t\in\R}
       \{\Var_y(f-tV_y)+2tr-t^2q\}.
\end{align*}
This form is useful because the inductive inequality can be applied to
the whole function $f-tV_y$, for every fixed real number $t$.
Let $p_z=\nabla_z f\in\R^{d-1}$ and define the scalar function
$c=f_y-p_z^TC^{-1}b$. The inductive inequality gives
\begin{align*}
 \Var_y(f-tV_y)+2tr-t^2q
 &\le\E_y[(p_z-tb)^TC^{-1}(p_z-tb)]+2tr-t^2q\\
 &=\E_y[p_z^TC^{-1}p_z]+2t\E_y c-t^2\E_y s.
\end{align*}
Taking the supremum over $t$ and completing the square once more yields
\[
 A_0+\frac{(r-B_0)^2}{q-C_0}
 \le\E_y[p_z^TC^{-1}p_z]+
                    \frac{(\E_y c)^2}{\E_y s}.
\]
Weighted Cauchy--Schwarz gives
$(\E_y c)^2\le\E_y(c^2/s)\E_y s$.
The last display is therefore at most
\[
 \E_y\left[p_z^TC^{-1}p_z+\frac{c^2}{s}\right]
 =\E_y[\nabla f^T(\nabla^2V)^{-1}\nabla f].
\]
The final equality is the block inverse formula of
Lemma~\ref{pre:schur}, with the coordinate order reversed if necessary.
Substitute this estimate into \eqref{eq:pre-BL-conditional-bound}
and average over $y$. This proves the induction step and hence the
inequality on every compact rectangular box.

\paragraph{Step 4: pass to the whole space.}
For a potential on $\R^d$, apply the box result to
$[-R,R]^d$ and let $R\to\infty$. The normalizing constants converge
to $Z_V$ by monotone convergence of $e^{-V}\one_{[-R,R]^d}$.
The first and second moments of $f$ converge by the assumed
integrability; the first moment is integrable by Cauchy--Schwarz.
The nonnegative integral on the right of the first inequality in
\eqref{eq:pre-BL} also converges, by monotone convergence of its
restriction to the boxes. Divide by the converging positive normalizers.
This proves the result on $\R^d$ and completes the proof.
\end{proof}

\begin{corollary}[Open boxes and endpoint limits]\label{pre:box-BL}
The same inequality holds on an open rectangular box
$\Omega=\prod_{j=1}^d(a_j,b_j)$ when $V:\Omega\to\R$ is $C^2$,
$\nabla^2V\succeq H\succ0$, and the density and the displayed
integrals are well defined. The endpoints may be infinite. The
potential may tend to $+\infty$ at finite endpoints.
\end{corollary}
\begin{proof}
Exhaust $\Omega$ by closed boxes lying strictly inside it and apply
Theorem~\ref{pre:BL} to the normalized restrictions. Their indicators
increase to $\one_\Omega$. Exactly the convergence argument in
Step~4 passes both sides to the limit. This avoids integration by parts
at a singular endpoint. The same proof applies after adding any linear
term to the potential, because a linear term does not change its Hessian.
\end{proof}

\begin{corollary}[Curvature controls covariance and exponential moments]
\label{pre:BL-mgf}
In the setting of Theorem~\ref{pre:BL} or
Corollary~\ref{pre:box-BL}, let $X\in\R^d$ have the stated density,
and suppose the law is tilted by any $h^TX$, $h\in\R^d$.
Then
\[
 \Cov_h(X)\preceq H^{-1}.
\]
Consequently, for $m=\E X\in\R^d$ and $u\in\R^d$,
\[
 \E e^{u^T(X-m)}\le e^{u^TH^{-1}u/2}.
\]
If the domain is invariant under $x\mapsto-x$ and $V(-x)=V(x)$,
then $m=0$.
\end{corollary}
\begin{proof}
The tilted potential is $V(x)-h^Tx$, with the same Hessian as $V$.
Apply the variance inequality to $f(x)=u^Tx$, whose gradient is the
constant column $u$. This gives
$u^T\Cov_h(X)u\le u^TH^{-1}u$ for every $u$, which is the
matrix inequality. Lemma~\ref{pre:mgf-tail} then proves the
exponential-moment bound. Symmetry makes every coordinate of the mean
the integral of an odd function, hence zero.
\end{proof}

\begin{corollary}[A fourth-moment bound]\label{pre:fourth}
Let $X\in\R^d$ have a centered law covered by
Corollary~\ref{pre:BL-mgf}, and put $C=H^{-1}\in\R^{d\times d}$.
For every $v\in\R^d$,
\begin{equation}\label{eq:pre-fourth}
 \E(v^TX)^2\le v^TCv,
 \qquad \E(v^TX)^4\le5(v^TCv)^2.
\end{equation}
\end{corollary}
\begin{proof}
The second-moment bound is the covariance inequality. For the fourth
moment apply \eqref{eq:pre-BL} to $f(x)=(v^Tx)^2$, whose gradient
is $2(v^Tx)v\in\R^d$. If $s=\E(v^TX)^2$, the result is
\[
 \E(v^TX)^4-s^2=\Var((v^TX)^2)
 \le4(v^TCv)s.
\]
Use $s\le v^TCv$ to obtain the claimed constant $5$.
The required moments are finite by Gaussian decay, or by boundedness
of the box. Limits from inner boxes handle the open-box formulation.
\end{proof}

\subsubsection{An upper curvature bound gives a lower covariance bound}

The Brascamp--Lieb inequality bounds covariance from above. At one
point we also need a lower bound. The following integration-by-parts
argument, often called a Cram\'er--Rao covariance comparison, proves
exactly the version we use.

\begin{lemma}\label{pre:cramer-rao}
Let $V:\R^d\to\R$ be $C^2$ and suppose
\[
 0\prec H\preceq\nabla^2V(x)\preceq K
 \qquad(x\in\R^d),
\]
where $H,K\in\R^{d\times d}$ are fixed symmetric positive-definite
matrices. If $X\in\R^d$ has density proportional to $e^{-V}$, then
\begin{equation}\label{eq:pre-cramer-rao}
 \Cov(X)\succeq K^{-1}.
\end{equation}
\end{lemma}
\begin{proof}
Let $m=\E X\in\R^d$ and introduce the score column
$S=\nabla V(X)\in\R^d$. The lower Hessian bound gives Gaussian
decay by \eqref{eq:pre-strong-convexity}; the upper bound implies
that $\nabla V$ grows at most linearly. Thus the following integrations
by parts have vanishing boundary terms. Integrating a first derivative
of the density gives $\E S=0$. Integrating the derivative of
$(x_i-m_i)e^{-V(x)}$ in coordinate $j$ gives
\[
 \E[(X-m)S^T]=I_d.
\]
Integrating the derivative of $(\partial_iV)e^{-V}$ in coordinate
$j$ gives
\[
 J:=\E[SS^T]=\E[\nabla^2V(X)]\preceq K,
 \qquad J\in\R^{d\times d}.
\]
For arbitrary deterministic $u,v\in\R^d$, expand the nonnegative
square $\E[(u^T(X-m)-v^TS)^2]$ to obtain
\[
 u^T\Cov(X)u\ge2u^Tv-v^TJv.
\]
Choose $v=K^{-1}u$. Since $J\preceq K$, the right side is at least
$2u^TK^{-1}u-u^TK^{-1}KK^{-1}u=u^TK^{-1}u$.
This holds in every direction $u$, proving
\eqref{eq:pre-cramer-rao}.
\end{proof}

The two curvature comparisons have complementary uses. A potential
whose curvature is bounded below confines its distribution and gives
an upper bound on fluctuations. A potential whose curvature is also
bounded above cannot concentrate arbitrarily tightly, and gives the
lower covariance bound just proved. Both statements compare quadratic
forms in all directions, which is stronger than controlling each
coordinate variance separately.

\subsection{A convex-support estimate and a lattice energy bound}
\label{sec:lattice}

The analytic argument must eventually compare the central interval of each
cosine with all of its other periods. In one dimension, a small-variance
Gaussian almost never reaches another period. In many dimensions, however,
the coordinates can be strongly correlated, and there are infinitely many
periodic cells to consider. We need an estimate that controls their
\emph{total} contribution without assuming independent coordinates.

This section develops that estimate in three steps. First, a bound on all
linear exponential moments gives a bound on the mass of a convex set.
Second, a scalar noisy observation can reveal its integer period with small
expected logarithmic loss. Third, an entropy argument converts this decoding
guarantee into a sum over all periodic cells. The entropy argument is the
reason correlations between coordinates cause no problem.

\subsubsection{Keeping the support when using an exponential-moment bound}

A centered Gaussian vector with covariance matrix $D$ has exponential
moments $\exp(u^TDu/2)$. A different distribution may satisfy the same upper
bound, even if it is discrete or supported on a lower-dimensional set.
The next lemma uses only that inequality. It also preserves the restriction
that the random vector lies in a specified convex set, which will be crucial
when the set is a distant periodic cell.

\begin{lemma}[An exponential-moment bound with a convex support restriction]
\label{lem:support}
Let $d\ge1$, let $X$ be an $\R^d$-valued random vector, and let
$D\in\R^{d\times d}$ be symmetric and positive definite. Suppose
\[
 \E e^{u^TX}\le e^{u^TDu/2}\qquad\text{for every }u\in\R^d.
\]
For every nonempty closed convex set $K\subseteq\R^d$ and every vector
$h\in\R^d$,
\begin{equation}\label{eq:support-expanded}
 \E\big[e^{h^TX}\one_{\{X\in K\}}\big]
 \le\exp\left\{\sup_{x\in K}
             \left(h^Tx-\tfrac12x^TD^{-1}x\right)\right\}.
\end{equation}
\end{lemma}

\begin{proof}
The function $x\mapsto h^Tx-x^TD^{-1}x/2$, from $\R^d$ to $\R$,
is strictly concave and tends to $-\infty$ as $\norm{x}\to\infty$.
It therefore attains its maximum on $K$ at a unique point $x_*\in K$.
For any $x\in K$, the segment $x_*+s(x-x_*)$, $0\le s\le1$, lies in
$K$. Differentiating along this segment at its maximizing endpoint gives
\[
 (h-D^{-1}x_*)^T(x-x_*)\le0.
\]
Set $u=D^{-1}x_*\in\R^d$. Rearranging the preceding inequality shows
that, for $x\in K$,
\[
 h^Tx\le(h-u)^Tx_*+u^Tx.
\]
Exponentiation, followed by expectation, now gives
\begin{align*}
 \E\big[e^{h^TX}\one_{\{X\in K\}}\big]
 &\le e^{(h-u)^Tx_*}\E e^{u^TX}\\
 &\le e^{(h-u)^Tx_*+u^TDu/2}
  =e^{h^Tx_*-x_*^TD^{-1}x_*/2}.
\end{align*}
This is the claimed supremum. The segment argument never uses an interior
point of $K$, so the proof also covers convex sets of smaller dimension.
\end{proof}

For example, when $h=0$, the lemma says that the probability of $K$ is at
most the exponential of minus its least quadratic energy
$x^TD^{-1}x/2$. The general source $h$ will allow us to compare a shifted
cell to a central reference distribution.

\subsubsection{A scalar Gaussian tail estimate}

Fix throughout the remaining proof
\[
 \beta=\frac9{10},\qquad \kappa_0=\frac34.
\]
We first record the elementary tail calculation used by the decoder.
The logarithm in this calculation, as everywhere else, is natural.

\begin{claim}[A Gaussian tail with a quadratic cost]
\label{cl:decoder-gaussian-tail}
If $Z$ is a real standard normal random variable, then for every real
$t\ge2$,
\begin{equation}\label{eq:decoder-gaussian-tail-expanded}
 \E\big[(Z^2+\log3)\one_{\{|Z|>t\}}\big]
 \le8e^{-t^2/4}.
\end{equation}
\end{claim}

\begin{proof}
Write $\varphi(z)=(2\pi)^{-1/2}e^{-z^2/2}$ for the density
$\varphi:\R\to(0,\infty)$. Since $\varphi'(z)=-z\varphi(z)$,
integration by parts gives
\[
 \int_t^\infty z^2\varphi(z)\,dz
 =t\varphi(t)+\int_t^\infty\varphi(z)\,dz.
\]
Also, for $t>0$,
\[
 \int_t^\infty\varphi(z)\,dz
 \le\frac1t\int_t^\infty z\varphi(z)\,dz
 =\frac{\varphi(t)}t.
\]
By symmetry, the left side of
\eqref{eq:decoder-gaussian-tail-expanded} is therefore at most
\[
 2\varphi(t)\left(t+\frac{1+\log3}{t}\right)
 \le4t\varphi(t).
\]
The last inequality uses $1+\log3<3\le t^2$. The function
$t\mapsto te^{-t^2/4}$ decreases for $t\ge2$ and has value $2/e<1$
at $t=2$. Consequently
\[
 4t\varphi(t)
 =\frac4{\sqrt{2\pi}}\big(te^{-t^2/4}\big)e^{-t^2/4}
 \le8e^{-t^2/4},
\]
as required. The generous constant $8$ leaves ample room in the later
numerical estimates.
\end{proof}

\subsubsection{A decoder that tolerates offsets within a period}

Suppose a real observation has the form $\pi k+x+N$, where $k\in\Z$
is a period index, $x\in[-\pi/2,\pi/2]$ is an unknown offset, and $N$
is small Gaussian noise. A deterministic nearest-period rule is unstable
when $x$ is close to $\pm\pi/2$. We instead assign a probability to
\emph{every} integer. Near a cell boundary, some ambiguity is acceptable:
the decoder's permitted loss includes a constant times $x^2$.

\begin{lemma}[A padded Gaussian decoder]\label{lem:decoder}
For each real $v$ with $0<v\le10^{-6}$, there is a function
$q_v:\Z\times\R\to(0,1]$, written $q_v(j\mid w)$, such that
$\sum_{j\in\Z}q_v(j\mid w)=1$ for every $w\in\R$ and
\begin{equation}\label{eq:decoder-expanded}
 \E[-\log q_v(k\mid\pi k+x+N)]
 \le\frac{\beta\kappa_0}{2}x^2+16e^{-10^{-5}/v}.
\end{equation}
Here $k\in\Z$ and $x\in[-\pi/2,\pi/2]$ are arbitrary fixed values,
and $N$ is any centered real Gaussian random variable of variance at most
$v$, including variance zero.
\end{lemma}

\begin{proof}
Define the distance function $d_0:\R\to[0,\infty)$ and the normalizing
function $D_v:\R\to(0,\infty)$ by
\[
 d_0(t)=\dist(t,[-1.6,1.6])=(|t|-1.6)_+,
 \qquad
 D_v(w)=\sum_{j\in\Z}e^{-d_0(w-\pi j)^2/v},
\]
where $(s)_+=\max\{s,0\}$ for a real number $s$. Set
\[
 q_v(j\mid w)=\frac{e^{-d_0(w-\pi j)^2/v}}{D_v(w)}.
\]
The padding interval has half-width $1.6$, slightly larger than $\pi/2$.
Thus the correct period receives weight one for every noiseless offset,
including a boundary offset.

\paragraph{Step 1: bound the denominator for every observation.}
The function $D_v$ is even and $\pi$-periodic. It is enough to consider
$0\le w\le\pi/2$. The term $j=0$ is one, and the term $j=1$ is at
most one. For $j=-1,-2,\ldots$, the distances to their padding intervals
are at least $1.5,1.5+\pi,\ldots$. The same lower bounds hold for
$j=2,3,\ldots$. Therefore
\begin{align*}
 1\le D_v(w)
 &\le2+2\sum_{\ell=0}^\infty e^{-(1.5+\ell\pi)^2/v}\\
 &\le2+2e^{-2.25/v}\sum_{\ell=0}^\infty e^{-\ell/v}
 \le2+4e^{-2.25/v}<3.
\end{align*}
For the middle inequality, $(1.5+\ell\pi)^2-2.25\ge\ell$.
For the next one, $e^{-1/v}<1/2$, so the geometric series is at most
$2$. This also proves convergence of $D_v$ and validates the probability
distribution just defined.

Translation by $\pi k$ reduces the rest of the proof to $k=0$.
For an observation $x+N$, its logarithmic loss is exactly
\begin{equation}\label{eq:decoder-loss-expanded}
 -\log q_v(0\mid x+N)
 =\frac{d_0(x+N)^2}{v}+\log D_v(x+N).
\end{equation}
Write $N=\sqrt{w_0}\,Z$, where $0\le w_0\le v$ and $Z$ is standard
normal. Since $x\in[-\pi/2,\pi/2]\subset[-1.6,1.6]$, distance to the
padding interval is at most $|N|$. Hence the loss is always at most
$Z^2+\log3$.

\paragraph{Step 2: offsets with $|x|\le1.5$.}
On the event $|N|\le0.01$, we have $|x+N|\le1.51$, so the true weight
is one. For each wrong index $j\ne0$,
\[
 d_0(x+N-\pi j)\ge\pi|j|-1.51-1.6
 >0.03+(|j|-1)\pi.
\]
Applying the same geometric-series estimate to the wrong weights gives
\[
 D_v(x+N)-1\le4e^{-0.0009/v},
 \qquad \log D_v(x+N)\le4e^{-0.0009/v}.
\]
Indeed $(0.03+\ell\pi)^2-0.0009\ge\ell$ for every integer
$\ell\ge0$. The first term of \eqref{eq:decoder-loss-expanded} vanishes
on this event. On its complement,
$|Z|>0.01/\sqrt v\ge10$, so Claim~\ref{cl:decoder-gaussian-tail}
bounds the expected loss there by $8e^{-0.000025/v}$. Thus the full
expected loss is at most
\[
 4e^{-0.0009/v}+8e^{-0.000025/v}
 \le16e^{-0.00001/v}.
\]
This is stronger than the desired bound for these offsets.

\paragraph{Step 3: the remaining offsets.}
For every $|x|\le\pi/2$, the true weight remains one on
$|N|\le0.029$, because $1.6-\pi/2>0.029$. On that event the denominator
bound gives loss at most
$\log2+2e^{-2.25/v}$, using $\log(1+s)\le s$ for $s\ge0$.
The complementary expected loss is at most
$8e^{-0.00021025/v}$ by the Gaussian-tail claim. Consequently the total
expected loss is bounded by
\[
 \log2+2e^{-2.25/v}+8e^{-0.00021025/v}<0.7.
\]
For completeness, when $v\le10^{-6}$ both exponential arguments exceed
$11$ in absolute value; the two exponential terms sum to less than
$10\cdot2^{-11}<0.005$, while $\log2<0.694$.
If $|x|>1.5$, the allowed quadratic cost is larger than
\[
 \frac{\beta\kappa_0}{2}(1.5)^2=0.759375>0.7.
\]
Combining Steps 2 and 3 proves \eqref{eq:decoder-expanded}.
\end{proof}

\subsubsection{From decoding to a sum over all cells}

The next lemma considers $d$ linear observations of one latent vector in
$\R^r$. The integers recording their periods can be dependent in any way.
Its conclusion says that the number of possible cells, weighted by the
energy needed to reach them, is small. The proof adds Gaussian noise to
the latent vector and asks how much information its noisy version can
reveal about the cell index.

\begin{lemma}[A lattice energy sum]\label{lem:lattice}
Let $r,d\ge1$ be integers. Let $b_1,\ldots,b_d\in\R^r$ satisfy
$\norm{b_i}^2\le q$ for a real number $q>0$, and let $\kappa\ge3/4$.
Define $Q_d=[-\pi/2,\pi/2]^d\subset\R^d$ and the function
$E:\Z^d\to[0,\infty]$ by
\[
 E(k)=E_k:=
 \inf_{\substack{z\in\R^r,\ x\in Q_d\\
                  b_i^Tz=\pi k_i+x_i\ (1\le i\le d)}}
       \frac12\bigl(\norm z^2+\kappa\norm x^2\bigr).
\]
An infeasible constraint has energy $+\infty$, with
$e^{-\infty}=0$. If $q/\beta\le10^{-6}$, then
\begin{align}
 \sum_{k\in\Z^d}e^{-\beta E_k}
 &\le\exp\bigl(16d e^{-10^{-5}\beta/q}\bigr),
 \label{eq:lattice-total-expanded}\\
 \sum_{k\in\Z^d\setminus\{0\}}e^{-E_k}
 &\le\varepsilon_d(q):=
 \exp\left(-\frac{\pi^2}{80q}+16d e^{-10^{-5}\beta/q}\right).
 \label{eq:lattice-tail-expanded}
\end{align}
For each integer $d\ge1$, this formula defines
$\varepsilon_d:(0,\infty)\to(0,\infty)$; we use its estimate only under
the displayed cutoff.
\end{lemma}

\begin{proof}
We first sum over finitely many feasible cells. We then bound the
entropy of their index in two complementary ways and pass to all cells.

\paragraph{Step 1: turn a finite energy sum into an entropy.}
Let $J\subseteq\Z^d$ be any finite nonempty set of feasible indices.
For every $k\in J$, choose a minimizing pair
$z_k\in\R^r$, $x_k\in Q_d$. A minimizer exists: the constraint is
closed, and a bounded-energy sublevel set bounds both $z$ and $x$, hence
is compact. Put
\[
 Z_J=\sum_{k\in J}e^{-\beta E_k},\qquad
 \Pp(K=k)=e^{-\beta E_k}/Z_J\quad(k\in J).
\]
Here $K$ is a random vector taking values in $J$, and $Z_J>0$ is a
scalar. Expanding the definition of its entropy gives the exact identity
\begin{equation}\label{eq:lattice-gibbs-expanded}
 H(K)=\beta\E E_K+\log Z_J.
\end{equation}

\paragraph{Step 2: bound information in a noisy latent vector.}
Independently of $K$, draw $Z\in\R^r$ with law $N(0,I_r/\beta)$ and
observe $V=z_K+Z\in\R^r$. Conditional on $K=k$, its distribution is
$N(z_k,I_r/\beta)$. The relative-entropy chain rule in the preliminaries
(\ref{pre:entropy}), applied against the reference distribution
$N(0,I_r/\beta)$, gives
\begin{align*}
 I(K;V)
 &\le\E_K D\bigl(N(z_K,I_r/\beta)\,\Vert\,N(0,I_r/\beta)\bigr)\\
 &=\frac\beta2\E\norm{z_K}^2.
\end{align*}
The inequality discards the nonnegative relative entropy of the marginal
law of $V$ against that reference. The equality follows directly by
dividing the two Gaussian densities; the expected log ratio for a shift
$z_k$ is $\beta\norm{z_k}^2/2$.

\paragraph{Step 3: bound the information still missing.}
For $1\le i\le d$, define the real observation
$W_i=b_i^TV$ and noise $N_i=b_i^TZ$. The cell constraint says
\[
 W_i=\pi K_i+x_{K,i}+N_i,
 \qquad \Var(N_i)=\norm{b_i}^2/\beta\le q/\beta.
\]
The vector of noises may be correlated, but it is independent of $K$.
Conditional-entropy subadditivity and the fact that $W_i$ is a function
of $V$ give
\[
 H(K\mid V)\le\sum_{i=1}^d H(K_i\mid V)
             \le\sum_{i=1}^d H(K_i\mid W_i).
\]
For each conditional distribution of $K_i$ given $W_i$, nonnegativity of
relative entropy gives the cross-entropy inequality
\[
 H(K_i\mid W_i)
 \le\E[-\log q_{q/\beta}(K_i\mid W_i)].
\]
Condition now on the entire index $K=k$. Its offset $x_{k,i}$ is fixed,
and Lemma~\ref{lem:decoder} applies to the independent Gaussian noise
$N_i$. Averaging over $k$ and summing over $i$ yields
\[
 H(K\mid V)
 \le\frac{\beta\kappa_0}{2}\E\norm{x_K}^2
       +16d e^{-10^{-5}\beta/q}
 \le\frac{\beta\kappa}{2}\E\norm{x_K}^2
       +16d e^{-10^{-5}\beta/q}.
\]
This is the precise place where the scalar decoder is used; no assertion
of independence between different observations appears.

\paragraph{Step 4: sum all cells and remove the central one.}
Adding the bounds on $I(K;V)$ and $H(K\mid V)$ gives
$H(K)\le\beta\E E_K+16d e^{-10^{-5}\beta/q}$.
Equation~\eqref{eq:lattice-gibbs-expanded} therefore bounds
$\log Z_J$ by the last error term. Take increasing finite sets of
feasible indices whose union contains every feasible index. Monotone
convergence of the nonnegative sums proves
\eqref{eq:lattice-total-expanded}.

If $k\ne0$ is feasible, choose a coordinate $i$ with $|k_i|\ge1$.
Every feasible pair has
$|b_i^Tz|=|\pi k_i+x_i|\ge\pi/2$. Cauchy--Schwarz gives
$q\norm z^2\ge\pi^2/4$, hence $E_k\ge\pi^2/(8q)$. Consequently
\[
 \sum_{k\ne0}e^{-E_k}
 =\sum_{k\ne0}e^{-(1-\beta)E_k}e^{-\beta E_k}
 \le e^{-\pi^2/(80q)}\sum_k e^{-\beta E_k}.
\]
This proves \eqref{eq:lattice-tail-expanded}, since $1-\beta=1/10$.
\end{proof}

The central cell has energy zero, witnessed by $z=0$ and $x=0$.
The last estimate therefore says that all other cells together have tiny
weight when $q$ is sufficiently small relative to the logarithm of the
number of observations. It is a bound relative to a fixed central scale,
not a count of cells in an ambient box. This is what permits its use with
highly correlated Gaussian projections and, later, discrete spin sums.

\subsection{A covariance bound for negative quadratic spin weights}
\label{sec:zero-field}

We next prove the matrix estimate that will make the joint spin--continuous
reference distribution well behaved. The discrete random vector in this
section is a spin vector $\sigma\in\{-1,1\}^d$. For a symmetric matrix
$A\in\R^{d\times d}$ with $A\succeq0$ and a field $h\in\R^d$, define
\[
 Z_A(h)=2^{-d}\sum_{\sigma\in\{-1,1\}^d}
          e^{-\sigma^TA\sigma/2+h^T\sigma},
 \qquad
 \mu_{A,h}(\sigma)=\frac{2^{-d}e^{-\sigma^TA\sigma/2+h^T\sigma}}{Z_A(h)}.
\]
Thus $Z_A:\R^d\to(0,\infty)$ is a normalizing function, and
$\mu_{A,h}:\{-1,1\}^d\to(0,1]$ is a probability mass function.
The negative quadratic term penalizes spin vectors in directions where
$A$ has a large eigenvalue. We want their covariance to reflect that
penalty, rather than merely being bounded by a constant multiple of the
identity matrix.

Our conclusion will be Proposition~\ref{prop:zero-field}:
if every diagonal entry of $A$ is at most $10^{-10}/\log(4d)$, then
\[
 \Cov_{\mu_{A,0}}\sigma\preceq(A+0.93I_d)^{-1}.
\]
This allows $A$ to have large eigenvalues: small diagonal entries do not
imply a small operator norm in high dimension. The inverse matrix on the
right is therefore an essential part of the conclusion.

The proof uses a Gaussian Fourier representation of the spin weights.
That representation involves a product of cosines, which can be negative.
We will compare it to a positive, globally convex reference density,
bound their relative errors, and then transfer a covariance bound back to
the spins. The lattice estimate from Section~\ref{sec:lattice} controls
all of the noncentral cosine periods.

\subsubsection{A relative bound on translated Gaussian cells}

For a positive-definite symmetric matrix $T\in\R^{d\times d}$, write
\[
 p_T(g)=\frac{\exp(-g^TT^{-1}g/2)}{(2\pi)^{d/2}\sqrt{\det T}}
 \qquad(g\in\R^d).
\]
This is the density $p_T:\R^d\to(0,\infty)$ of $N(0,T)$.
Recall that $Q_d=[-\pi/2,\pi/2]^d\subset\R^d$.

\begin{lemma}[Gaussian periodic tails relative to the central cell]
\label{lem:periodic-gaussian}
Let $T\in\R^{d\times d}$ be symmetric positive definite and suppose
$\max_iT_{ii}\le q$ with $q/\beta\le10^{-6}$. For $1\le i\le d$,
let $f_i:\R\to[0,\infty)$ be an even $C^2$ function satisfying
$f_i(0)=0$ and $f_i''(t)\ge3/4$ for every $t\in\R$.
For each index $k\in\Z^d$, define the positive scalar
\[
 I_k=\int_{Q_d}p_T(\pi k+x)e^{-\sum_{i=1}^d f_i(x_i)}\,dx.
\]
Then
\begin{equation}\label{eq:periodic-gaussian-expanded}
 \sum_{k\in\Z^d\setminus\{0\}}I_k\le\varepsilon_d(q)I_0.
\end{equation}
\end{lemma}

\begin{proof}
Normalize the integrand of $I_0$ to a probability density on $Q_d$, and
let $X\in Q_d$ have this distribution, denoted by $\nu_0$.
Its potential has Hessian at least $T^{-1}+(3/4)I_d$.
The density and box are invariant under $x\mapsto-x$, so $\E X=0$.
The Brascamp--Lieb MGF consequence for boxes, established in the
preliminaries, gives the covariance proxy
\[
 C=(T^{-1}+(3/4)I_d)^{-1}\in\R^{d\times d},
 \qquad \E_{\nu_0}e^{u^TX}\le e^{u^TCu/2}\quad(u\in\R^d).
\]
Dividing the translated Gaussian density by the central one gives
\[
 \frac{I_k}{I_0}
 =e^{-\pi^2k^TT^{-1}k/2}\,
   \E_{\nu_0}e^{-\pi k^TT^{-1}X}.
\]
Apply Lemma~\ref{lem:support} with the source
$h=-\pi T^{-1}k\in\R^d$ and support $Q_d$. Its exponent, including
the prefactor just displayed, is
\begin{align*}
 &-\tfrac12\pi^2k^TT^{-1}k-\pi k^TT^{-1}x-\tfrac12x^TC^{-1}x\\
 &\hspace{20pt}=-\tfrac12\left((\pi k+x)^TT^{-1}(\pi k+x)
                                     +\tfrac34\norm x^2\right).
\end{align*}
It follows that $I_k/I_0\le e^{-E_k}$ for the energy
\[
 E_k=\frac12\min_{x\in Q_d}
       \left((\pi k+x)^TT^{-1}(\pi k+x)+\tfrac34\norm x^2\right).
\]
To identify this with Lemma~\ref{lem:lattice}, let
$e_i\in\R^d$ be the $i$th standard basis vector and put
$b_i=T^{1/2}e_i\in\R^d$. Then $\norm{b_i}^2=T_{ii}\le q$.
The constraints $b_i^Tz=\pi k_i+x_i$ say
$T^{1/2}z=\pi k+x$, and hence
$\norm z^2=(\pi k+x)^TT^{-1}(\pi k+x)$.
Lemma~\ref{lem:lattice}, with latent dimension $r=d$ and
$\kappa=3/4$, now proves \eqref{eq:periodic-gaussian-expanded}.
\end{proof}

The estimate is relative to $I_0$, which may itself be very small.
An absolute Gaussian tail bound would not be enough after we divide by a
small normalizing constant. This relative form is the main benefit of
retaining the support in Lemma~\ref{lem:support}.

\subsubsection{A positive envelope that agrees with cosine near zero}

The function $-\log\cos x$ is strongly convex on $(-\pi/2,\pi/2)$,
but diverges at its endpoints. We need a smooth global potential with a
bounded second derivative, because later we will use both upper and lower
Hessian bounds. Keep $-\log\cos x$ near zero and continue it quadratically.

Set $a_*=1/4$ and $L=\sec^2(a_*)$. Define $f:\R\to[0,\infty)$ by
\[
 f(x)=\begin{cases}
 -\log\cos x,&|x|\le a_*,\\[2pt]
 -\log\cos a_*+\tan a_*(|x|-a_*)+\dfrac L2(|x|-a_*)^2,
       &|x|\ge a_*.
 \end{cases}
\]
The values and first two derivatives match at $\pm a_*$, so $f$ is
even and $C^2$. Inside the interval, $f''(x)=\sec^2x$; outside it,
$f''(x)=L$. Consequently
\begin{equation}\label{eq:capped-properties-expanded}
 1\le f''(x)\le L<\frac{16}{15},\qquad
 \frac{x^2}{2}\le f(x)\le\frac L2x^2.
\end{equation}
The quadratic bounds follow by integrating the second-derivative bounds
twice, since $f(0)=f'(0)=0$. To check the numerical upper bound on $L$,
use $\cos(1/2)>1-(1/2)^2/2=7/8$, and hence
$\cos^2(1/4)=(1+\cos(1/2))/2>15/16$.
For $a_*\le x<\pi/2$, the second derivative of $-\log\cos x$ is at
least $L$; its value and first derivative match those of $f$ at $a_*$.
Two integrations therefore also give
\begin{equation}\label{eq:capped-majorant-expanded}
 f(x)\le-\log\cos x\qquad(|x|<\pi/2).
\end{equation}
Thus $e^{-f(x)}$ is an upper bound on the nonnegative central cosine,
and is exactly that cosine on $[-1/4,1/4]$.

For $T\succ0$ of size $d\times d$, define the scalar $\mathcal I(T)>0$
and the probability measure $\nu_T$ on $\R^d$ by
\[
 \mathcal I(T)=\int_{\R^d}p_T(g)e^{-\sum_i f(g_i)}\,dg,
 \qquad
 \nu_T(dg)=\mathcal I(T)^{-1}p_T(g)e^{-\sum_i f(g_i)}\,dg.
\]
As a function of $T$ on the positive-definite symmetric matrices,
$\mathcal I$ takes values in $(0,1]$. The potential of $\nu_T$, up to
an additive constant, is the function $V_T:\R^d\to\R$ given by
$V_T(g)=g^TT^{-1}g/2+\sum_i f(g_i)$. Its Hessian satisfies
\[
 T^{-1}+I_d\preceq\nabla^2V_T(g)\preceq T^{-1}+LI_d.
\]
The distribution is centered by $g\mapsto-g$. By the MGF consequence
of Brascamp--Lieb, its covariance and centered exponential moments have
proxy
\[
 C_T=(T^{-1}+I_d)^{-1}\in\R^{d\times d},
 \qquad \Cov_{\nu_T}G\preceq C_T\preceq T.
\]
Here and below $G\in\R^d$ denotes the vector being integrated; the
subscript on an expectation specifies whether its distribution is
$N(0,T)$, $\nu_T$, or a restriction of $\nu_T$.

Put $q=\max_iT_{ii}>0$, and define the inner box
$K=[-1/4,1/4]^d\subset\R^d$ and the scalar bounds
\[
 \Delta=2d e^{-1/(32q)},\qquad \alpha=\Delta+\varepsilon_d(q).
\]
Each coordinate under $\nu_T$ has variance proxy at most $q$.
The exponential-moment tail bound from the preliminaries, applied at
$1/4$ and summed over the coordinates, gives $\nu_T(K^c)\le\Delta$.

\subsubsection{Comparing normalizers without treating signed weights as probabilities}

For $t\in\R$, let $[t]_\pi\in[-\pi/2,\pi/2)$ be its centered
remainder modulo $\pi$. This defines a map
$[\,\cdot\,]_\pi:\R\to[-\pi/2,\pi/2)$; explicitly,
$[t]_\pi=t-\pi\lfloor t/\pi+1/2\rfloor$.
Define functions $e,e_{\rm per}:\R^d\to(0,1]$ by
\[
 e(g)=e^{-\sum_i f(g_i)},\qquad
 e_{\rm per}(g)=e^{-\sum_i f([g_i]_\pi)}.
\]
The first is the global envelope and the second repeats its central part
periodically. The following pointwise inequality separates two errors:
\begin{equation}\label{eq:pointwise-envelope-expanded}
 \left|\prod_i\cos g_i-e(g)\right|
 \le\one_{K^c}(g)e(g)+\one_{Q_d^c}(g)e_{\rm per}(g).
\end{equation}
On $K$ the two functions agree. On $Q_d\setminus K$, the cosine product
is between zero and $e(g)$, by \eqref{eq:capped-majorant-expanded}.
Outside $Q_d$, its absolute value is at most $e_{\rm per}(g)$; the
triangle inequality then gives the displayed bound.

The Gaussian Fourier identity in the preliminaries gives
\[
 Z_T(0)=\E_{\sigma\sim\Unif(\{-1,1\}^d)}e^{-\sigma^TT\sigma/2}
       =\E_{G\sim N(0,T)}\prod_i\cos G_i>0.
\]
For this tail comparison assume $q/\beta\le10^{-6}$.
Integrating \eqref{eq:pointwise-envelope-expanded}, the first error is
at most $\Delta\mathcal I(T)$. Decompose the second into noncentral
periodic cells and apply Lemma~\ref{lem:periodic-gaussian} with $f_i=f$.
Its central integral is at most $\mathcal I(T)$, so
\begin{equation}\label{eq:normalizer-error-expanded}
 |Z_T(0)-\mathcal I(T)|\le\alpha\mathcal I(T).
\end{equation}
Cell boundaries have Lebesgue measure zero; choosing half-open cells
makes the decomposition disjoint. This normalizer estimate does not
claim the cosine product is nonnegative outside the central cell.

\subsubsection{A relative second-moment error}

We need the same comparison after multiplying by a quadratic function of
$g$. A probability tail alone does not bound that extra factor. The next
argument absorbs it into a small change of Gaussian covariance.
Assume from now until the final proposition that $2q/\beta\le10^{-6}$,
so that the periodic estimate is available at both $q$ and $2q$.

Define the symmetric matrix $\mathcal M_T\in\R^{d\times d}$ by
\[
 \mathcal M_T=
 \frac{\E_{G\sim N(0,T)}GG^T\prod_i\cos G_i}{Z_T(0)}.
\]
It is a signed normalized moment, not the covariance of a probability
measure. The next claim states the precise comparison we need.

\begin{claim}[Relative comparison of signed second moments]
\label{cl:signed-moment}
With $T,q,\alpha,\Delta$, and $\nu_T$ as defined above, assume
$\alpha<1$. Define the nonnegative scalar
\[
 \zeta=\frac{\alpha+\sqrt{5\Delta}+4\sqrt2\,\varepsilon_d(2q)}{1-\alpha}.
\]
Then
\begin{equation}\label{eq:signed-moment-error-expanded}
 -\zeta T\preceq\mathcal M_T-\Cov_{\nu_T}G\preceq\zeta T.
\end{equation}
\end{claim}

\begin{proof}

Fix a nonzero test vector $v\in\R^d$ and scalars
$w=v^TTv>0$, $t=1/(2w)>0$. Define the $d\times d$ positive-definite
matrix
\[
 T_t=(T^{-1}-tvv^T)^{-1}
     =T+\frac{t}{1-tw}Tvv^TT.
\]
The inverse formula is checked by multiplication. Positivity follows
from $(v^Tx)^2\le w x^TT^{-1}x$ for $x\in\R^d$, since $tw=1/2$.
Also $(Tv)_i^2\le T_{ii}w$, so $(T_t)_{ii}\le2T_{ii}\le2q$.
The matrix $T^{-1/2}T_tT^{-1/2}$ has eigenvalue $2$ in the direction
$T^{1/2}v$ and eigenvalue $1$ on its orthogonal complement. Thus
$\det T_t=2\det T$, and the Gaussian density change is exactly
\begin{equation}\label{eq:rank-one-density-expanded}
 p_T(g)e^{t(v^Tg)^2/2}=\sqrt2\,p_{T_t}(g).
\end{equation}

Let $I_0(T)=\int_{Q_d}p_T(g)e(g)\,dg>0$, and let
$\nu_{T,Q_d}$ be the probability measure with this normalized integrand
on $Q_d$. The box version of Brascamp--Lieb gives the same MGF proxy
$C_T$ for this centered measure. Using an independent standard normal
real variable $Z$ to represent a scalar Gaussian exponential, we obtain
\begin{align*}
 \E_{\nu_{T,Q_d}}e^{t(v^TG)^2/2}
 &=\E_Z\E_{\nu_{T,Q_d}}e^{\sqrt t Z v^TG}\\
 &\le\E_Z e^{tZ^2v^TC_Tv/2}
  =(1-tv^TC_Tv)^{-1/2}\le\sqrt2.
\end{align*}
All integrands before the last Gaussian calculation are nonnegative, so
the order of integration is valid. Combining this with
\eqref{eq:rank-one-density-expanded} shows that
\begin{equation}\label{eq:central-normalizer-expanded}
 I_0(T_t)=\frac{I_0(T)}{\sqrt2}
     \E_{\nu_{T,Q_d}}e^{t(v^TG)^2/2}
 \le I_0(T)\le\mathcal I(T).
\end{equation}
This is why the covariance change does not create a troublesome new
normalizer.

For $u\ge0$, $u\le e^u$. Apply this with
$u=(v^Tg)^2/(4w)$ to obtain
$(v^Tg)^2\le4w e^{t(v^Tg)^2/2}$.
Equations \eqref{eq:rank-one-density-expanded} and
\eqref{eq:central-normalizer-expanded}, followed by the periodic lemma
at parameter $2q$, now give
\begin{align}
 &\int_{Q_d^c}(v^Tg)^2p_T(g)e_{\rm per}(g)\,dg \notag\\
 &\quad\le4\sqrt2\,w\int_{Q_d^c}p_{T_t}(g)e_{\rm per}(g)\,dg
 \le4\sqrt2\,\varepsilon_d(2q)\mathcal I(T)w.
 \label{eq:noncentral-moment-expanded}
\end{align}

For the global envelope, put $s=v^TC_Tv\le w$.
Brascamp--Lieb applied to $g\mapsto(v^Tg)^2$, a function from
$\R^d$ to $[0,\infty)$, gives
\[
 \Var_{\nu_T}((v^TG)^2)
 \le4s\E_{\nu_T}(v^TG)^2\le4s^2.
\]
Since the second moment is at most $s$, its fourth moment is at most
$5s^2$. Cauchy--Schwarz on $K^c$ therefore gives
\begin{equation}\label{eq:global-moment-expanded}
 \E_{\nu_T}(v^TG)^2\one_{K^c}
 \le\sqrt{\E_{\nu_T}(v^TG)^4\,\nu_T(K^c)}
 \le\sqrt{5\Delta}\,w.
\end{equation}

To compare the signed normalized moment with a genuine covariance,
define the two real scalars
\[
 a_v=\E_{G\sim N(0,T)}(v^TG)^2\prod_i\cos G_i,
 \qquad b_v=\mathcal I(T)\E_{\nu_T}(v^TG)^2.
\]
Multiplying \eqref{eq:pointwise-envelope-expanded} by $(v^Tg)^2$ and
using \eqref{eq:noncentral-moment-expanded}--\eqref{eq:global-moment-expanded}
gives
\[
 |a_v-b_v|\le\mathcal I(T)
      \bigl(\sqrt{5\Delta}+4\sqrt2\,\varepsilon_d(2q)\bigr)w.
\]
If $\alpha<1$, then $Z_T(0)\ge(1-\alpha)\mathcal I(T)$.
Also $0\le b_v/\mathcal I(T)\le w$. Subtract the normalized moments as
\[
 \frac{a_v}{Z_T(0)}-\frac{b_v}{\mathcal I(T)}
 =\frac{a_v-b_v}{Z_T(0)}
   +\frac{b_v}{\mathcal I(T)}\frac{\mathcal I(T)-Z_T(0)}{Z_T(0)}.
\]
The first term is controlled by the moment error, the second by the
normalizer error. Their sum in absolute value is at most
$\zeta w=\zeta v^TTv$. The inequality is trivial when $v=0$, so it
holds for every $v\in\R^d$ and proves
\eqref{eq:signed-moment-error-expanded} by the definition of PSD order.
The global-envelope contribution on all of $K^c$ includes its contribution
outside $Q_d$. The periodic term separately controls the signed cosine
there. Thus neither contribution has been omitted.
\end{proof}

\subsubsection{Transfer the Gaussian moment back to the spins}

The Hessian upper bound on $V_T$ now provides a \emph{lower} bound on
the envelope covariance. Together with the signed-moment comparison,
this yields an estimate directly on the discrete spins.

\begin{claim}[Spin covariance with an explicit error]
\label{cl:spin-precision-error}
For the positive-definite matrix $T\in\R^{d\times d}$ defined above,
assume $2q/\beta\le10^{-6}$ and $\alpha<1$. With the same scalar
$\zeta$ as in Claim~\ref{cl:signed-moment},
\begin{equation}\label{eq:precision-with-error-expanded}
 \Cov_{\mu_{T,0}}\sigma
 \preceq(T+L^{-1}I_d)^{-1}+\zeta T^{-1}.
\end{equation}
\end{claim}

\begin{proof}
We first give the details of the Cram\'er--Rao comparison. Let
$S:\R^d\to\R^d$, $S(g)=\nabla V_T(g)$, be the score function, and define the
$d\times d$ matrix $J=\E_{\nu_T}S(G)S(G)^T$.
Integration by parts gives
\[
 \E_{\nu_T}G S(G)^T=I_d,
 \qquad J=\E_{\nu_T}\nabla^2V_T(G)\preceq T^{-1}+LI_d.
\]
These integrations have no boundary terms: the density has Gaussian
decay and $S$ grows at most linearly by the bounds on $f''$.
Also $J\succeq T^{-1}+I_d\succ0$. The expectation of the positive
semidefinite matrix
$(G-J^{-1}S(G))(G-J^{-1}S(G))^T$ is
$\Cov_{\nu_T}G-J^{-1}$, because $\E_{\nu_T}G=0$.
Therefore inverse order gives
\begin{equation}\label{eq:cramer-rao-expanded}
 \Cov_{\nu_T}G\succeq J^{-1}\succeq(T^{-1}+LI_d)^{-1}.
\end{equation}

For a fixed spin vector $\sigma\in\{-1,1\}^d$, the Gaussian
integration-by-parts identity from the preliminaries reads
\[
 \E_{G\sim N(0,T)}GG^Te^{i\sigma^TG}
 =(T-T\sigma\sigma^TT)e^{-\sigma^TT\sigma/2}.
\]
Average this identity over uniform spins and use
$\E_\sigma e^{i\sigma^Tg}=\prod_i\cos g_i$.
The law $\mu_{T,0}$ has mean zero by $\sigma\mapsto-\sigma$,
so its second moment equals its covariance. Dividing by $Z_T(0)$ gives
\[
 \mathcal M_T=T-T\Cov_{\mu_{T,0}}(\sigma)T,
 \qquad
 \Cov_{\mu_{T,0}}\sigma=T^{-1}-T^{-1}\mathcal M_TT^{-1}.
\]
Use the lower bound on $\mathcal M_T$ from
\eqref{eq:signed-moment-error-expanded}, followed by
\eqref{eq:cramer-rao-expanded}, to conclude
\[
 \Cov_{\mu_{T,0}}\sigma
 \preceq T^{-1}-T^{-1}(T^{-1}+LI_d)^{-1}T^{-1}+\zeta T^{-1}.
\]
The first two terms sum to $(T+L^{-1}I_d)^{-1}$.
This identity can be checked on an eigenvalue $s>0$
of $T$:
$s^{-1}-s^{-2}(s^{-1}+L)^{-1}=(s+L^{-1})^{-1}$.
This proves \eqref{eq:precision-with-error-expanded}.
\end{proof}

We have obtained the desired inverse dependence, plus an exponentially
small error. It remains to show that this error fits into the small
margin between $L^{-1}>0.9375$ and the target constant $0.93$.

\subsubsection{Uniform constants and absorption of the error}

\begin{proposition}[Zero-field precision bound]\label{prop:zero-field}
Let $d\ge1$ be an integer, and let $A\in\R^{d\times d}$ be symmetric
and positive semidefinite. Set $\gamma=10^{-10}$. If
\[
 \rho:=\max_{1\le i\le d}A_{ii}\le\frac{\gamma}{\log(4d)},
\]
then
\begin{equation}\label{eq:zero-field-expanded}
 \Cov_{\mu_{A,0}}\sigma\preceq(A+0.93I_d)^{-1}.
\end{equation}
\end{proposition}

\begin{proof}
If $\rho=0$, positivity implies $|A_{ij}|^2\le A_{ii}A_{jj}=0$, so
$A=0$. The spins are independent and their covariance is $I_d$, which
is at most $(0.93I_d)^{-1}$.
Suppose henceforth that $\rho>0$, and take $T=A+\rho I_d\succ0$.
Because $\sigma_i^2=1$, adding a diagonal matrix changes
$\sigma^TA\sigma$ only by a constant independent of $\sigma$.
Thus $\mu_{T,0}=\mu_{A,0}$.
We have $T\succeq\rho I_d$, $q=\max_iT_{ii}\le2\rho$, and
every eigenvalue $\lambda$ of $A$ satisfies
$0\le\lambda\le\tr A\le d\rho$.

\paragraph{Step 1: check every cutoff and bound both periodic errors.}
The diagonal hypothesis implies
\[
 d\le\frac14 e^{\gamma/\rho},\qquad
 0<\rho\le\frac{\gamma}{\log4}<10^{-10}.
\]
In particular $4\rho/\beta<5\cdot10^{-10}<10^{-6}$, so all earlier
applications at parameters $q$ and $2q$ satisfy their decoder cutoffs.
For any $0<u\le4\rho$,
\[
 16d e^{-10^{-5}\beta/u}
 \le4\exp\left(-\frac{(9/4)10^{-6}-\gamma}{\rho}\right)<1.
\]
The exponent's positive numerator divided by $\rho$ exceeds $20000$,
which certainly makes the last expression smaller than one. Since
$\pi^2>9$ and $\rho<1/80$, the definition of $\varepsilon_d$ gives
\begin{equation}\label{eq:epsilon-small-expanded}
 \varepsilon_d(u)
 \le\exp\left(-\frac9{320\rho}+1\right)
 \le e^{-1/(64\rho)}.
\end{equation}
For the last inequality, subtract the exponents: it is precisely
$1\le1/(80\rho)$.

\paragraph{Step 2: bound the complete relative moment error.}
Using $q\le2\rho$ and the dimension bound,
\[
 \Delta\le\frac12e^{-(1/64-\gamma)/\rho}
          \le\frac12e^{-1/(128\rho)},
 \qquad
 \alpha\le\frac32e^{-1/(128\rho)}<\frac38.
\]
The comparison $1/64-\gamma\ge1/128$ is immediate from
$\gamma=10^{-10}$. The final strict inequality follows, for example,
from $1/(128\rho)>2$ and $e^2>4$.
Inserting these estimates and \eqref{eq:epsilon-small-expanded} into
\eqref{eq:signed-moment-error-expanded} yields
\[
 \zeta\le
 2\left(\frac32+\sqrt{5/2}+4\sqrt2\right)e^{-1/(256\rho)}
 <20e^{-1/(256\rho)}.
\]
We used $(1-\alpha)^{-1}<2$, and weakened every exponential to the
slowest one, $e^{-1/(256\rho)}$.

\paragraph{Step 3: absorb the dimension and the smallest eigenvalue.}
The error in \eqref{eq:precision-with-error-expanded} contains
$T^{-1}$, whose eigenvalues can be as large as $1/\rho$.
The target inverse difference also involves the largest eigenvalue
$d\rho$. Both are controlled in the following one bound:
\begin{align}
 \zeta(d\rho+3+2/\rho)
 &\le20(\rho/4+3+2/\rho)e^{-(1/256-\gamma)/\rho}\notag\\
 &\le\frac{120}{\rho}e^{-1/(512\rho)}
 \le720\cdot512^3\rho^2<10^{-9}<10^{-3}.
 \label{eq:absorption-expanded}
\end{align}
For the second line, $\rho<1$ gives
$\rho/4+3+2/\rho\le6/\rho$, and
$1/256-\gamma\ge1/512$. The next inequality uses only
$e^x\ge x^3/6$ at $x=1/(512\rho)$.
Finally,
$720\cdot512^3\cdot10^{-20}=9.663676416\cdot10^{-10}<10^{-9}$.
This makes the estimate uniform over every allowed dimension.

\paragraph{Step 4: compare the inverse matrices eigenvalue by eigenvalue.}
Write $c=0.93$. Since $L^{-1}>15/16$, for any eigenvalue
$\lambda\ge0$ of $A$,
\[
 \frac1{\lambda+c}-\frac1{\lambda+\rho+L^{-1}}
 =\frac{\rho+L^{-1}-c}{(\lambda+c)(\lambda+\rho+L^{-1})},
 \qquad \rho+L^{-1}-c>0.0075.
\]
To compare the remaining error $\zeta/(\lambda+\rho)$ with this gap,
observe that
\begin{align*}
 \frac{(\lambda+c)(\lambda+\rho+L^{-1})}{\lambda+\rho}
 &=\lambda+c+L^{-1}\frac{\lambda+c}{\lambda+\rho}\\
 &\le\lambda+c+L^{-1}+\frac{cL^{-1}}\rho
 \le\lambda+3+2/\rho.
\end{align*}
Here $0<c<1$ and $0<L^{-1}\le1$.
Multiplication by $\zeta$ and \eqref{eq:absorption-expanded} bound the
last expression by $0.001$, because $\lambda\le d\rho$.
This is smaller than the numerator margin $0.0075$. Therefore
\[
 \frac1{\lambda+\rho+L^{-1}}+\frac\zeta{\lambda+\rho}
 \le\frac1{\lambda+c}.
\]
All matrices in this last comparison are functions of $A$, since
$T=A+\rho I_d$; they share an orthonormal eigenbasis.
Combining the scalar comparison with
\eqref{eq:precision-with-error-expanded} proves
\eqref{eq:zero-field-expanded}.
\end{proof}

The proof has used a small diagonal bound, not a small operator-norm
bound. It therefore retains suppression in every high-energy direction
of $A$, even in very large dimension. The next section extends this
zero-field estimate to arbitrary real fields without weakening the
constant $0.93$.

\subsection{Extending the covariance estimate to every external field}
\label{sec:all-field}

The preceding section treated a spin distribution with no preferred direction: replacing every spin by its negative did not change its probability. We now need a version that remains true after introducing an arbitrary preference for the signs. This is essential later, because conditioning on our continuous auxiliary variables will create such a preference.

Recall the notation. For a positive integer $d$, a real symmetric positive-semidefinite matrix $N\in\R^{d\times d}$, and a field $h\in\R^d$, the probability mass function $\mu_{N,h}:\{\pm1\}^d\to(0,1]$ is
\[
\mu_{N,h}(\sigma)
 =\frac{2^{-d}\exp(-\sigma^TN\sigma/2+h^T\sigma)}{Z_N(h)},
\qquad
Z_N(h)=2^{-d}\sum_{\sigma\in\{\pm1\}^d}
             \exp(-\sigma^TN\sigma/2+h^T\sigma).
\]
Here $Z_N:\R^d\to(0,\infty)$ is the normalizing function. The matrix term assigns a cost to a spin vector; the linear term can favor either sign in each coordinate. The covariance matrix is $d$ by $d$, even when the field makes its mean nonzero.

Our method uses two independent samples from this distribution. Their difference has a useful symmetry that either sample may lack. Before using that observation, we prove a matrix inequality that will average the resulting bounds.

\subsubsection{Averaging inverses supported on random coordinate sets}

If $T\subseteq[d]$, write $N_T\in\R^{T\times T}$ for the principal submatrix obtained by keeping the rows and columns indexed by $T$. Its identity matrix is $I_T$. To \emph{embed} a matrix indexed by $T$ means to place it in a $d$-by-$d$ matrix and fill all other entries with zeros. For $T=\varnothing$, the embedded matrix is the zero matrix. These conventions let us compare matrices arising from different coordinate sets.

\begin{lemma}[Averaging principal inverses]\label{lem:principal-inverses}
Let $N\in\R^{d\times d}$ be symmetric and positive semidefinite, let $a>0$, and let $0<p\le1$. Suppose $T$ is a random subset of $[d]$ such that $\Pp(i\in T)\le p$ for every $i\in[d]$. Then
\[
\E_T\operatorname{embed}\bigl((N_T+aI_T)^{-1}\bigr)
 \preceq (N+(a/p)I_d)^{-1}.
\]
The membership events $\{i\in T\}$ may be arbitrarily dependent.
\end{lemma}

\begin{proof}
Fix an arbitrary vector $z\in\R^d$. To prove the matrix inequality, we will prove the corresponding scalar inequality after multiplying on the left by $z^T$ and on the right by $z$.

For each realized set $T$, the variational formula from Lemma~\ref{pre:variational} gives
\[
z^T\operatorname{embed}\bigl((N_T+aI_T)^{-1}\bigr)z
 =\max_{\substack{w\in\R^d\\w_i=0\text{ for }i\notin T}}
       \bigl(2z^Tw-w^TNw-a\norm{w}^2\bigr).
\]
There are only finitely many possible sets $T$. We may therefore choose the maximizing vector separately for each set, giving a function $w:2^{[d]}\to\R^d$ with $w(T)$ supported on $T$. Equivalently, the expected quadratic form is the maximum, over all such choices, of
\[
\E\bigl[2z^Tw(T)-w(T)^TNw(T)-a\norm{w(T)}^2\bigr].
\]
We bound this expression for an arbitrary choice of $w(T)$.

Put $\bar w=\E w(T)\in\R^d$. Positive semidefiniteness gives
\[
\E w(T)^TNw(T)-\bar w^TN\bar w
 =\E\bigl[(w(T)-\bar w)^TN(w(T)-\bar w)\bigr]\ge0.
\]
We also gain a factor $1/p$ in the ordinary squared norm. Indeed, $w_i(T)$ vanishes outside the event $i\in T$, so Cauchy--Schwarz gives
\[
\bar w_i^2
 =\bigl(\E[w_i(T)\one_{\{i\in T\}}]\bigr)^2
 \le \Pp(i\in T)\,\E w_i(T)^2
 \le p\,\E w_i(T)^2.
\]
This argument includes an inclusion probability of zero: in that case $w_i(T)=0$ almost surely and $\bar w_i=0$. Summing the coordinate inequalities yields
\[
\E\norm{w(T)}^2\ge \frac1p\norm{\bar w}^2.
\]
Consequently every allowed choice has objective value at most
\[
2z^T\bar w-\bar w^TN\bar w-\frac ap\norm{\bar w}^2
 \le \max_{v\in\R^d}
       \left(2z^Tv-v^TNv-\frac ap\norm v^2\right)
 =z^T(N+(a/p)I_d)^{-1}z.
\]
The last equality is again the variational formula. Since $z$ was arbitrary, this proves the claim.
\end{proof}

The reason this lemma is useful is quite concrete. A vector supported on a rarely available coordinate must fluctuate more in that coordinate to have a prescribed average. The resulting extra squared-norm cost becomes the extra diagonal term $a/p$ in the inverse. No statement about independence of the available coordinates enters this argument.

\subsubsection{Two independent copies remove the field from their difference}

\begin{proposition}[Covariance bound for every field]\label{prop:all-field}
Let $\gamma=10^{-10}$. Suppose $N\in\R^{d\times d}$ is symmetric and positive semidefinite and
\[
2\max_{i\in[d]}N_{ii}\le\frac{\gamma}{\log(4d)}.
\]
Then, for every $h\in\R^d$,
\[
\Cov_{\mu_{N,h}}(\sigma)\preceq(N+.93I_d)^{-1}.
\]
\end{proposition}

\begin{proof}
Let $\sigma,\tau\in\{\pm1\}^d$ be independent random vectors, each with law $\mu_{N,h}$. Define random vectors $u,v\in\R^d$ and a random subset $T\subseteq[d]$ by
\[
u=\frac{\sigma+\tau}{2},\qquad
v=\frac{\sigma-\tau}{2},\qquad
T=\{i\in[d]:\sigma_i\ne\tau_i\}.
\]
If $i\in T$, then $u_i=0$ and $v_i\in\{\pm1\}$. If $i\notin T$, then $v_i=0$ and $u_i\in\{\pm1\}$. Thus $T$ records where the two samples disagree, and $v$ records the signs of those disagreements.

\paragraph{Step 1: identify the conditional distribution.}
Substitute $\sigma=u+v$ and $\tau=u-v$ into their joint exponential weight. Symmetry of $N$ gives
\[
\begin{aligned}
&-\frac12(u+v)^TN(u+v)-\frac12(u-v)^TN(u-v)
       +h^T(u+v)+h^T(u-v)\\
&\hspace{25mm}=-u^TNu-v^TNv+2h^Tu.
\end{aligned}
\]
The mixed quadratic terms cancel, and the field has disappeared from the $v$ term.

Fix a realized set $T$. Each choice of signs $u_{T^c}\in\{\pm1\}^{T^c}$ and $v_T\in\{\pm1\}^T$ determines exactly one ordered pair $(\sigma,\tau)$ with this disagreement set. Its conditional weight factors into a function of $u_{T^c}$ and a function of $v_T$. Therefore these two sign vectors are conditionally independent, and the conditional law of $v_T$ is exactly the zero-field law with matrix $2N_T$:
\[
\Pp(v_T=w\mid T)\ \propto\ \exp(-w^TN_Tw),
\qquad w\in\{\pm1\}^T.
\]
In particular, $\E(v_T\mid T)=0$, because this mass function is unchanged by $w\mapsto-w$.

\paragraph{Step 2: express covariance using the difference.}
Independence and identical distribution of $\sigma$ and $\tau$ imply
\[
\Cov(\sigma)
 =\frac12\E\bigl[(\sigma-\tau)(\sigma-\tau)^T\bigr]
 =2\E[vv^T].
\]
This identity does not require $\E\sigma=0$: the two mixed moments are exactly $(\E\sigma)(\E\sigma)^T$.

For a nonempty set $T$, Proposition~\ref{prop:zero-field} applies to the $|T|$-by-$|T|$ matrix $2N_T$. Its diagonal entries satisfy the required bound, since
\[
\max_{i\in T}(2N_T)_{ii}
 \le\frac{\gamma}{\log(4d)}
 \le\frac{\gamma}{\log(4|T|)}.
\]
Hence
\[
2\E[vv^T\mid T]
 \preceq 2\operatorname{embed}\bigl((2N_T+.93I_T)^{-1}\bigr)
 =\operatorname{embed}\bigl((N_T+.465I_T)^{-1}\bigr).
\]
For the empty set, both sides are zero. Averaging gives the corresponding inequality without conditioning.

\paragraph{Step 3: bound the inclusion probabilities.}
Write $m_i=\E\sigma_i\in[-1,1]$. A single coordinate equals $1$ with probability $(1+m_i)/2$. Since the two copies are independent,
\[
\Pp(i\in T)
 =2\frac{1+m_i}{2}\frac{1-m_i}{2}
 =\frac{1-m_i^2}{2}\le\frac12.
\]
Apply Lemma~\ref{lem:principal-inverses} with $a=.465$ and $p=1/2$. It gives
\[
\Cov(\sigma)
 \preceq\E_T\operatorname{embed}\bigl((N_T+.465I_T)^{-1}\bigr)
 \preceq(N+.93I_d)^{-1},
\]
as required.
\end{proof}

Notice the exact balance of constants. Passing to two copies multiplies the conditional quadratic matrix by two, while the disagreement vector represents one-half of their difference. Each coordinate disagrees with probability at most one-half. The principal-inverse inequality combines these three facts without losing a factor in the final $.93$.

\subsection{A positive reference function that closely follows cosine}
\label{sec:cosine-reference}

Our eventual quantity is an average of products of cosines. Such a product can be negative, which prevents us from treating it as a probability weight. We construct a positive function $F_\lambda$ that majorizes cosine on its central interval and approximates it extremely closely on $[-1,1]$. The parameter $\lambda$ controls the approximation error. We will also arrange that $F_\lambda$ is a Gaussian factor times the moment-generating function of a positive probability measure. That representation is what will let us use the covariance estimates proved above.

\subsubsection{The scalar probability measure and its guarantees}

Fix a real number $\lambda\ge2$. Define the Gaussian density $p_{2\lambda}:\R\to(0,\infty)$ and the positive endpoint $b_\lambda$ by
\[
p_{2\lambda}(t)=\frac1{\sqrt{4\pi\lambda}}e^{-t^2/(4\lambda)},
\qquad b_\lambda=\pi\lambda.
\]
Thus $p_{2\lambda}$ is the density of a centered real Gaussian of variance $2\lambda$. Define
\[
Z_c=\int_{-b_\lambda}^{b_\lambda}
          p_{2\lambda}(t)\cos\left(\frac{t}{2\lambda}\right)\,dt,
\qquad
q_\lambda(t)=\frac1{Z_c}p_{2\lambda}(t)
          \cos\left(\frac{t}{2\lambda}\right)
          \one_{\{|t|<b_\lambda\}}.
\]
This defines a density $q_\lambda:\R\to[0,\infty)$ and its
probability law $\nu_\lambda(dt)=q_\lambda(t)\,dt$.
The cosine in this density is positive on the open interval of integration. We will prove that $Z_c\ge7/8$, so this is a well-defined probability measure. Its support is contained in the compact interval $[-b_\lambda,b_\lambda]$.

For a scalar random variable $T$ with law $\nu_\lambda$, put
\[
F_\lambda(x)=e^{-\lambda x^2}\E_{\nu_\lambda}e^{xT},
\qquad x\in\R.
\]
This defines a function $F_\lambda:\R\to(0,\infty)$. The expectation is finite for every real $x$ because $T$ is bounded. In this section $T$ is a single real random variable; the next section will use several independent copies as the coordinates of a vector.

\begin{lemma}[Positive cosine reference]\label{lem:positive-reference}
The measure and function defined above satisfy $7/8\le Z_c\le1$. The function $F_\lambda$ is even, and
\begin{align}
F_\lambda(x)&\ge\cos x
 &&\text{for }|x|\le\pi/2,\label{eq:expanded-reference-majorant}\\
0\le\log F_\lambda(x)-\log\cos x
 &\le\delta_\lambda:=4\sqrt\lambda\,e^{-\lambda/4}
 &&\text{for }|x|\le1,\label{eq:expanded-reference-error}\\
F_\lambda(x)&\ge e^{-\lambda x^2}
 &&\text{for every }x\in\R.\label{eq:expanded-reference-lower}
\end{align}
On the open interval $(-b_\lambda,b_\lambda)$, define the prior potential $U_\lambda:(-b_\lambda,b_\lambda)\to[0,\infty)$ by
\[
U_\lambda(t)=\frac{t^2}{4\lambda}
                -\log\cos\left(\frac{t}{2\lambda}\right).
\]
The density of $\nu_\lambda$ is a positive constant times $e^{-U_\lambda(t)}$ on that interval, and
\[
U_\lambda''(t)\ge H_\lambda:=\frac{2\lambda+1}{4\lambda^2}>0.
\]
\end{lemma}

\begin{proof}
We divide the proof into the properties that will serve different purposes later.

\paragraph{Step 1: normalization, symmetry, and curvature.}
Define $g:\R\to[0,1]$ by $g(u)=\cos u$ for $|u|\le\pi/2$ and $g(u)=0$ otherwise. The inequality
\[
\cos u\ge1-\frac{u^2}{2}
\]
follows from $1-\cos u=2\sin^2(u/2)\le u^2/2$. On the central interval it therefore gives $g(u)\ge1-u^2/2$. Outside that interval, $|u|\ge\pi/2>\sqrt2$, so $1-u^2/2\le0=g(u)$. We have proved this lower bound on the entire real line.

Let $G\in\R$ be a centered Gaussian of variance $2\lambda$. Then
\[
Z_c=\E g\left(\frac{G}{2\lambda}\right)
 \ge1-\frac{\E G^2}{8\lambda^2}
 =1-\frac1{4\lambda}\ge\frac78.
\]
The upper bound $Z_c\le1$ follows from $g\le1$. The density is even, so $T$ and $-T$ have the same law. It follows that $F_\lambda$ is even and
\[
\E e^{xT}=\E\cosh(xT)\ge1.
\]
Multiplying by $e^{-\lambda x^2}$ proves \eqref{eq:expanded-reference-lower}.

Differentiating the potential on its open domain gives
\[
U_\lambda'(t)=\frac{t}{2\lambda}
                   +\frac1{2\lambda}\tan\left(\frac{t}{2\lambda}\right),
\qquad
U_\lambda''(t)=\frac1{2\lambda}
                   +\frac1{4\lambda^2}\sec^2\left(\frac{t}{2\lambda}\right)
                   \ge H_\lambda.
\]
The extra term $1/(4\lambda^2)$ in this lower bound will be useful when we couple $T$ to the spins.

\paragraph{Step 2: derive the differential equation, keeping its boundary terms.}
Let $w:[-b_\lambda,b_\lambda]\to[0,\infty)$ denote the unnormalized density
\[
w(t)=p_{2\lambda}(t)\cos\left(\frac{t}{2\lambda}\right),
\]
and define $h:\R\to(0,\infty)$ by
\[
h(x)=\int_{-b_\lambda}^{b_\lambda}w(t)e^{xt}\,dt.
\]
The integral is over a compact interval, so it may be differentiated any number of times. In particular, $h'(x)$ and $h''(x)$ are obtained by inserting factors $t$ and $t^2$. Also,
\[
F_\lambda(x)=Z_c^{-1}e^{-\lambda x^2}h(x).
\]
Direct differentiation of $w$ gives
\[
w'(t)=p_{2\lambda}(t)
 \left[-\frac{t}{2\lambda}\cos\left(\frac{t}{2\lambda}\right)
       -\frac1{2\lambda}\sin\left(\frac{t}{2\lambda}\right)\right].
\]
A second differentiation, followed by collection of the cosine and sine terms, yields
\[
4\lambda^2w''(t)+4\lambda t w'(t)
                 +(t^2+2\lambda+1)w(t)=0.
\]
At the endpoints,
\[
w(\pm b_\lambda)=0,
\qquad
w'(b_\lambda)=-\frac{p_{2\lambda}(b_\lambda)}{2\lambda},
\qquad
w'(-b_\lambda)=\frac{p_{2\lambda}(b_\lambda)}{2\lambda}.
\]
These nonzero derivatives are important. Although the density itself vanishes at the endpoints, its derivative contributes when we integrate twice by parts.

Multiply the differential identity by $e^{xt}$ and integrate over $[-b_\lambda,b_\lambda]$. The two derivative integrals are
\[
\begin{aligned}
\int_{-b_\lambda}^{b_\lambda} w''(t)e^{xt}\,dt
 &=\bigl[w'(t)e^{xt}\bigr]_{-b_\lambda}^{b_\lambda}
   -x\bigl[w(t)e^{xt}\bigr]_{-b_\lambda}^{b_\lambda}+x^2h(x)\\
 &=-\frac{p_{2\lambda}(b_\lambda)}{\lambda}\cosh(b_\lambda x)
   +x^2h(x),\\
\int_{-b_\lambda}^{b_\lambda}t w'(t)e^{xt}\,dt
 &=\bigl[t w(t)e^{xt}\bigr]_{-b_\lambda}^{b_\lambda}
     -\int_{-b_\lambda}^{b_\lambda}(1+xt)w(t)e^{xt}\,dt\\
 &=-h(x)-xh'(x).
\end{aligned}
\]
Substitution therefore gives
\[
h''(x)-4\lambda xh'(x)
 +(4\lambda^2x^2-2\lambda+1)h(x)
 =4\lambda p_{2\lambda}(b_\lambda)\cosh(b_\lambda x).
\]
Differentiating $F_\lambda(x)=Z_c^{-1}e^{-\lambda x^2}h(x)$ twice shows that this is equivalent to
\begin{equation}\label{eq:expanded-reference-ode}
F_\lambda''(x)+F_\lambda(x)=R_\lambda(x),
\qquad
R_\lambda(x):=
 \frac{4\lambda p_{2\lambda}(b_\lambda)}{Z_c}
       e^{-\lambda x^2}\cosh(b_\lambda x)>0.
\end{equation}
Here $R_\lambda:\R\to(0,\infty)$ is an explicit positive function. Its positivity, including the sign inherited from the boundary terms, will prove the majorant property.

\paragraph{Step 3: compare with cosine.}
We have $F_\lambda(0)=1$ and $F_\lambda'(0)=0$, by normalization and evenness. Cosine has the same initial values and solves $f''+f=0$. Consequently
\begin{equation}\label{eq:expanded-reference-integral}
F_\lambda(x)-\cos x
 =\int_0^x\sin(x-y)R_\lambda(y)\,dy
 \qquad(x\ge0).
\end{equation}
This formula can be checked directly by differentiating its right-hand side twice and checking its value and first derivative at zero. Uniqueness here is elementary: the difference $D$ between two proposed solutions satisfies $D''+D=0$ with zero initial values, so $(D^2+(D')^2)'=0$ forces $D=0$. For $0\le y\le x\le\pi/2$, the sine factor is nonnegative. Since $R_\lambda$ is positive, \eqref{eq:expanded-reference-integral} proves $F_\lambda(x)\ge\cos x$ on this half-interval. Evenness proves \eqref{eq:expanded-reference-majorant} on the entire central interval.

\paragraph{Step 4: make the error exponentially small.}
Use $Z_c\ge1/2$, $\cosh u\le e^{|u|}$, and $b_\lambda=\pi\lambda$ in the formula for $R_\lambda$. Completing a scalar square gives
\[
\begin{aligned}
R_\lambda(x)
 &\le\frac4{\sqrt\pi}\sqrt\lambda\,
       \exp\left(-\frac{\pi^2\lambda}{4}-\lambda x^2
                       +\pi\lambda|x|\right)\\
 &=\frac4{\sqrt\pi}\sqrt\lambda\,
       e^{-\lambda(\pi/2-|x|)^2}
 \le4\sqrt\lambda\,e^{-\lambda/4}=\delta_\lambda
 \qquad(|x|\le1).
\end{aligned}
\]
The last inequality uses $\pi/2-1>1/2$. Formula~\eqref{eq:expanded-reference-integral} now implies
\[
0\le F_\lambda(x)-\cos x
 \le\delta_\lambda\int_0^{|x|}\sin(|x|-y)\,dy
 =\delta_\lambda(1-\cos x)
 \qquad(|x|\le1).
\]
On this interval $\cos x\ge\cos1>1/2$, so $(1-\cos x)/\cos x\le1$. Divide by $\cos x$ and apply $\log(1+u)\le u$ for $u\ge0$:
\[
0\le\log\frac{F_\lambda(x)}{\cos x}
 \le\frac{F_\lambda(x)-\cos x}{\cos x}
 \le\delta_\lambda.
\]
This is \eqref{eq:expanded-reference-error}.
\end{proof}

The construction has two complementary features. Its density is everywhere nonnegative, so it can be used to define a genuine probability distribution. Its discrepancy from the desired cosine is controlled by an explicit boundary contribution, which is exponentially small in $\lambda$ on $[-1,1]$. We have not approximated cosine by a fixed-degree polynomial; the error estimate holds uniformly throughout a fixed interval.

\subsection{A joint covariance estimate for spins and auxiliary variables}
\label{sec:joint}

We now put the discrete covariance estimate and the positive cosine reference together. The product of reference functions will be the marginal weight of a larger probability distribution. Working with that larger distribution exposes a simple block matrix that controls both kinds of variables at once.

Let $r,s\ge1$ be integers and let $B\in\R^{r\times s}$ be a real matrix. Assume that every row and every column has squared Euclidean norm at most a positive number $\rho$. Fix $\lambda\ge2$ and suppose
\begin{equation}\label{eq:expanded-joint-cutoff}
4\lambda\rho\le\frac{\gamma}{\log(4r)},
\qquad \gamma=10^{-10}.
\end{equation}
There will be $r$ discrete spins and $s$ continuous auxiliary coordinates. The distinction matters: the logarithm in \eqref{eq:expanded-joint-cutoff} depends on the number $r$ of spins.

\subsubsection{Defining the joint law and identifying its marginal}

Start with independent random variables $\sigma\in\{\pm1\}^r$ and $T=(T_1,\ldots,T_s)^T\in\R^s$. The vector $\sigma$ is uniform, and each $T_j$ has law $\nu_\lambda$ from Section~\ref{sec:cosine-reference}. Write $Y=B^T\sigma\in\R^s$. We change the joint law by multiplying its mass and density by the positive weight
\[
\exp\bigl(-\lambda\norm{B^T\sigma}^2+\sigma^TBT\bigr).
\]
More precisely, define the joint mass--density function
$p_\mu:\{\pm1\}^r\times\R^s\to[0,\infty)$ by
\begin{equation}\label{eq:expanded-joint-law}
p_\mu(\sigma,t)
 =\frac{2^{-r}}{Z_F}\exp\bigl(-\lambda\norm{B^T\sigma}^2+\sigma^TBt\bigr)
       \prod_{j=1}^s q_\lambda(t_j),
\qquad t\in\R^s.
\end{equation}
Here $q_\lambda$ is the scalar density defined in the preceding section,
and $Z_F>0$ is chosen so that summing over $\sigma$ and integrating over
$t$ gives one. This defines the probability law $\mu$. The normalizer
is finite because the spin space is finite and the auxiliary variables
lie in a compact box.

Integrating out the independent auxiliary coordinates explains this choice. For a fixed spin vector and its associated $Y\in\R^s$,
\[
\begin{aligned}
\E_{T\sim\nu_\lambda^{\otimes s}}
       e^{-\lambda\norm Y^2+Y^TT}
 &=\prod_{j=1}^s\left(e^{-\lambda Y_j^2}
                           \E_{\nu_\lambda}e^{Y_jT_j}\right)
 =\prod_{j=1}^sF_\lambda(Y_j).
\end{aligned}
\]
Consequently
\begin{equation}\label{eq:expanded-reference-normalizer}
Z_F=\E_{\sigma\sim\Unif}\prod_{j=1}^sF_\lambda((B^T\sigma)_j),
\qquad
\mu_\sigma(\sigma)=\frac{2^{-r}}{Z_F}
                  \prod_{j=1}^sF_\lambda((B^T\sigma)_j).
\end{equation}
The second formula is the marginal probability mass function of $\sigma$: it is obtained by forgetting $T$ after sampling from $\mu$.

Set
\[
c=.93,\qquad
\kappa=2\lambda-H_\lambda^{-1}
      =\frac{2\lambda}{2\lambda+1}\ge\frac45>\frac34.
\]
Define a real symmetric $(r+s)$-by-$(r+s)$ matrix by
\[
\Pi=
\begin{pmatrix}
2\lambda BB^T+cI_r&-B\\
-B^T&H_\lambda I_s
\end{pmatrix}.
\]
Its upper-left block is $r$ by $r$, its lower-right block is $s$ by $s$, and the other two blocks are $r$ by $s$ and $s$ by $r$.

\begin{proposition}[Joint precision bound]\label{prop:joint}
The matrix $\Pi$ is positive definite. For every pair of fields $h_\sigma\in\R^r$ and $h_T\in\R^s$, tilt $\mu$ by the positive factor $e^{h_\sigma^T\sigma+h_T^TT}$ and renormalize. Under the resulting probability law, the covariance of the concatenated vector $(\sigma,T)\in\R^{r+s}$ is at most $\Pi^{-1}$ in positive-semidefinite order. In particular, the original law satisfies
\begin{equation}\label{eq:expanded-joint-mgf}
\E_\mu e^{h^T(\sigma,T)}
 \le e^{h^T\Pi^{-1}h/2}
 \qquad\text{for every }h\in\R^{r+s}.
\end{equation}
\end{proposition}

\begin{proof}
Fix arbitrary fields $h_\sigma\in\R^r$ and $h_T\in\R^s$. We first restrict $T$ to the closed box
\[
[-b_\lambda+\eta,b_\lambda-\eta]^s,
\qquad 0<\eta<b_\lambda,
\]
and renormalize. On this box the potential $U_\lambda$ is smooth with finite derivatives. Thus the box version of the Brascamp--Lieb inequality from the preliminaries applies directly. All bounds below will be independent of $\eta$; we remove the restriction at the end.

\paragraph{Step 1: condition on the continuous coordinates.}
For fixed $t\in\R^s$ in the box, the conditional mass function of $\sigma$ is proportional to
\[
\exp\left(-\frac12\sigma^TA\sigma+(h_\sigma+Bt)^T\sigma\right),
\qquad A:=2\lambda BB^T\in\R^{r\times r}.
\]
The matrix $A$ is positive semidefinite. Its $i$th diagonal entry is $2\lambda$ times the squared norm of the $i$th row of $B$, so
\[
2\max_{i\in[r]}A_{ii}\le4\lambda\rho
 \le\frac{\gamma}{\log(4r)}.
\]
Proposition~\ref{prop:all-field} therefore gives
\begin{equation}\label{eq:expanded-conditional-covariance}
C(t):=\Cov(\sigma\mid T=t)
 \preceq M:=(A+cI_r)^{-1}.
\end{equation}
Both $C(t)$ and $M$ are $r$-by-$r$ matrices. More explicitly, the finite spin formula defines $C:\R^s\to\R^{r\times r}$, with symmetric positive-semidefinite values. The estimate holds for every $t$ and every field $h_\sigma$; this is precisely why we needed the all-field result.

\paragraph{Step 2: compute the curvature of the continuous marginal.}
Sum over the spins. Up to a positive constant independent of $t$, the marginal density of $T$ on the box is
\[
\exp\left(-\sum_{j=1}^sU_\lambda(t_j)+h_T^Tt\right)
          Z_A(h_\sigma+Bt).
\]
Its negative logarithm, up to an additive constant, is the function $V:(-b_\lambda,b_\lambda)^s\to\R$ given by
\[
V(t)=\sum_{j=1}^sU_\lambda(t_j)-h_T^Tt-\log Z_A(h_\sigma+Bt).
\]
The tilting derivative identity from the preliminaries says that the Hessian of $\log Z_A$ is the spin covariance. The chain rule therefore gives the exact $s$-by-$s$ Hessian
\[
\nabla^2V(t)
 =\diag\bigl(U_\lambda''(t_1),\ldots,U_\lambda''(t_s)\bigr)
       -B^TC(t)B
 \succeq H_\lambda I_s-B^TC(t)B.
\]
The loss of curvature caused by summing over the spins is thus explicitly controlled by their covariance.

We next prove the matrix bound
\[
B^TMB\preceq\frac1{2\lambda}I_s.
\]
For clarity, this can be checked without assuming that $B$ is square
or invertible. Put $D_s=2\lambda B^TB+cI_s\in\R^{s\times s}$,
which is positive definite. The identity
$(2\lambda BB^T+cI_r)B=BD_s$ implies $MB=BD_s^{-1}$.
Consequently
\[
 B^TMB=B^TBD_s^{-1}
       =\frac1{2\lambda}(I_s-cD_s^{-1})
       \preceq\frac1{2\lambda}I_s.
\]
The last inequality holds because $cD_s^{-1}\succeq0$. It follows that
\begin{equation}\label{eq:expanded-curvature-gap}
H_\lambda I_s-B^TC(t)B
 \succeq H_\lambda I_s-B^TMB
 \succeq\frac1{4\lambda^2}I_s\succ0.
\end{equation}
This also proves $\Pi\succ0$: its upper-left block is $M^{-1}\succ0$, and its Schur complement is $H_\lambda I_s-B^TMB\succ0$.

\paragraph{Step 3: bound every linear combination jointly.}
Fix test vectors $u\in\R^r$ and $v\in\R^s$. Define the conditional mean function $m:\R^s\to\R^r$ by
\[
m(t)=\E(\sigma\mid T=t).
\]
The finite conditional spin formula defines this smooth function on all of $\R^s$, even though we currently integrate only over a box. Differentiation with respect to the field gives its Jacobian $C(t)B\in\R^{r\times s}$. Consequently
\[
\nabla_t\bigl(u^Tm(t)+v^Tt\bigr)=B^TC(t)u+v\in\R^s.
\]
The law of total variance separates the variation inside a conditional spin law from the variation of its conditional mean:
\[
\Var(u^T\sigma+v^TT)
 =\E[u^TC(T)u]+\Var_T\bigl(u^Tm(T)+v^TT\bigr).
\]
Apply the pointwise-Hessian form of Brascamp--Lieb to the second term. The inverse-order rule and \eqref{eq:expanded-curvature-gap} give
\[
\begin{aligned}
\Var(u^T\sigma+v^TT)\le\E\bigl[&u^TC(T)u\\
 &+(v+B^TC(T)u)^T(H_\lambda I_s-B^TC(T)B)^{-1}
                         (v+B^TC(T)u)\bigr].
\end{aligned}
\]
It would be unsafe to replace separate occurrences of $C(T)$ by $M$ without considering the whole expression. Instead, for any symmetric matrix $C\in\R^{r\times r}$ with $0\preceq C\preceq M$, the expression inside the expectation equals
\begin{equation}\label{eq:expanded-joint-variational}
\sup_{z\in\R^s}
 \left((u+Bz)^TC(u+Bz)+2v^Tz-H_\lambda\norm z^2\right).
\end{equation}
To verify this identity, expand the square. The terms depending on $z$ become
\[
2(v+B^TCu)^Tz-z^T(H_\lambda I_s-B^TCB)z,
\]
whose supremum is the inverse quadratic form appearing above.

For each fixed $z$, increasing $C$ in positive-semidefinite order increases \eqref{eq:expanded-joint-variational}. We can therefore replace $C$ by $M$ in the entire expression. The resulting supremum is
\[
u^TMu+(v+B^TMu)^T(H_\lambda I_s-B^TMB)^{-1}(v+B^TMu)
 =(u,v)^T\Pi^{-1}(u,v),
\]
by the block-inverse formula. We have proved the desired bound for every test vector $(u,v)\in\R^{r+s}$, and hence for the full covariance matrix.

\paragraph{Step 4: remove the temporary box restriction.}
As $\eta\downarrow0$, the boxes increase to $(-b_\lambda,b_\lambda)^s$. The original prior gives this open box probability one. Relative to the original product measure, the tilted joint weight is a continuous bounded function on the finite spin space times the compact auxiliary box. Every first and second coordinate moment is bounded there as well. Dominated convergence therefore passes the normalizing constants, means, and second moments to their limits. The full tilted normalizing constant is positive, so the covariance matrices converge. The same bound by $\Pi^{-1}$ survives in the limit. This proves the assertion for every pair of fields without imposing a boundary condition on the original density.

Finally, with both fields zero, the transformation $(\sigma,T)\mapsto(-\sigma,-T)$ preserves the joint law. Thus its mean is zero. Applying the preliminary integration-of-tilts argument to the covariance bound in every field proves \eqref{eq:expanded-joint-mgf}.
\end{proof}

\subsubsection{A useful consequence for each projected spin sum}

The upper-left block of $\Pi^{-1}$ is, by the Schur complement of its lower-right block,
\[
\bigl(2\lambda BB^T+cI_r-H_\lambda^{-1}BB^T\bigr)^{-1}
 =(cI_r+\kappa BB^T)^{-1}\preceq c^{-1}I_r.
\]
For each $j\in[s]$, let $b_j\in\R^r$ be the $j$th column of $B$. Then $Y_j=b_j^T\sigma$. Insert the source $(t b_j,0)\in\R^{r+s}$ into \eqref{eq:expanded-joint-mgf}, where $t\in\R$ is a scalar. We obtain
\[
\E_\mu e^{tY_j}
 \le\exp\left(\frac{t^2}2
          b_j^T(cI_r+\kappa BB^T)^{-1}b_j\right)
 \le e^{t^2\rho/(2c)}.
\]
Thus each $Y_j$ is centered sub-Gaussian with variance proxy at most $\rho/c$. The scalar tail bound and a union bound give
\begin{equation}\label{eq:expanded-joint-tail}
\Pp_\mu\left(\max_{j\in[s]}|Y_j|>1\right)
 \le2s e^{-c/(2\rho)}.
\end{equation}
The coordinates $Y_j$ need not be independent. The row-norm assumption on $B$ was used to control the conditional spin law; the column-norm assumption is what controls these individual projections.

\subsection{Controlling the other periods and proving uniform positivity}
\label{sec:uniform}

The joint covariance estimate says that the positive reference distribution usually keeps every projected spin sum $Y_j$ near zero. This alone would not justify discarding the other periods of cosine: the positive reference normalizer $Z_F$ can itself be small. We need to bound the unwanted contribution \emph{relative to $Z_F$}. The convex-set support bound and the lattice energy lemma provide exactly that comparison.

Throughout the first three subsections, retain the integers $r,s$, matrix $B\in\R^{r\times s}$, and parameters $\lambda,\rho,c,\kappa,\Pi$ from Section~\ref{sec:joint}. In particular, $Y=B^T\sigma\in\R^s$. Expectations with subscript $\sigma$ alone will refer to a uniform spin vector in $\{\pm1\}^r$, whereas expectations with subscript $\mu$ refer to the joint reference law. Recall also $\beta=9/10$ from the lattice lemma. The joint cutoff ensures that its small-variance condition is available:
\[
\frac{\rho}{\beta c}
 \le\frac{\gamma}{8\beta c\log4}<10^{-6}.
\]
Here we used $\lambda\ge2$ and $r\ge1$.

\subsubsection{Translate a periodic cell without changing the prior}

Let $Q_s=[-\pi/2,\pi/2]^s\subseteq\R^s$. For each integer vector $\ell\in\Z^s$, the translated box $\pi\ell+Q_s$ is a periodic cell. Define its nonnegative reference weight by
\[
J_\ell=\E_\sigma\left[
 \prod_{j=1}^sF_\lambda(Y_j-\pi\ell_j)
          \one_{\{Y\in\pi\ell+Q_s\}}\right].
\]
These weights define a function $J:\Z^s\to[0,\infty)$, with value $J_\ell$ at $\ell$. The scalar $J_\ell$ is a weighted mass of this cell, not its probability under the original uniform spin law.

Write $v=\pi\ell\in\R^s$. For each fixed spin vector, the integral representation of $F_\lambda$ gives
\[
\prod_{j=1}^sF_\lambda(Y_j-v_j)
 =\E_{T\sim\nu_\lambda^{\otimes s}}
        \exp\bigl(-\lambda\norm{Y-v}^2+(Y-v)^TT\bigr).
\]
Expand the exponent and compare it to the defining weight of $\mu$:
\[
-\lambda\norm{Y-v}^2+(Y-v)^TT
 =-\lambda\norm Y^2+Y^TT
     -\lambda\norm v^2+2\lambda v^TY-v^TT.
\]
Therefore
\begin{equation}\label{eq:expanded-cell-translation}
\frac{J_\ell}{Z_F}
 =e^{-\lambda\norm v^2}
   \E_\mu\left[
       e^{(2\lambda Bv,-v)^T(\sigma,T)}
       \one_{\{B^T\sigma\in v+Q_s\}}\right].
\end{equation}
The source vector $(2\lambda Bv,-v)$ lies in $\R^{r+s}$. Notice that we have translated the arguments of $F_\lambda$, but have not translated the auxiliary vector $T$. Its prior remains the same positive probability measure with the same normalizer.

\subsubsection{The joint precision turns translation into energy}

Define the closed convex set
\[
K_\ell=\{(z,t)\in\R^r\times\R^s:B^Tz\in v+Q_s\}
       \subseteq\R^{r+s}.
\]
If this set is empty, then the cell has zero mass and all bounds below are immediate. Otherwise apply Lemma~\ref{lem:support} to the random vector $(\sigma,T)\in\R^{r+s}$, with its moment-generating-function proxy $\Pi^{-1}$ from Proposition~\ref{prop:joint}. Equation~\eqref{eq:expanded-cell-translation} is at most the exponential of the supremum, over $(z,t)\in K_\ell$, of
\[
-\lambda\norm v^2+(2\lambda Bv,-v)^T(z,t)
                     -\frac12(z,t)^T\Pi(z,t).
\]
This supremum allows every $z\in\R^r$ and $t\in\R^s$ satisfying the displayed constraint. It thus enlarges both the discrete spin support and the compact auxiliary support, which is legitimate for an upper bound.

For such a point, write $B^Tz=v+x$, where $x\in Q_s$. Expanding the block quadratic form shows why the particular matrix $\Pi$ is useful:
\[
\begin{aligned}
&-\lambda\norm v^2+(2\lambda Bv,-v)^T(z,t)
                     -\frac12(z,t)^T\Pi(z,t)\\
&=-\lambda\norm v^2+2\lambda v^TB^Tz-\lambda\norm{B^Tz}^2
       -\frac c2\norm z^2+(B^Tz-v)^Tt-\frac{H_\lambda}{2}\norm t^2\\
&=-\frac c2\norm z^2-\lambda\norm x^2
                  +x^Tt-\frac{H_\lambda}{2}\norm t^2.
\end{aligned}
\]
The large translated terms have combined into $-\lambda\norm x^2$. For each fixed $x$, completing the square in $t$ gives
\[
x^Tt-\frac{H_\lambda}{2}\norm t^2
 =\frac{\norm x^2}{2H_\lambda}
       -\frac{H_\lambda}{2}\norm{t-x/H_\lambda}^2.
\]
Its maximum occurs at $t=x/H_\lambda$. Because $\kappa=2\lambda-H_\lambda^{-1}$, we conclude
\begin{equation}\label{eq:expanded-cell-energy}
J_\ell\le Z_F e^{-E_\ell},
\qquad
E_\ell=\inf_{\substack{z\in\R^r,\ x\in Q_s\\B^Tz=\pi\ell+x}}
       \left(\frac c2\norm z^2+\frac\kappa2\norm x^2\right).
\end{equation}
For an infeasible constraint, define $E_\ell=+\infty$ and $e^{-E_\ell}=0$. Thus the energy function has domain $\Z^s$ and range $[0,\infty]$.

To recognize the lattice lemma, change variables to $u=\sqrt c\,z\in\R^r$. Then the constraint and energy become
\[
(B/\sqrt c)^Tu=\pi\ell+x,
\qquad \frac12\bigl(\norm u^2+\kappa\norm x^2\bigr).
\]
Each column of $B/\sqrt c\in\R^{r\times s}$ has squared norm at most $\rho/c$, and $\kappa>3/4$. Lemma~\ref{lem:lattice}, with $d=s$ and $q=\rho/c$, therefore gives
\begin{equation}\label{eq:expanded-relative-cell-tail}
\sum_{\ell\in\Z^s\setminus\{0\}}J_\ell
 \le Z_F\varepsilon_s(\rho/c),
\qquad
\varepsilon_s(q)=
 \exp\left(-\frac{\pi^2}{80q}+16s e^{-10^{-5}\beta/q}\right).
\end{equation}
This is the required relative estimate. It does not assume that the coordinates of $Y$ are independent. It also does not require $B^TB$ to be invertible: the minimization retains the original variable $z\in\R^r$ and works equally well for a singular matrix $B$.

\subsubsection{Separate the central positive contribution from the tail}

Let $A$ be the event $\max_{j\in[s]}|Y_j|\le1$. On $A$, every cosine is positive, and the reference error bound gives
\[
\prod_{j=1}^s\cos Y_j
 \ge e^{-s\delta_\lambda}\prod_{j=1}^sF_\lambda(Y_j).
\]
Using the spin marginal in \eqref{eq:expanded-reference-normalizer}, followed by \eqref{eq:expanded-joint-tail}, we obtain
\[
\begin{aligned}
\E_\sigma\left[\prod_{j=1}^s\cos Y_j\,\one_A\right]
 &\ge e^{-s\delta_\lambda}Z_F\Pp_\mu(A)\\
 &\ge Z_F(1-2s e^{-c/(2\rho)})e^{-s\delta_\lambda}.
\end{aligned}
\]
The part of the central cell $Q_s$ outside $A$ contributes nonnegatively, because all of its individual cosine factors are nonnegative.

For a noncentral cell, put $x=Y-\pi\ell\in Q_s$. The identity $\cos(Y_j)=(-1)^{\ell_j}\cos x_j$ gives
\[
\left|\prod_{j=1}^s\cos Y_j\right|
 =\prod_{j=1}^s\cos x_j
 \le\prod_{j=1}^sF_\lambda(x_j).
\]
Thus the absolute contribution of this cell is at most $J_\ell$. We may partition space into half-open cells to avoid counting boundary points twice. Using the closed cells in the upper bounds can only increase the sum of the nonnegative $J_\ell$. Alternatively, on every cell boundary at least one cosine factor is zero, so boundaries contribute nothing to the original cosine product.

Subtracting the relative tail bound from the central contribution proves
\begin{equation}\label{eq:expanded-cosine-bracket}
\E_\sigma\prod_{j=1}^s\cos Y_j
 \ge Z_F\left[(1-2s e^{-c/(2\rho)})e^{-s\delta_\lambda}
                       -\varepsilon_s(\rho/c)\right].
\end{equation}
We also need a numerical lower bound on $Z_F$. The bound $F_\lambda(x)\ge e^{-\lambda x^2}$ and Jensen's inequality give
\begin{equation}\label{eq:expanded-Z-lower}
Z_F\ge\E_\sigma e^{-\lambda\norm{B^T\sigma}^2}
    \ge e^{-\lambda\E_\sigma\norm{B^T\sigma}^2}
    =e^{-\lambda\norm{B}_F^2}.
\end{equation}
In this formula $\norm{B}_F^2$ denotes the squared Frobenius norm $\sum_{i=1}^r\sum_{j=1}^sB_{ij}^2$. The last equality follows by expanding the square and using $\E\sigma_i\sigma_{i'}=0$ for $i\ne i'$ and $\E\sigma_i^2=1$.
For the middle inequality, only the elementary exponential form of
Jensen's inequality is needed. If $U$ is any integrable real random
variable with finite exponential moment, then
$e^u\ge e^m(1+u-m)$ with $m=\E U$ gives
$\E e^U\ge e^{\E U}$ after taking expectations. Here
$U=-\lambda\norm{B^T\sigma}^2$ is bounded because the spin space is finite.

\subsubsection{Choose the parameters and check every error term}

\begin{proof}[Proof of Theorem~\ref{thm:uniform}]
Fix an integer $k\ge1$ and a sign matrix $S\in\{\pm1\}^{k\times k}$. Put
\[
a=10^{-7},\qquad L_k=\log(4k)>1,\qquad
\theta=\frac{a}{\sqrt k\,L_k},\qquad
\lambda=64L_k,\qquad B=\theta S\in\R^{k\times k}.
\]
We apply the preceding estimates with $r=s=k$. Every row and every column of $B$ has squared norm
\[
\rho=k\theta^2=\frac{a^2}{L_k^2}.
\]
First verify the hypotheses. We have $\lambda>64>2$, and
\[
4\lambda\rho=\frac{256a^2}{L_k}
 \le\frac{\gamma}{\log(4k)},
\qquad 256a^2=2.56\cdot10^{-12}<10^{-10}=\gamma.
\]
Thus the conditional spin covariance cutoff holds in dimension $k$. Also $\rho/(\beta c)<2\cdot10^{-14}<10^{-6}$, as required for the lattice estimate.

We now check the three error terms in \eqref{eq:expanded-cosine-bracket}. Each of the constants
\[
\frac{c}{2a^2},\qquad
\frac{c\pi^2}{80a^2},\qquad
\frac{10^{-5}\beta c}{a^2}
\]
is greater than $2$. Since $L_k^2\ge L_k$, the first gives
\[
2k e^{-c/(2\rho)}
 \le2k e^{-2L_k}=\frac1{8k}\le\frac14.
\]
For the reference error, substitute $\lambda=64L_k$ into $\delta_\lambda=4\sqrt\lambda e^{-\lambda/4}$. Because $k=e^{L_k}/4$,
\[
k\delta_\lambda
 =8\sqrt{L_k}\,e^{-15L_k}
 \le8e^{-14L_k}\le2^{-25}<\log(3/2).
\]
Here $\sqrt{L_k}\le e^{L_k}$ and $L_k\ge\log4$ were used. Consequently $e^{-k\delta_\lambda}\ge2/3$.
For an elementary numerical comparison,
$\log(3/2)=\int_1^{3/2}t^{-1}\,dt\ge1/3>2^{-25}$.

Finally, the second and third constants above bound the lattice error by
\[
\begin{aligned}
\log\varepsilon_k(\rho/c)
 &=-\frac{c\pi^2L_k^2}{80a^2}
      +16k\exp\left(-\frac{10^{-5}\beta cL_k^2}{a^2}\right)\\
 &\le-2L_k+16k e^{-2L_k}
  =-2L_k+\frac1k\le-2L_k+1.
\end{aligned}
\]
Therefore
\[
\varepsilon_k(\rho/c)\le\frac{e}{16k^2}
                         \le\frac1{4k}\le\frac14.
\]
The middle inequality uses $e<4$ and $k\ge1$.
The bracket in \eqref{eq:expanded-cosine-bracket} is thus at least
\[
\left(1-\frac14\right)\frac23-\frac14=\frac14.
\]
Since the squared Frobenius norm of $B$ is $k\rho=a^2k/L_k^2$, the normalizer bound gives
\[
\E_\sigma\prod_{j=1}^k\cos((B^T\sigma)_j)
 \ge\frac14Z_F
 \ge\frac14\exp\left(-\frac{64a^2k}{L_k}\right)>0.
\]

It remains to identify this with the characteristic function stated in the theorem. Let $\sigma,\tau\in\{\pm1\}^k$ be independent uniform spin vectors. Conditional on $\sigma$, the coordinates of $\tau$ are independent, so
\[
\begin{aligned}
\E_\tau e^{i\theta\sigma^TS\tau}
 &=\prod_{j=1}^k\E_{\tau_j}
         e^{i\tau_j(B^T\sigma)_j}\\
 &=\prod_{j=1}^k
       \frac{e^{i(B^T\sigma)_j}+e^{-i(B^T\sigma)_j}}2
 =\prod_{j=1}^k\cos((B^T\sigma)_j).
\end{aligned}
\]
Averaging over $\sigma$ proves the required bound for $\phi_S(\theta)$, with $L_k=\log(4k)$. In particular the characteristic function is real and strictly positive. Every estimate used only the row and column norms of $B$, so the result holds for every sign matrix $S$, with no averaging or genericity assumption on its signs.
\end{proof}

\section{Explanation of \cref{fact:subgaussian}}\label{sec:appendixSubgaussian}

Let $B_1^n=\{x\in\R^n:\|x\|_1\leq1\}$. The rows of $A$ are independent uniform sign vectors. Their distribution is isotropic, since $\E\langle a,x\rangle^2=\|x\|_2^2$, and is subgaussian with an absolute constant: for every $\lambda\in\R$,
\[
\E e^{\lambda\langle a,x\rangle}
=\prod_{i=1}^n\cosh(\lambda x_i)
\leq e^{\lambda^2\|x\|_2^2/2}.
\]
Thus, the ``moreover'' part of Corollary 2.5 in \cite{mendelson2005reconstruction}, applied with $T=B_1^n$, $\Gamma=A$, and $k=D$, gives absolute constants $C,c>0$ such that, with probability at least $1-e^{-cD}$,
\[
\operatorname{diam}(\ker A\cap B_1^n)
\leq \frac{C\ell_*(B_1^n)}{\sqrt D}.
\]
Here the diameter is Euclidean, and, for independent standard Gaussian variables $g_1,\dots,g_n$,
\[
\ell_*(B_1^n)=\E\sup_{x\in B_1^n}|\langle g,x\rangle|
=\E\max_{i\in[n]}|g_i|
\leq\sqrt{2\log(2n)}.
\]
The last bound follows from $\E e^{\lambda\max_i|g_i|}\leq2n e^{\lambda^2/2}$ and Jensen's inequality, with $\lambda=\sqrt{2\log(2n)}$.

On this event, every nonzero $v\in\ker A$ satisfies $v/\|v\|_1\in\ker A\cap B_1^n$. Since $0$ also belongs to this set,
\[
\frac{\|v\|_2}{\|v\|_1}
\leq\operatorname{diam}(\ker A\cap B_1^n)
\leq C\sqrt{\frac{2\log(2n)}{D}}.
\]
The same conclusion is immediate for $v=0$. Taking $C_0=\sqrt2 C$ and $c_0=c$ proves \cref{fact:subgaussian} simultaneously for all $v\in\ker A$.

\section{A Uniform Decoder for Streaming Connectivity}\label{sec:uniformDecoder}

In this section, we detail a \emph{uniform} version of the streaming decoder for connectivity. 

In the proof of \cref{thm:streaming}, we showed that the information stored by the streaming algorithm uniquely determines the connected components of the final graph $G$. We now give an \emph{explicit} decoding procedure which uses only $O(n\log n)$ additional bits of working space beyond that stored sketch. In particular, we do not enumerate candidate graphs consistent with the sketch, since storing the adjacency matrix of such a candidate graph would require $\Theta(n^2)$ bits. 

We now briefly recall the notation from \cref{sec:streaming}. We denote the adjacency matrix of the final graph $G$ by $A_G\in\{0,1\}^{n\times n}$, and partition the vertex set $V$ into $W$ groups $C_1,\ldots,C_W$. The algorithm stores, for every vertex $v\in V$ and group $C_j$, the number of edges $|E_G[v,C_j]|$, along with the sparse-recovery sketch $P\cdot\Gamma(v)$ from \cref{clm:sparseRecoverySketch} with parameters $k=2W$ and $q=n$. Furthermore, the stored buffers and compressed sketches define the function $f$ from the proof of \cref{thm:streaming}, which satisfies
\begin{equation}\label{eq:stream-decoding-error}
  \left|f(T)-|E_G[T]|\right|
  \le 2\cdot\frac{n^{1+\varepsilon}}{W}\,\tau\cdot |T|
  \qquad\text{for every }T\subseteq V.
\end{equation}
The value $f(T)$ can be computed directly from the retained buffers and the stored functions $f_i^+,f_i^-$, without reconstructing $G$.

We use the same choice
\[
  W:=3n^{2/3+\varepsilon/3}\tau^{1/3}
\]
as in the proof of \cref{thm:streaming}, with the number of groups rounded up to an
integer. As in that proof, if $W>n/2$ then we store the entire graph, so we
may assume $W\le n/2$. We take the groups to be balanced, in which case
\begin{equation}\label{eq:stream-decoding-group-size}
  |C_j|\le\left\lceil\frac nW\right\rceil
  \le\frac{3n}{2W},
  \qquad
  W^3\ge27n^{2+\varepsilon}\tau.
\end{equation}
The slightly stronger constants in \eqref{eq:stream-decoding-group-size}
allow us to account for integer rounding, otherwise they play no role. The key claim we will show is that \emph{for any cut}, we can determine (in small additional space) whether the cut is empty in the resulting graph $G$ from the stream:

\begin{claim}\label{claim:stream-decoding-cut}
Given the information stored in the proof of \cref{thm:streaming}, and the fixed matrix $P$ used to define the sparse-recovery sketches, one can determine, for any $S\subseteq V$, whether $\delta_G(S)=0$, using only $O(n+W+\log n)$ additional bits of working space.
\end{claim}

\begin{proof}
Fix an arbitrary set $S\subseteq V$. Let $J_S\in\{0,1\}^{n\times n}$ be
the diagonal matrix with $(J_S)_{v,v}=1$ if and only if $v\in S$, as in \cref{sec:n74}. For each group $C_j$, define the vector
\begin{equation}\label{eq:stream-decoding-vector}
  d^{(j)}:=J_SA_G\mathbf{1}_{C_j}
            -A_G\mathbf{1}_{S\cap C_j}\in\mathbb Z^n.
\end{equation}
Crucially, this vector satisfies the same coordinate
formula for the edges crossing the cut:
\begin{equation}\label{eq:stream-decoding-coordinates}
  d^{(j)}_u=
  \begin{cases}
     |E_G[u,C_j\cap\overline S]|, & u\in S,\\
     -|E_G[u,C_j\cap S]|, & u\in\overline S.
  \end{cases}
\end{equation}
In particular, just as in \cref{sec:n18}, every coordinate has absolute value at most $|C_j|\le n$.

Although the decoder does not know the vector $d^{(j)}$, it can compute
$P\cdot d^{(j)}$ using only the stored information. Indeed, since
$A_G\mathbf{1}_{C_j}$ is the vector whose $u$th coordinate is
$|E_G[u,C_j]|$, and since $A_G\mathbf{1}_{\{v\}}=\Gamma(v)$, we have
\begin{align}
  P\cdot d^{(j)}
   &=P J_S A_G\mathbf{1}_{C_j}
     -P A_G\mathbf{1}_{S\cap C_j}\notag\\
   &=\sum_{u\in S}|E_G[u,C_j]|\,P\mathbf{1}_{\{u\}}
     -\sum_{v\in S\cap C_j}P\cdot\Gamma(v).
     \label{eq:stream-decoding-compute}
\end{align}
The right-hand side uses only the stored values $|E_G[u,C_j]|$, the stored
sketches $P\cdot\Gamma(v)$, and the fixed columns of $P$. Thus, the decoder
can determine whether $P\cdot d^{(j)}=0$ for every $j\in[W]$ without
constructing any of the vectors $d^{(j)}$.

Our decoder for the cut $(S,\overline S)$ is the following:
\begin{enumerate}
  \item For every $j\in[W]$, compute $P\cdot d^{(j)}$ using
  \eqref{eq:stream-decoding-compute}. If any of these sketches is nonzero,
  report that $\delta_G(S)>0$.
  \item Otherwise, enumerate every set $U$ which is a union of groups
  $C_j$. For each such $U$, let $X=S\cap U$ and $Y=\overline S\cap U$.
  If
  \begin{equation}\label{eq:stream-decoding-test}
    f(U)-f(X)-f(Y)
      >4\cdot\frac{n^{1+\varepsilon}}{W}\,\tau\cdot |U|,
  \end{equation}
  report that $\delta_G(S)>0$.
  \item If neither of the preceding steps succeeds, report that
  $\delta_G(S)=0$.
\end{enumerate}
We prove that this decoder is correct.

First, suppose that $\delta_G(S)=0$. By
\eqref{eq:stream-decoding-coordinates}, all the vectors $d^{(j)}$ are zero,
so the first step never reports a nonempty cut. Moreover, for every union
$U$ of groups, with $X=S\cap U$ and $Y=\overline S\cap U$, it must be that
$|E_G[X,Y]|=0$. Because
\[
  |E_G[U]|=|E_G[X]|+|E_G[Y]|+|E_G[X,Y]|,
\]
the accuracy guarantee \eqref{eq:stream-decoding-error} gives
\begin{align*}
  \bigl|f(U)-f(X)-f(Y)\bigr|
  &\le2\cdot\frac{n^{1+\varepsilon}}W\,\tau
    \bigl(|U|+|X|+|Y|\bigr)\\
  &=4\cdot\frac{n^{1+\varepsilon}}W\,\tau\cdot|U|.
\end{align*}
Thus \eqref{eq:stream-decoding-test} never holds, and the decoder correctly
reports that the cut is empty.

Conversely, suppose that $\delta_G(S)=b>0$. If the first step finds a
nonzero sketch $P\cdot d^{(j)}$, then the cut must be nonempty, by
\eqref{eq:stream-decoding-coordinates}, so its answer is correct. We may
therefore assume that
\begin{equation}\label{eq:stream-decoding-zero}
  P\cdot d^{(j)}=0\qquad\text{for every }j\in[W].
\end{equation}
As in \cref{clm:blueEdgeVector}, we call a group $C_j$ \emph{active} if $d^{(j)}\ne0$.
Let $B_j$ denote the number of incidences of edges crossing
$(S,\overline S)$ at vertices of $C_j$, counting an edge twice when both of
its endpoints lie in $C_j$. By
\eqref{eq:stream-decoding-coordinates}, we have
\[
  B_j=\|d^{(j)}\|_1,
  \qquad 2b=\sum_{j=1}^W B_j.
\]
By \cref{clm:sparseRecoverySketch}, \eqref{eq:stream-decoding-zero}, and the fact that the
coordinates of $d^{(j)}$ have magnitude at most $n$, every active group
satisfies $|\operatorname{Supp}(d^{(j)})|>2W$, and hence $B_j>2W$.

Let $g$ be the number of active groups, and let $U$ be their union. Since
$b>0$, we have $g\ge1$, and the preceding bounds give
\begin{equation}\label{eq:stream-decoding-active}
  2b=\sum_{j=1}^W B_j>2Wg,
  \qquad\text{so}\qquad b>Wg.
\end{equation}
Also, by \eqref{eq:stream-decoding-group-size},
\begin{equation}\label{eq:stream-decoding-U}
  |U|\le\frac{3n}{2W}\,g.
\end{equation}
Crucially, both endpoints of every edge crossing $(S,\overline S)$ lie in
active groups. Indeed, if $u\in S$, $v\in\overline S$, and $(u,v)$ is an
edge with $u\in C_j$, then \eqref{eq:stream-decoding-coordinates} implies
$d^{(j)}_v<0$, so $C_j$ is active. If we repeat this same argument with $u$ and $v$
interchanged, we can also see that the group containing $v$ is active. Consequently,
if we let $X=S\cap U$ and $Y=\overline S\cap U$, then
\begin{equation}\label{eq:stream-decoding-all-crossing}
  |E_G[X,Y]|=b.
\end{equation}

We now show that this particular set $U$ passes the test
\eqref{eq:stream-decoding-test}. From
\eqref{eq:stream-decoding-U} and our choice of $W$, we have
\[
  8\cdot\frac{n^{1+\varepsilon}}{W}\,\tau\cdot|U|
   \le\frac{12n^{2+\varepsilon}\tau}{W^2}\,g
   <Wg<b.
\]
Here the strict inequality uses $W^3\ge27n^{2+\varepsilon}\tau$.
On the other hand, by \eqref{eq:stream-decoding-error} and
\eqref{eq:stream-decoding-all-crossing},
\begin{align*}
  f(U)-f(X)-f(Y)
  &\ge |E_G[X,Y]|
       -4\cdot\frac{n^{1+\varepsilon}}W\,\tau\cdot|U|\\
  &>4\cdot\frac{n^{1+\varepsilon}}W\,\tau\cdot|U|.
\end{align*}
Thus $U$ satisfies \eqref{eq:stream-decoding-test}, and so the second step
reports that the cut is nonempty. Since the decoder enumerates all unions
of groups, it will examine this set $U$ even though it \emph{does not know} $U$ in advance or know which groups are active.

Finally, we bound the additional working space. The cut $S$ can be
represented by $n$ bits, and each union $U$ of groups by $W$ bits (simply specifying which subset of the $W$ groups to include).
For \eqref{eq:stream-decoding-compute}, we can compute one coordinate of
$P\cdot d^{(j)}$ at a time, using only an index and an integer
accumulator. All entries of $P$, all stored counts, and all intermediate
sums have magnitude $\operatorname{poly}(n)$ and thus require only
$O(\log n)$ bits each. Likewise, each evaluation of $f(T)$ can be
performed by reading the compressed sketches and scaling up the number of edges inside the set $T$,
using $O(\log n)$ additional bits for the relevant counters. The total additional
working space is therefore $O(n+W+\log n)$ bits. We can also store the matrix $P$ itself \emph{explicitly} in just $\widetilde O(nW)$ bits, within the space
bound already allowed in \cref{thm:streaming}.
\end{proof}

We can now recover the full connected-component partition using the cut
decoder.

\begin{claim}\label{claim:stream-decoding-components}
The stored information from \cref{thm:streaming} suffices to report the connected
components of $G$ using $O(n\log n)$ additional bits of working space.
Consequently the space bound in \cref{thm:streaming} remains
$\widetilde O(n^{15/8+3\varepsilon/8})$ bits, including the decoding
procedure.
\end{claim}

\begin{proof}
For any two vertices $u,v\in V$, we claim that they belong to the same
connected component of $G$ if and only if there is no set $S\subseteq V$
with $u\in S$, $v\notin S$, and $\delta_G(S)=0$. Indeed, if there is a
path from $u$ to $v$, then every cut separating them is crossed by at
least one edge of that path. Conversely, if there is no path, the
connected component containing $u$ determines an empty cut which
separates them.

Thus, for each pair $u,v$, we can decide whether they are in the same
connected component by enumerating all sets $S\subseteq V$ which
separate them and invoking Claim~\ref{claim:stream-decoding-cut}. To
produce the entire partition, for every vertex $v$ we find the smallest
vertex $u$ which belongs to the same connected component as $v$,
and output $u$ as the label of $v$. We need only store the current pair
$u,v$, an $n$-bit representation of $S$, and, if we want to retain all
the labels before outputting them, $O(n\log n)$ bits for the labels.
By Claim~\ref{claim:stream-decoding-cut}, every cut test reuses only
$O(n+W+\log n)$ additional workspace. Hence the total additional
space is $O(n\log n)$ bits.

Finally, the streaming state already uses
$\widetilde O(n^{15/8+3\varepsilon/8})$ bits by \cref{thm:streaming}, and the
additional decoding workspace is smaller than this bound. This
concludes the proof.
\end{proof}

\end{document}